\documentclass{aa}  

\usepackage{graphicx}
\usepackage{txfonts}
\usepackage{xcolor} 
\newcommand{\hl}[1]{{\textcolor{black}{#1}}}
\newcommand{\edt}[1]{{\textcolor{black}{#1}}}
\newcommand{\edttwo}[1]{{\textcolor{black}{#1}}}
\usepackage{bm}
\usepackage{subcaption}
\usepackage{threeparttable}
\usepackage{alphalph}
\usepackage{float}
\begin{document}

   \title{Implementation of disequilibrium chemistry to spectral retrieval code ARCiS and application to \edt{sixteen exoplanet} transmission spectra}
   \titlerunning{Implementation of disequilibrium chemistry to ARCiS and application to 16 exoplanet spectra}

   \subtitle{\hl{\edt{Indication of} disequilibrium chemistry for HD~209458b and WASP-39b}}

   \author{Yui Kawashima
          \inst{1,2}
          \and
          Michiel Min\inst{2}
          }

   \institute{Cluster for Pioneering Research, RIKEN, 2-1 Hirosawa, Wako, Saitama 351-0198, Japan\\
              \email{yui.kawashima@riken.jp}
   \and
   SRON Netherlands Institute for Space Research, Sorbonnelaan 2, 3584 CA Utrecht, The Netherlands\\
             }

   \date{Received September 15, 1996; accepted March 16, 1997}

% \abstract{}{}{}{}{} 
% 5 {} token are mandatory
 
  \abstract %(up to 300 words)
   {The retrieval approach is currently a standard method for %when 
   deriving atmospheric properties from observed spectra of exoplanets. However, the approach ignores disequilibrium chemistry in most current retrieval codes, which can lead to misinterpretation of the metallicity or elemental abundance ratios of the atmosphere.}
   {We have implemented the disequilibrium effect of vertical mixing or quenching for the major species in hydrogen/helium-dominated atmospheres, namely $\mathrm{CH_4}$, $\mathrm{CO}$, $\mathrm{H_2O}$, $\mathrm{NH_3}$, $\mathrm{N_2}$, and $\mathrm{CO_2}$, for the spectral retrieval code ARCiS %\citep{2020A&A...642A..28M} 
   with a physical basis.}
   {We used the chemical relaxation method %with the chemical timescale derived 
   %by \citet{2018ApJ...862...31T} 
   and developed a module to compute the profiles of molecular abundances taking the disequilibrium effect into account. Then, using ARCiS updated with this module, we have performed retrievals of the observed transmission spectra of 16 exoplanets with sizes ranging from Jupiter to mini-Neptune.}
   {We find \edt{indications of} disequilibrium chemistry for HD~209458b \hl{($\geq 4.1\sigma$)} and WASP-39b \hl{($\geq 2.7\sigma$)}.
   %For HD~209458b, we retrieve the value of the eddy diffusion coefficient $K_{zz}^\mathrm{Ch}$ used in the chemistry calculation as large as $\log_{10}{\left( K_{zz}^\mathrm{Ch}~[\mathrm{cm}^2~\mathrm{s}^{-1}] \right)} = 13.21^{+1.32}_{-3.79}$ from our standard retrieval while $\log_{10}{\left( K_{zz}~[\mathrm{cm}^2~\mathrm{s}^{-1}] \right)} = 4.17^{+1.15}_{-1.05}$ when assuming the same $K_{zz}$ value for both chemistry and cloud structure calculations.
   %Despite this difference, also due to the different thermal structure, in both cases, disequilibrium chemistry plays an important role in determining the molecular abundance profiles in its atmosphere.
   The retrieved spectrum of HD~209458b exhibits a strong $\mathrm{NH_3}$ absorption feature at 10.5~$\mu$m accessible by {JWST} owing to an enhanced abundance of $\mathrm{NH_3}$ due to the quenching effect. This feature is absent in the spectrum retrieved assuming equilibrium chemistry, which makes HD~209458b an ideal target for studying disequilibrium chemistry in exoplanet atmospheres.
   Moreover, for HAT-P-11b \edt{and GJ~436b}, we obtain relatively different results than for the retrieval with the equilibrium assumption, such as a $2.9\sigma$ difference for the C/O ratio.
   We have also examined the retrieved eddy diffusion coefficient, but could not identify a trend over the equilibrium temperature, possibly due to the limits of the current observational precision.
   %This demonstrates the importance of taking the quenching effect into account for the spectral retrieval 
   }
   {We have demonstrated that the assumption of equilibrium chemistry can lead to a misinterpretation of the observed data, showing that spectral retrieval with a consideration of disequilibrium chemistry is essential in the era of {JWST} and Ariel.}


   \maketitle
%
%-------------------------------------------------------------------

\section{Introduction} \label{sec:intro}
Since the first discovery of an exoplanet in 1995 \citep{1995Natur.378..355M}, atmospheric spectra have been observed for dozens of exoplanets via transmission, emission, and direct imaging spectroscopy by both space- and ground-based telescopes. From these spectra,
%This has lead to the 
%constraints on 
the abundances of several chemical species have been determined for some of the observed planets, most of which are expected to be abundant in hydrogen/helium-dominated atmospheres such as $\mathrm{H_2O}$, $\mathrm{CH_4}$, $\mathrm{CO}$, and $\mathrm{CO_2}$
%, for some of those observed planets 
\citep[e.g.,][]{2014ApJ...783...70L}.
%Also, recent observations has revealed detailed properties of clouds \citep[e.g.,][]{2018ApJ...859...34O, 2018ApJ...860...18P, 2018ApJ...863..165G, 2019A&A...622A.121O, 2019A&A...631A..79H, 2019ApJ...887..170P} and haze \citep[e.g.,][]{2017ApJ...847...32L, 2018ApJ...853....7K, 2019ApJ...874...61A, 2019ApJ...876L...5K, 2019ApJ...877..109K, 2019ApJ...878..118L}.

Recently, spectral retrieval models have come 
%into use 
to be used 
as a standard when deriving the atmospheric properties from observed spectra \citep[e.g.,][]{2008JQSRT.109.1136I, 2012ApJ...753..100B, 2013ApJ...775..137L, 2015ApJ...802..107W, 2018MNRAS.481.4698F, 2019A&A...627A..67M, 2020A&A...642A..28M}. Since the computational cost of the spectral retrieval codes is quite high, some assumptions and/or simplifications have to be made for atmospheric chemistry. 
For simplicity, most of the current spectral retrieval models assume constant abundances of chemical species throughout the atmosphere or their chemical equilibrium abundances, ignoring any disequilibrium chemistry.
%full kinetic chemistry, namely thermo- and photo-chemical reactions, and transport due to atmospheric circulation, cannot be included. Instead, 

In reality, however, the abundances of chemical species in the atmospheres cannot always be determined from the equilibrium chemistry, as has been studied by various 1D (photo-)chemical models \citep[e.g.,][]{1985JGR....9010497K, 2011ApJ...737...15M, 2012A&A...546A..43V, 2012ApJ...761..166H, 2014MNRAS.439.2386G, 2016ApJS..224....9R, 2017ApJS..228...20T, 2018ApJ...853....7K} and 2D or 3D atmospheric circulation models with chemistry \citep[e.g.,][]{2006ApJ...649.1048C, 2012A&A...548A..73A, 2014A&A...564A..73A, 2018ApJ...855L..31D, 2018ApJ...869...28D, 2018ApJ...869..107M, 2020A&A...636A..68D}.
Among the chemical disequilibrium processes such as photochemistry, the vertical quenching effect due to eddy diffusion transport is regarded as one of the most important %ones. This is 
because it significantly affects the atmospheric region where we can probe via observation.
Indeed, \citet{2021A&A...648A.127B} recently analyzed the transit depths of 49 gas giants measured at the 3.6 and 4.5~$\mu$m bands of {\textit{Spitzer}} and found evidence of disequilibrium chemistry, which can be explained by the quenching effect of $\mathrm{CH_4}$.
Quenching happens at altitudes where the thermochemical reaction timescale $\tau_\mathrm{chem}$ is equal to that of the eddy diffusion $\tau_\mathrm{diff}$.
Below this altitude, where $\tau_\mathrm{chem} < \tau_\mathrm{diff}$, the volume mixing ratio of a chemical species is consistent with that in thermochemical equilibrium due to high temperature and number density.
On the other hand, above that altitude, where $\tau_\mathrm{diff} < \tau_\mathrm{chem}$, the abundance is ``frozen'' to the value at the quenching altitude since vertical transport due to eddy diffusion tends to smooth out the gradient of the volume mixing ratio.
Thus, deriving the metallicity or elemental abundance ratios of the atmosphere from those frozen molecular abundances without considering the quenching effect can lead to over- or under-estimates of the metallicity or elemental abundance ratios.
Because of the longer chemical timescale at lower temperatures, this quenching effect is especially important for cooler ($\lesssim 1000~$K) atmospheres, which are the primary targets for upcoming characterizations of exoplanet atmospheres.

\citet{2017AJ....153...86M} introduced a quenching pressure as a parameter for spectral retrieval, assuming the same quenching pressures for all species. This treatment of the quenching effect was also adopted in \citet{2020A&A...640A.131M}.
%while they let the eddy diffusion coefficient used in their cloud simulation retrieved. 
However, the quenching pressure should be different for each species because the chemical timescale is different for each species, especially between $\mathrm{CH_4}$/CO and $\mathrm{NH_3}$/$\mathrm{N_2}$ \citep[e.g.,][]{2014RSPTA.37230073M}.
\edt{Also, as a more general approach for capturing discontinuities in the abundance profiles,
%, which could be 
caused not only by vertical mixing but also by photodissociation and the formation of clouds and haze, \citet{2019ApJ...886...39C} recently proposed a ``two-layer'' retrieval approach that allows the atmosphere to have different constant abundances in the upper and lower atmospheres.}

Predicting the quenching pressure for each species with a physical basis involves calculating chemical reactions under the effect of eddy diffusion transport.
It is, however, unrealistic to couple full kinetic chemistry to spectral retrieval codes even for 1D modeling due to its high computational cost.
%They applied their model to the forward spectrum model calculated with the chemical profiles from photochemical model of \citet{2012ApJ...745....3M}, which includes the effects of vertical mixing and photochemistry.
%While they demonstrated that their "two-layer" approach can recover the input abundances, that approach cannot treat the }
%
As such, we can refer to previous works that attempted to couple chemistry with computationally expensive 2D or 3D atmospheric circulation models to study the disequilibrium effect due to atmospheric circulation.
\citet{2006ApJ...649.1048C} adopted the chemical relaxation method, followed by further studies \citep[e.g.,][]{2018ApJ...855L..31D, 2018ApJ...869..107M, 2018ApJ...869...28D}.
The chemical relaxation scheme replaces numerous chemical production and loss terms in the continuity equation of a species with a single term given by the deviation of the abundance from the equilibrium value divided by its chemical timescale \citep{1998Icar..132..176S, 2006ApJ...649.1048C}.
%There are also some o
Other ways to reduce the computational cost have also been
adopted, %so far, 
such as simplifying the atmospheric dynamics model \citep{2012A&A...548A..73A, 2014A&A...564A..73A} or using the reduced chemical network \citep{2020A&A...636A..68D}.

When using the chemical relaxation method, adopting an appropriate chemical timescale for each species is important.
The chemical timescale can be approximated by that of the rate-limiting or the slowest reaction along the conversion pathway from one species to another.
Previous works have investigated the rate-limiting reaction for the conversion, such as that from $\mathrm{CH_4}$ to $\mathrm{CO}$ and/or that from $\mathrm{NH_3}$ to $\mathrm{N_2}$ \citep[e.g.,][]{1977Sci...198.1031P, 1988Icar...73..516Y, 2010Icar..209..602V, 2011ApJ...737...15M, 2014ApJ...797...41Z}.
To revisit the chemical relaxation method, \citet{2018ApJ...862...31T} derived the chemical timescales of the major species in hydrogen/helium-dominated atmospheres valid for the wide pressure and temperature ranges of currently observable exoplanet atmospheres, namely temperatures from 500~K to 3000~K and pressures from 0.1~mbar to 1000 bar.

In this paper, we adopt the chemical relaxation method to incorporate the effect of vertical mixing or quenching for each major species in hydrogen/helium-dominated atmospheres into the spectral retrieval code ARCiS \citep{2020A&A...642A..28M} with a physical basis.
This enables us to \edt{directly} retrieve the eddy diffusion coefficient of exoplanet atmospheres, unlike for the various previous spectral retrieval codes.
\edt{We note that we consider the quenching effect only as a disequilibrium process since this is what most affects the composition of the atmospheric region that we can probe by transmission spectroscopy in the optical and infrared. We neglect photochemistry, which is important in the upper atmosphere.}
We use the chemical timescales derived by \citet{2018ApJ...862...31T} as they were validated for the broad pressure and temperature ranges important for exoplanet atmospheres.
Then, using the updated ARCiS with the consideration of disequilibrium chemistry, we perform retrievals of the observed transmission spectra of exoplanets with sizes ranging from Jupiter to mini-Neptune and investigate the trend of disequilibrium chemistry in exoplanet atmospheres.

The rest of this paper is organized as follows. In Section~\ref{sec:method}, we describe our method and simulation setup. In Section~\ref{sec:results}, we present the results. In Section~\ref{sec:discussion}, we discuss several points that should be addressed in future works. Finally, we conclude this paper in Section~\ref{sec:conclusion}.

\section{Method} \label{sec:method}
\subsection{Model description}
The one-dimensional continuity-transport equation for the number density of species $i$, $n_{i}$, is written as
\begin{equation}
    \frac{\partial n_{i}}{\partial t} = P_{i} - L_{i} - \frac{\partial \Phi_{i}}{\partial z},
    \label{eq:cont}
\end{equation}
where $t$ and $z$ are the time and altitude, respectively, $P_{i}$ and $L_{i}$ are the production and loss rates of $n_i$ due to chemical reactions, respectively, and $\Phi_{i}$ is the vertical transport flux. Assuming that eddy diffusion is the dominant transport mechanism, which is usually the case for the atmospheric region of exoplanets that we can observe in the optical and infrared, $\Phi_{i}$ is given as
\begin{equation}
    \Phi_{i}= - K_\mathrm{zz} N \frac{\partial f_{i}}{\partial z}.
\end{equation}
Here, $K_\mathrm{zz}$ is the eddy diffusion coefficient, which we assume to be constant throughout the atmosphere for simplicity, and $N$ and $f_{i}$ are the total number density of the gaseous species and the volume mixing ratio of species $i$, namely $f_{i} = n_{i} / N$, respectively.

The chemical relaxation method replaces the chemical source and sink terms of $P_{i}- L_{i}$ by the deviation from the equilibrium number density divided by the chemical timescale \citep{1998Icar..132..176S, 2006ApJ...649.1048C}.
In that case, Eq.~(\ref{eq:cont}) can be rewritten as
\begin{equation}
    \frac{\partial n_{i}}{\partial t} = - \frac{n_{i}- n_{i, \mathrm{eq}}}{\tau_{i, \mathrm{chem}}} - \frac{\partial \Phi_{i}}{\partial z}.
    \label{eq:relax}
\end{equation}
Here, $n_{i, \mathrm{eq}}$ and $\tau_{i, \mathrm{chem}}$ denote the equilibrium number density and chemical timescale of species $i$, respectively.

If we assume a steady-state condition, Eq.~(\ref{eq:relax}) can be further transformed as
\begin{equation}
    n_{i}+ \tau_{i, \mathrm{chem}} \frac{\partial \Phi_{i}}{\partial z} = n_{i, \mathrm{eq}}.
    \label{eq:steady}
\end{equation}

Next, we discretize Eq.~(\ref{eq:steady}) using the subscript $j$ for the $j$th altitude layer, which we assume to have the same thickness of $\Delta z$ as the other layers, as
\begin{equation}
    n_{i, j}+ \tau_{i, j, \mathrm{chem}} \frac{\Phi_{i, j+\frac{1}{2}} - \Phi_{i, j-\frac{1}{2}}}{\Delta z} = n_{i, j, \mathrm{eq}}.
    \label{eq:disc}
\end{equation}
Here, we approximate \edt{the diffusion fluxes at the boundary between the $j+1$th and $j$th layers and that between the $j$th and $j-1$th layers,} $\Phi_{i, j+\frac{1}{2}}$ and $\Phi_{i, j-\frac{1}{2}}$\edt{,} as
\begin{equation}
\Phi_{i, j+\frac{1}{2}} = - K_\mathrm{zz} N_{j+\frac{1}{2}} \frac{f_{i, j+1} - f_{i, j}}{\Delta z}
\end{equation}
and
\begin{equation}
\Phi_{i, j-\frac{1}{2}} = - K_\mathrm{zz} N_{j-\frac{1}{2}} \frac{f_{i, j} - f_{i, j-1}}{\Delta z}.
\end{equation}
\edt{$N_{j+\frac{1}{2}}$ and $N_{j-\frac{1}{2}}$ are the total number densities at those boundaries.}

Finally, we can convert Eq.~(\ref{eq:disc}) into the matrix form as
\begin{equation}
    (\bm{E} + \bm{T_i M_i}) \bm{n_i} = \bm{n_{i, \mathrm{eq}}}.
    \label{eq:matrix}
\end{equation}
Here, $\bm{E}$ is the identity matrix, and the other matrices are given as follows. $\mathcal{N}$ is the number of the altitude layers and ``diag'' indicates a diagonal matrix.
\begin{equation}
\bm{n_i} = \left(n_{i, 1}, n_{i, 2}, n_{i, 3}, \cdots, n_{i, j}, \cdots, n_{i, \mathcal{N}} \right)^\mathrm{T}
\end{equation}

\begin{equation}
\bm{n_{i, \mathrm{eq}}} = \left(n_{i, 1, \mathrm{eq}}, n_{i, 2, \mathrm{eq}}, n_{i, 3, \mathrm{eq}}, \cdots, n_{i, j, \mathrm{eq}}, \cdots, n_{i, \mathcal{N}, \mathrm{eq}} \right)^\mathrm{T}
\end{equation}

\begin{equation}
\bm{T_i} = \mathrm{diag} \left(\tau_{i, 1, \mathrm{chem}}, \tau_{i, 2, \mathrm{chem}}, \tau_{i, 3, \mathrm{chem}}, \cdots, \tau_{i, j, \mathrm{chem}}, \cdots, \tau_{i, \mathcal{N}, \mathrm{chem}} \right)
\end{equation}

\newpage
\begin{strip}

\begin{equation}
\bm{M_i} = - K_\mathrm{zz}
\left(
\begin{array}{ccccccccc}
0 & 0 & 0 & \cdots & \cdots & \cdots & \cdots & \cdots & 0 \\
\frac{1}{{\Delta z_{2}}^2} \frac{N_{\frac{3}{2}}}{N_1} & - \frac{1}{{\Delta z_{2}}^2} \frac{N_{\frac{3}{2}} + N_{\frac{5}{2}}}{N_2} & \frac{1}{{\Delta z_{2}}^2} \frac{N_{\frac{5}{2}}}{N_3} & 0 & \cdots & \cdots & \cdots & \cdots & 0 \\
0 & \frac{1}{{\Delta z_{3}}^2} \frac{N_{\frac{5}{2}}}{N_2} & - \frac{1}{{\Delta z_{3}}^2} \frac{N_{\frac{5}{2}} + N_{\frac{7}{2}}}{N_3} & \frac{1}{{\Delta z_{3}}^2} \frac{N_{\frac{7}{2}}}{N_4} & 0 & \cdots & \cdots & \cdots & 0 \\
\vdots & & & \ddots & & & & & \vdots \\
0 & \cdots & 0 & \frac{1}{{\Delta z_{j}}^2}\frac{N_{j - \frac{1}{2}}}{N_{j-1}} & - \frac{1}{{\Delta z_{j}}^2} \frac{N_{j - \frac{1}{2}} + N_{j + \frac{1}{2}}}{N_j} & \frac{1}{{\Delta z_{j}}^2} \frac{N_{j + \frac{1}{2}}}{N_{j+1}} & 0 & \cdots & 0 \\
\vdots & & & & & \ddots & & & \vdots \\
\vdots & & & & & & \ddots & & \vdots \\
\vdots & & & & & & & \ddots & \vdots \\
0 & \cdots & \cdots & \cdots & \cdots & \cdots & 0 & \frac{1}{{\Delta z_{\mathcal{N}}}^2} \frac{N_{\mathcal{N} - \frac{1}{2}}}{N_{\mathcal{N}-1}} & - \frac{1}{{\Delta z_{\mathcal{N}}}^2} \frac{N_{\mathcal{N} - \frac{1}{2}}}{N_\mathcal{N}} 
\end{array}
\right)
\end{equation}
\end{strip}
In the above, we have adopted the lower boundary condition of the chemical equilibrium $n_{i, 1} = n_{i, 1, \mathrm{eq}}$ because of the short chemical timescale in the deeper atmosphere.
For the upper boundary condition, we have adopted zero flux, namely without atmospheric escape.
By multiplying both sides of Eq.~(\ref{eq:matrix}) by the inverse matrix of $(\bm{E} + \bm{T_i M_i})$, we finally get the solution as
\begin{equation}
    \bm{n_i} = (\bm{E} + \bm{T_i M_i})^{-1} \bm{n_{i, \mathrm{eq}}}.
    \label{eq:solution}
\end{equation}
\hl{We note that this equation leads to $n_i \sim n_{i, \mathrm{eq}}$, namely in equilibrium, when $K_{zz}$ has a negligible value.}

\subsection{Implementation to ARCiS} \label{sec:application}
We have implemented this module to the spectral retrieval code ARCiS \citep[ARtful modelling Code for exoplanet Science;][]{2020A&A...642A..28M} using the chemical timescales of $\mathrm{CH_4}$, $\mathrm{CO}$, $\mathrm{H_2O}$, $\mathrm{NH_3}$, and $\mathrm{N_2}$ from \citet{2018ApJ...862...31T}. 
\edt{We note that for the chemical timescales of $\mathrm{CH_4}$, $\mathrm{CO}$, and $\mathrm{H_2O}$, their different expressions above and below the C/O ratio of unity are used.}
$\mathrm{CO_2}$ \hl{tends to remain in pseudo-equilibrium even after its related molecules CO and $\mathrm{H_2O}$ are quenched \citep{2011ApJ...737...15M, 2018ApJ...862...31T}. Thus, for the calculation of $\mathrm{CO_2}$ abundance, instead of using Eq.~(\ref{eq:solution}),} we \hl{adopt the following} pseudo-equilibrium abundance formula \hl{of \citet{2018ApJ...862...31T},
\begin{equation}
    n_\mathrm{CO_2} = \frac{n_\mathrm{CO} n_\mathrm{H_2O}}{n_\mathrm{CO, eq} n_\mathrm{H_2O, eq}} n_\mathrm{CO_2, eq},
\end{equation}
which modifies the $\mathrm{CO_2}$ number density based on the quenched abundances of CO and $\mathrm{H_2O}$ when they are quenched}.
All the species except for the above are assumed to have their equilibrium abundances.

ARCiS uses GGchem \citep{2018A&A...614A...1W} for the calculation of thermochemical equilibrium abundances.
The original version of GGchem did not include $\mathrm{CH_2OH}$, $\mathrm{CH_3OH}$, and $\mathrm{N_2H_3}$, the abundances of which are needed to calculate the chemical timescale of the species {mentioned above}. Taking their thermodynamic data from \citet{osti_925269}\footnote{http://garfield.chem.elte.hu/Burcat/burcat.html}, we have added these species to GGchem, %which is also 
available on the GitHub page of GGchem \footnote{https://github.com/pw31/GGchem}.

\subsection{Application to the observed transmission spectra of 16 planets} \label{sec:application}
With the updated ARCiS, we performed spectral retrievals \hl{where disequilibrium chemistry is allowed to affect the profiles of the molecular abundances} (hereafter, ``disequilibrium retrieval'').
%Here we note that our disequilibrium retrieval can also treat the equilibrium case. In that case, the eddy diffusion coefficient should be retrieved as a negligible value.
To investigate the effect of the inclusion of disequilibrium chemistry to the retrieval code, we also performed retrievals imposing equilibrium chemistry (hereafter, ``equilibrium retrieval'') and compared the results of the two retrievals.

For the planet samples, we selected ten hot Jupiters and six Neptunes compiled in \citet{Sing:2016hi} and \citet{2017AJ....154..261C}, respectively.
High-precision transmission spectra were observed by the \textit{Hubble} Space Telescope (HST) and {\textit{Spitzer}} Space Telescope for those planets.
The samples are listed in Table~\ref{table:obsref}, along with the references for their observed transmission spectrum data used in the retrieval.
These planet samples range from clear to cloudy atmospheres, and we note that the hot Jupiter samples and their spectral data are the same as those used in \citet{2020A&A...642A..28M} except for HAT-P-12b, for which we adopt {the} observation data recently analyzed by \citet{2020AJ....159..234W}.

\begin{table}
\caption{References for the observational data}             % title of Table
\label{table:obsref}      % is used to refer this table in the text
\centering                          % used for centering table
\begin{threeparttable}
\begin{tabular}{l l}        % centered columns (4 columns)
\hline\hline                 % inserts double horizontal lines
Planet & Reference \\    % table heading 
\hline                        % inserts single horizontal line
GJ~436b & \edt{\citet{2011ApJ...735...27K}\tnote{1}} \\
& \edt{\citet{2014Natur.505...66K}} \\
& \edt{\citet{2015ApJ...802..117M}\tnote{2}} \\
& \edt{\citet{2018AJ....155...66L}} \\
GJ~1214b & \citet{2013ApJ...765..127F}\tnote{3} \\
& \citet{2014Natur.505...69K}\tnote{4} \\
%GJ~3470b & \citet{2013ApJ...768..154D} \\
%& \citet{2014AA...570A..89E}\tnote{4} \\
GJ~3470b & \edt{\citet{2019NatAs...3..813B}} \\
HAT-P-1b\tnote{*} & \citet{2013MNRAS.435.3481W} \\
& \citet{2014MNRAS.437...46N} \\
HAT-P-11b & \citet{2019AJ....158..244C} \\
HAT-P-12b & \citet{2020AJ....159..234W} \\
HAT-P-26b & \citet{2017Sci...356..628W} \\
HD~97658b & \citet{2020AJ....159..239G}\tnote{5} \\
HD~189733b\tnote{*} & \citet{2013MNRAS.432.2917P} \\
& \citet{2014ApJ...791...55M} \\
& \citet{Sing:2016hi} \\
HD~209458b\tnote{*} & \citet{Sing:2016hi} \\
%GJ~436b & \citet{2019NatAs...3..813B} \\
WASP-6b & \citet{2020MNRAS.494.5449C} \\
WASP-12b\tnote{*} & \citet{2013MNRAS.436.2956S} \\
& \citet{2015ApJ...814...66K} \\
& \citet{Sing:2016hi} \\
WASP-17b & \citet{Sing:2016hi} \\
WASP-19b\tnote{*} & \citet{2013MNRAS.434.3252H} \\
& \citet{Sing:2016hi} \\
WASP-31b & \citet{2015MNRAS.446.2428S} \\
& \citet{Sing:2016hi} \\
WASP-39b & \citet{Sing:2016hi} \\
& \citet{2016ApJ...827...19F} \\
& \citet{2018AJ....155...29W} \\
\hline                                   %inserts single line
\end{tabular}
\begin{tablenotes}
\item[*] Planet samples with masses larger than half Jupiter mass, for which disequilibrium retrieval with fixed $T_\mathrm{int}$ is additionally performed.
\edt{\item[1] The average of three measurements at the {\textit{Spitzer}} 8.0~$\mu$m band is used, excluding the observed transit on UT 2009 February 2 due to the possible stellar activity effect.
\item[2] The results presented in Table~6 are used.}
\item[3] The ``simultaneous'' fit results are used.
\item[4] The ``divide-white'' fit results are used.
\item[5] Following the approach of this study, we do not include the HST/STIS data points for our retrieval due to their large discrepancy at long wavelengths. Also, the results derived using the ``logarithmic visit-long trend'' are used.
%\item[4] Their results with "out-of transit data correction" are used.
\end{tablenotes}
\end{threeparttable}
\end{table}

%The equilibrium retrieval with ARCiS was already performed for all of our hot Jupiter samples in \citet{2020A&A...642A..28M}.
%\hl{Thus, while running disequilibrium retrieval for all of our samples,} we run the equilibrium retrieval only for the six Neptune samples, namely HAT-P-11b, HAT-P-26b, HD~97658b, GJ~436b, GJ~1214b, and GJ~3470b, in addition to HAT-P-12b, for which, again, we use recently published observed data of \citet{2020AJ....159..234W}.
%For this purpose, 
We adopt the same settings as the ``constrained retrieval'' of \citet{2020A&A...642A..28M}, which considered the thermal and cloud structures with the models of \citet{2010A&A...520A..27G} and \citet{2019A&A...622A.121O}, respectively, and adopted the pre-computed opacities presented in \citet{2021A&A...646A..21C}.
The retrieval parameters and the employed ranges and priors of those parameters are presented in Table~\ref{table:param}. They are also the same as in \citet{2020A&A...642A..28M}, except for the eddy diffusion coefficient used in the disequilibrium chemistry module $K_{zz}^\mathrm{Ch}$, which was not considered in {the previous} retrieval.
In this study, as a first step to consider disequilibrium chemistry in spectral retrieval, we define a ``standard'' disequilibrium retrieval as a retrieval treating the eddy diffusion coefficient used in the chemistry calculation $K_{zz}^\mathrm{Ch}$ and that used in the cloud simulation $K_{zz}^\mathrm{Cl}$ independently.
This is because the atmospheric region where clouds exist and the region where chemical species affecting the spectra are located are usually different, and hence the two eddy diffusion coefficients reflect the coefficients at different parts of the atmosphere.
Moreover, we also examine the effect of imposing the condition that the two eddy diffusion coefficients are the same, namely $K_{zz}^\mathrm{Ch} = K_{zz}^\mathrm{Cl}$ \edt{ (hereafter, ``same-$K_{zz}$ disequilibrium retrieval'')}.
%This is to examine how this procedure would affect our results.
Retrieval with a non-constant coefficient along the altitude is left as a future work.
We note that for the retrieval of WASP-12b, in the same way as \citet{2020A&A...642A..28M}, we adopt an additional parameter, which allows the scaling of the data from \citet{2015ApJ...814...66K} to match the remaining data points due to the existing slight offset of their data when compared to the other data \citep{2013MNRAS.436.2956S, Sing:2016hi}. See \citet{2020A&A...642A..28M} for a detailed treatment of this additional parameter.

\begin{table*}
\caption{Retrieval parameters}             % title of Table
\label{table:param}      % is used to refer this table in the text
\centering                          % used for centering table
\begin{threeparttable}
\begin{tabular}{l c c l l}        % centered columns (4 columns)
\hline\hline                 % inserts double horizontal lines
Parameter & Symbol & Range & Unit & Prior \\    % table heading 
\hline                        % inserts single horizontal line
\edt{Ratio of the visible to infrared opacity} & $\gamma$ & $10^{-2}$--$10^2$ & & flat log \\
Irradiation parameter & $f_\mathrm{irr}$ & 0.00--0.25 & & flat linear \\
Infrared opacity & $\kappa_\mathrm{IR}$ & $10^{-4}$--$10^4$ & $\mathrm{cm}^2$~$\mathrm{g}^{-1}$ & flat log \\
Intrinsic temperature\tnote{1} & $T_\mathrm{int}$ & 10--3000 & K & flat log \\
C/O ratio & C/O & 0.1--1.3 & & flat linear \\
Si/O ratio & Si/O & 0.0--0.3 & & flat linear \\
N/O ratio & N/O & 0.0--0.3 & & flat linear \\
Metallicity & $Z$ & $-$1--3 & dex & flat linear \\
Reference radius at 10~bar pressure & $R_\mathrm{ref}$ & 5$\sigma$ around the literature planetary radius & Jupiter radius & flat linear \\
Planetary gravity & $\log_{10}{g}$ & 5$\sigma$ around the literature value & $\mathrm{cm}$~$\mathrm{s}^{-2}$ for $g$ & Gaussian \\
Diffusion coefficient for cloud\tnote{2} & $K_{zz}^\mathrm{Cl}$ & $10^5$--$10^{12}$ & $\mathrm{cm}^2$~$\mathrm{s}^{-1}$ & flat log \\
Cloud nucleation rate & $\dot{\Sigma}$ & $10^{-17}$--$10^{-7}$ & g~$\mathrm{cm}^{-2}$~$\mathrm{s}^{-1}$ & flat log \\
Diffusion coefficient for chemistry\tnote{3} & $K_{zz}^\mathrm{Ch}$ & $10^0$--$10^{15}$ & $\mathrm{cm}^2$~$\mathrm{s}^{-1}$ & flat log \\
\hline                                   %inserts single line
\end{tabular}
\begin{tablenotes}
\item[1] Excluded for the disequilibrium retrieval with fixed $T_\mathrm{int}$
\item[2] Not used in the same-$K_{zz}$ retrieval. Instead, $K_{zz}^\mathrm{Cl}=K_{zz}^\mathrm{Ch}$ is assumed.
\item[3] Used only for the disequilibrium retrievals
\end{tablenotes}
\end{threeparttable}
\end{table*}

\hl{Our disequilibrium retrieval can also treat the equilibrium case since a negligible value of $K_{zz}$ leads to the molecular abundances in equilibrium, as indicated in Eq.~(\ref{eq:solution}).
We have confirmed that the lower bound of $K_{zz}^\mathrm{Ch}$ we adopted, namely $10^0$~$\mathrm{cm}^2$~$\mathrm{s}^{-1}$, is small enough to give molecular abundances consistent with the equilibrium values for the hotter planets of our samples.
For some cooler planets, however, a few species deviate from their equilibrium abundances even for $K_{zz}^\mathrm{Ch} = 10^0$~$\mathrm{cm}^2$~$\mathrm{s}^{-1}$.
This is because the chemical timescale becomes exceedingly long at lower temperatures \citep[see Figure~4 of][]{2018ApJ...862...31T}.
Considering a $\sim$ Gyr planetary age timescale, $K_{zz}^\mathrm{Ch} = 10^0$~$\mathrm{cm}^2$~$\mathrm{s}^{-1}$ is approximately the limit for the vertical transport by eddy diffusion to affect molecular abundances within the planetary age. Here, we estimate the diffusion timescale as $\tau_\mathrm{diff} \sim {H^2}/{K_{zz}}$ using the atmospheric scale height $H$ of each of our samples.
Thus, we set the lower bound of $K_{zz}^\mathrm{Ch}$ to be  $10^0$~$\mathrm{cm}^2$~$\mathrm{s}^{-1}$ even though the exploration range of $K_{zz}^\mathrm{Ch}$ does not include the ``true'' equilibrium regime for a few minor species in relatively cool planets such as GJ~436b, GJ~1214b, GJ~3470b, HAT-P-1b, HAT-P-11b, HAT-P-12b, HAT-P-26b, HD~97658b, HD~209458b, WASP-31b, and WASP-39b.
We have estimated this using the best-fit parameter set that yields the minimum $\chi^2$ \edt{compared to the observed data} from our standard disequilibrium retrieval calculations for each of our samples.
}

Finally, in addition to the aforementioned two disequilibrium retrievals, we also conduct a disequilibrium retrieval fixing one of the highly uncertain parameters, namely the intrinsic temperature $T_\mathrm{int}$, to an estimated value based on current knowledge.
%We refer to this retrieval as "disequilibrium retrieval with fixed $T_\mathrm{int}$" in this study.
This is done based on the expectation that quenching often happens in the deeper atmosphere, especially for the nitrogen species, which means that the retrieval of $K_{zz}^\mathrm{Ch}$ can degenerate with $T_\mathrm{int}$.
We assume the value of $T_\mathrm{int}$ for each of our planet samples using its relation to the equilibrium temperature \hl{$T_\mathrm{eq}$} at a thermal equilibrium inferred by \citet{2019ApJ...884L...6T}, who derived the fraction \hl{$\epsilon$} of the incident flux on the planet \hl{$F$} that heats the interior sufficiently to reproduce the observed radii of their hot Jupiter samples \hl{as a function of $F$}.
When assuming thermal equilibrium for the planet interiors, which will be reached in as little as tens of megayears, that fraction directly relates \hl{$T_\mathrm{eq}$ and $T_\mathrm{int}$ as} \citep{2018AJ....155..214T}
\hl{
\begin{equation}
    \begin{split}
    T_\mathrm{int} &= \epsilon \left( F \right)^{1/4} T_\mathrm{eq} \\
    &\sim 0.39 T_\mathrm{eq} \exp{\left( - \frac{\left( \log_{10}{F} - 0.14 \right)^2 }{1.095} \right)},
    \end{split}
\end{equation}
where $F = 4 \sigma T_\mathrm{eq}^4$ in units of Gerg $\mathrm{cm^{-2}}$ $\mathrm{s^{-1}}$ with the Stefan--Boltzmann constant $\sigma$.
}
We adopt this assumption for this retrieval.
%with the heating models of \citet{2019ApJ...874L..31T}
Since their formula was inferred from planet samples with masses larger than half Jupiter mass \citep{2018AJ....155..214T}, we apply this type of disequilibrium retrieval only to the planets that satisfy this criterion, namely HAT-P-1b, HD~189733b, HD~209458b, WASP-12b, and WASP-19b.
For those planets, the adopted values of $T_\mathrm{int}$ are 476, 376, 561, 517, and 653~K, respectively.

\section{Results} \label{sec:results}
\subsection{Standard disequilibrium retrieval in comparison with equilibrium retrieval} \label{sec:diseq}
In this section, we present the results of the standard disequilibrium retrieval and their differences from the equilibrium retrieval.
\hl{As we {have} mentioned in \S~\ref{sec:application}, for hotter planet samples, our disequilibrium retrieval can also treat the equilibrium cases.}
In Fig.~\ref{fig:spectra}, we show the results of the spectra from the standard disequilibrium and equilibrium retrievals along with those of the disequilibrium retrieval with the same $K_{zz}$ and that with fixed $T_\mathrm{int}$ (only shown for the planets with masses larger than half Jupiter mass), which we discuss in \S~\ref{sec:same} and \ref{sec:tint}, respectively.
Also, the retrieved pressure--temperature structure and the abundance profiles %for the best-fit parameters 
\edt{for four representative planets} are shown in the left and right panels of Fig.~\ref{fig:profile}, respectively.
\edt{The profiles for the other planets are presented in Fig.~\ref{fig:profile2} of Appendix~\ref{sec:profiles}.}
In addition, plots of the posterior distributions of the parameters from the sampling are presented in~Fig.~\ref{fig:corner} {of Appendix~\ref{append:corner}}.

\begin{figure*}
 \begin{minipage}{0.5\hsize}
  \centering
  \includegraphics[width=\hsize]{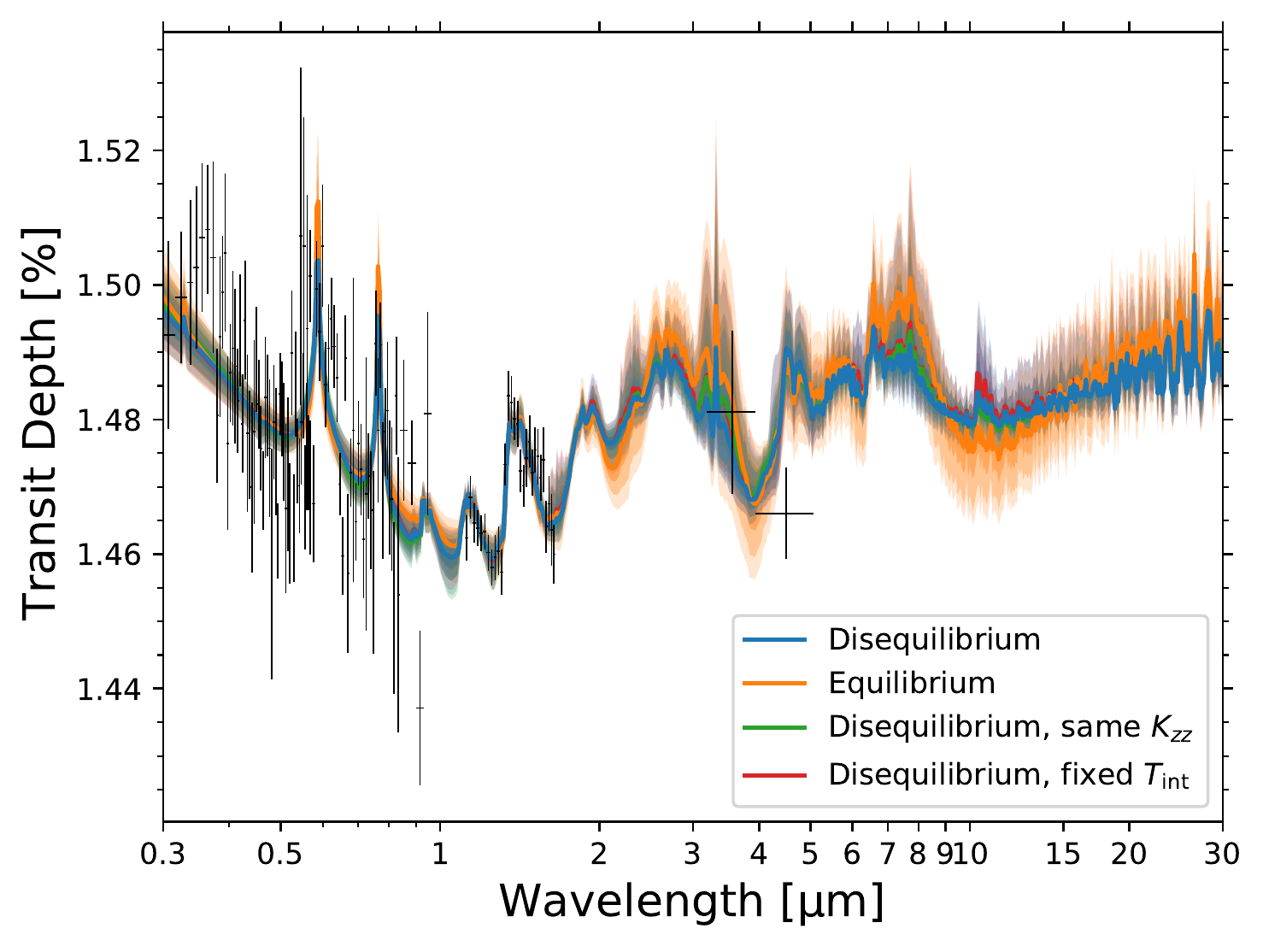}
  \subcaption{HD~209458b}
 \end{minipage}
 \begin{minipage}{0.5\hsize}
  \centering
  \includegraphics[width=\hsize]{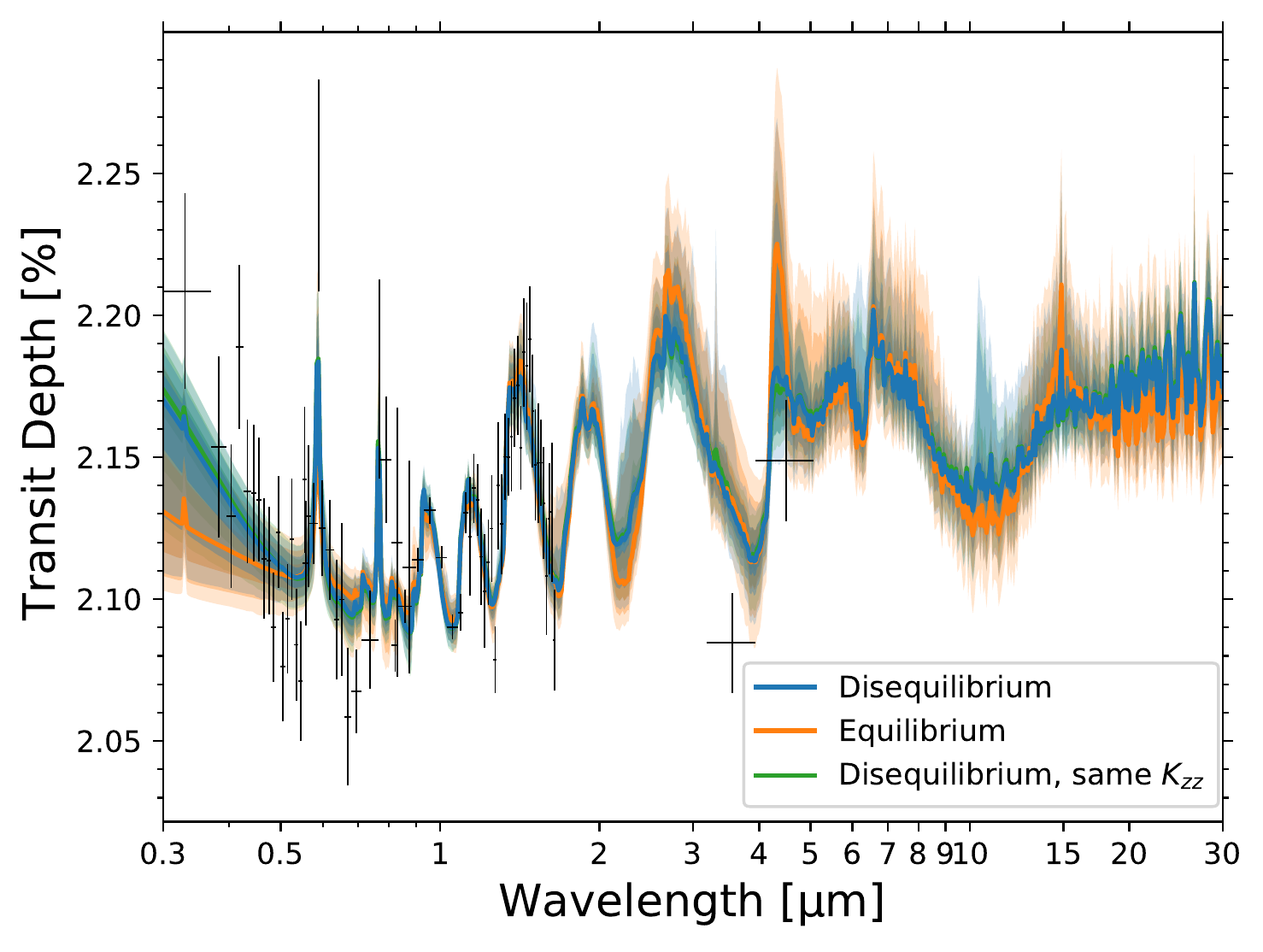}
  \subcaption{WASP-39b}
 \end{minipage}
 \begin{minipage}{0.5\hsize}
  \centering
  \includegraphics[width=\hsize]{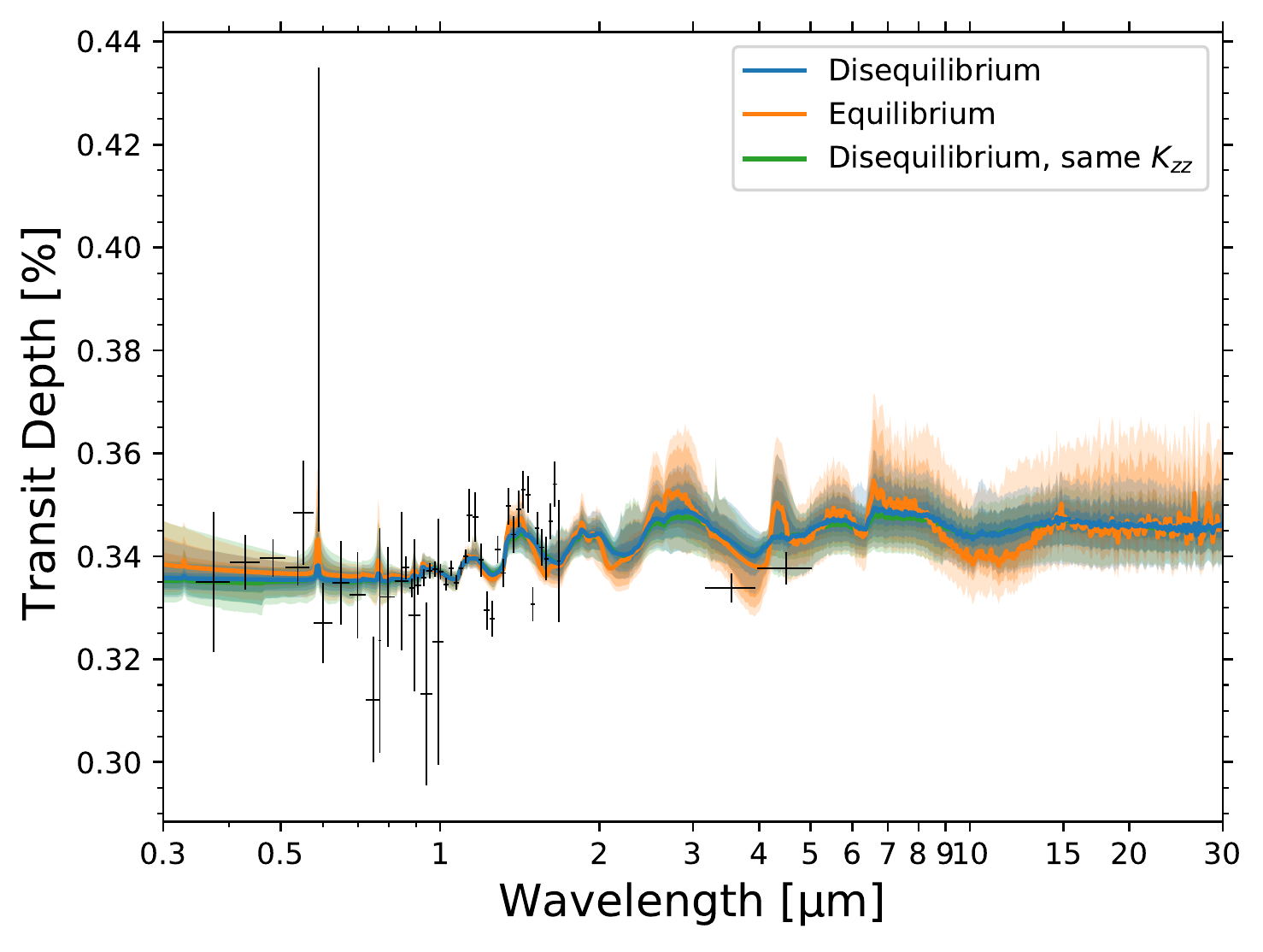}
  \subcaption{HAT-P-11b}
 \end{minipage}
 \begin{minipage}{0.5\hsize}
  \centering
  \includegraphics[width=\hsize]{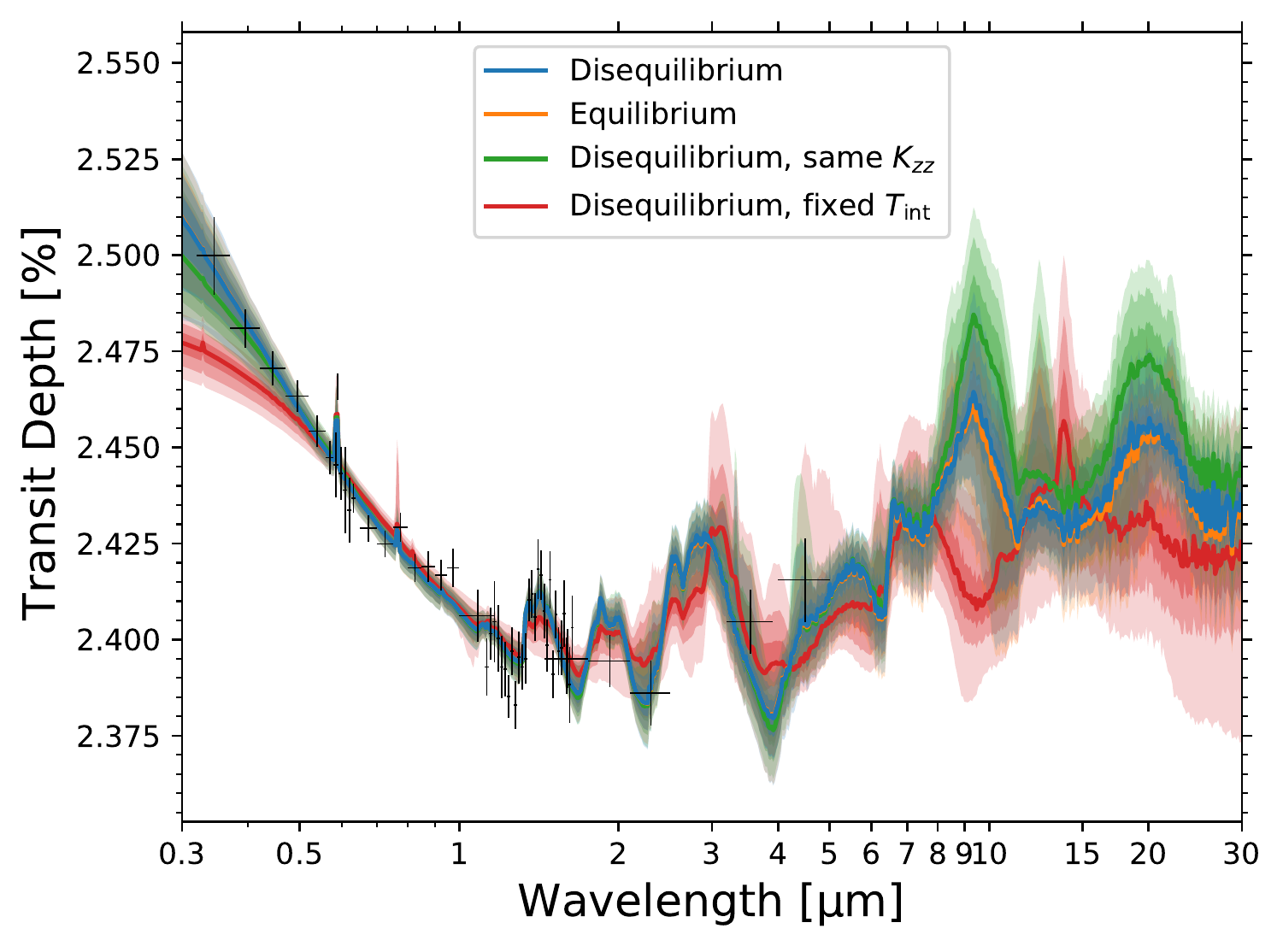}
  \subcaption{HD~189733b}
 \end{minipage} 
 \begin{minipage}{0.5\hsize}
   \centering
   \includegraphics[width=\hsize]{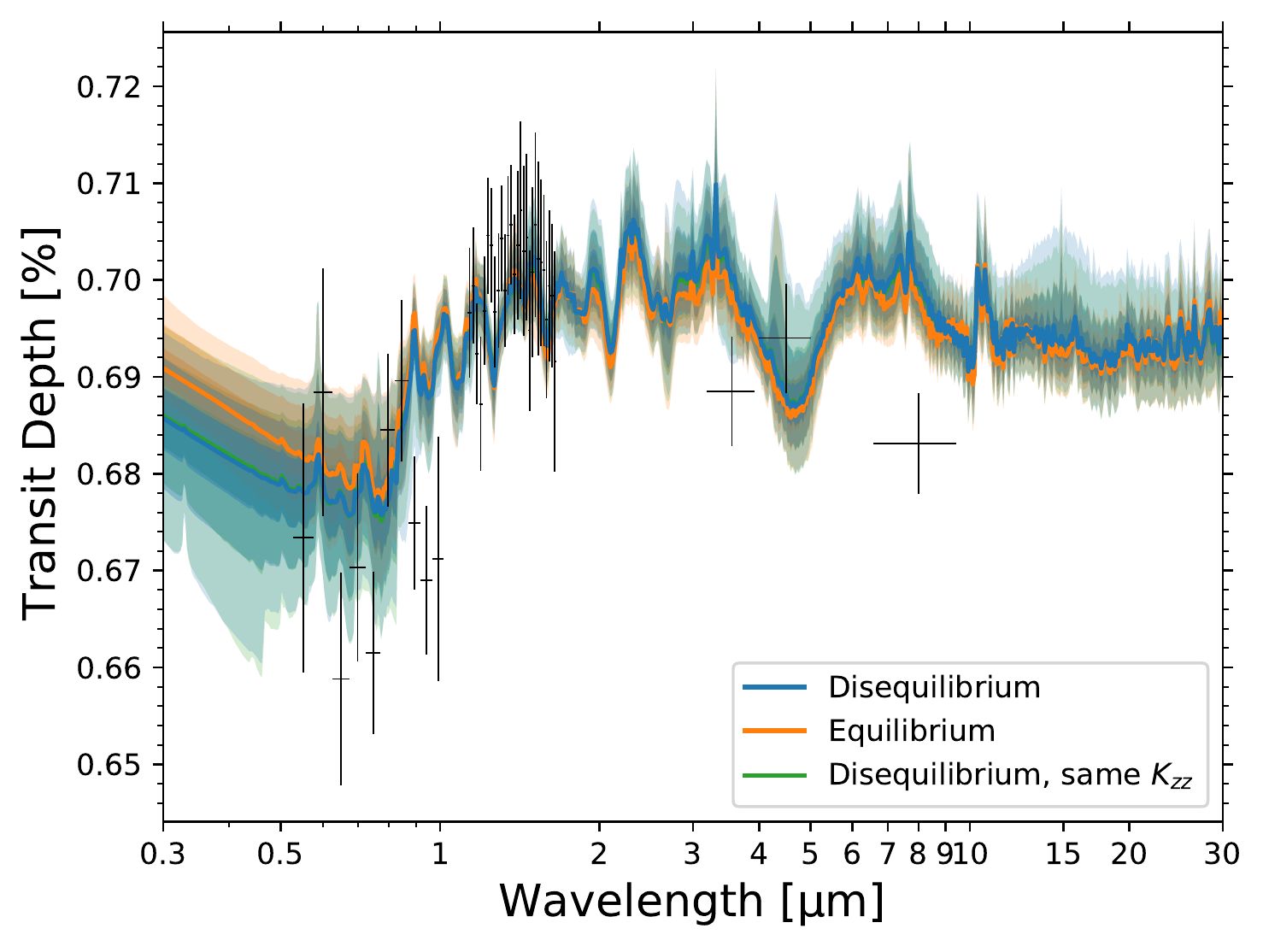}
   \subcaption{GJ~436b}
 \end{minipage}
 \begin{minipage}{0.5\hsize}
   \centering
   \includegraphics[width=\hsize]{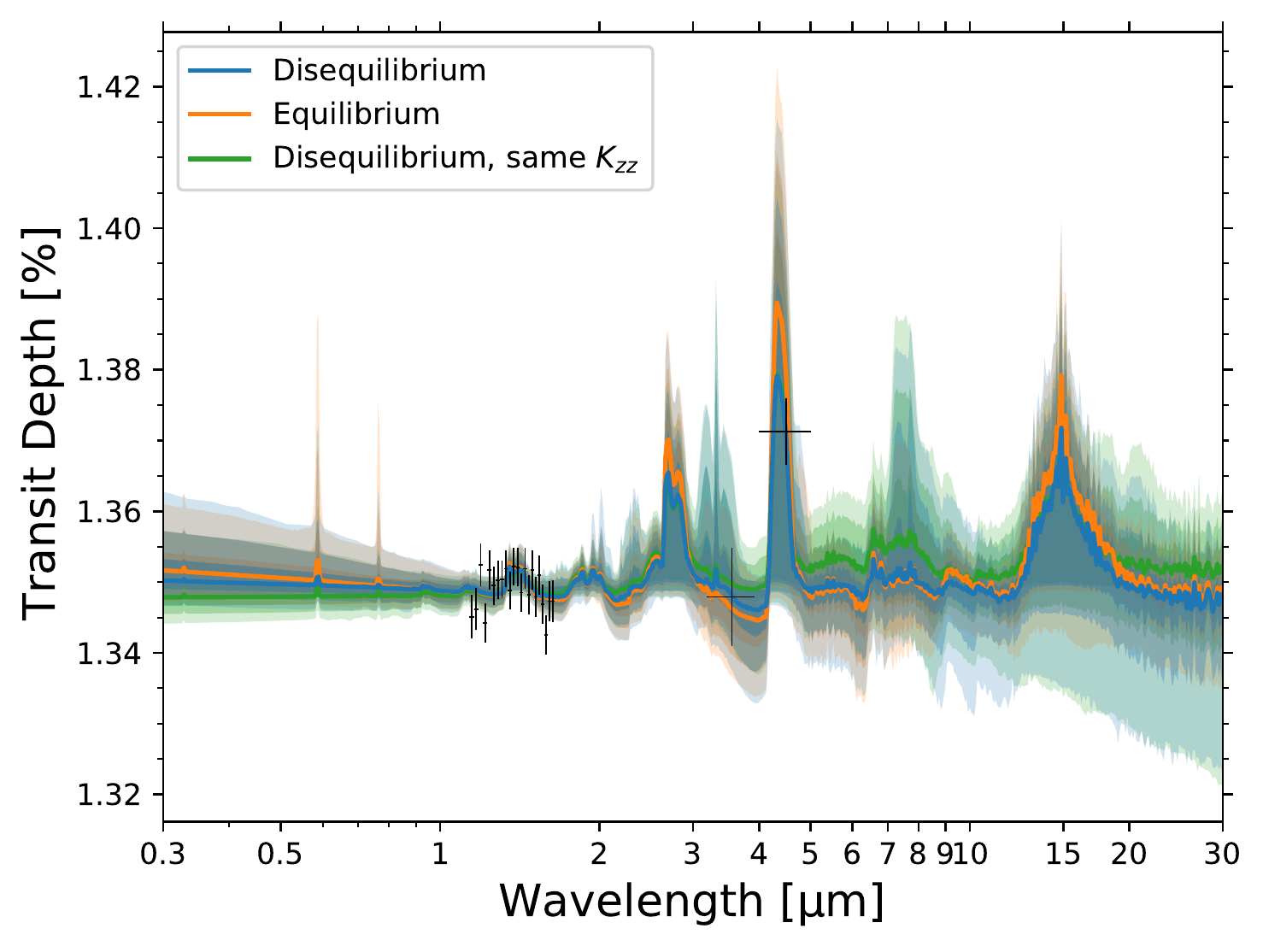}
   \subcaption{GJ~1214b}
  \end{minipage}
 \caption{Transmission spectra retrieved from the standard disequilibrium retrieval (blue), equilibrium retrieval (orange), and disequilibrium retrievals with the same $K_{zz}$ (green) and with fixed $T_\mathrm{int}$ (red; only shown for planets with masses larger than half Jupiter mass).
 The solid lines show the \hl{median} models while the shaded regions indicate the 1, 2, and 3$\sigma$ confidence intervals.
 The black points denote the observed data used.}
 \label{fig:spectra}
\end{figure*}
\begin{figure*}
\ContinuedFloat
 \begin{minipage}{0.5\hsize}
  \centering
  \includegraphics[width=\hsize]{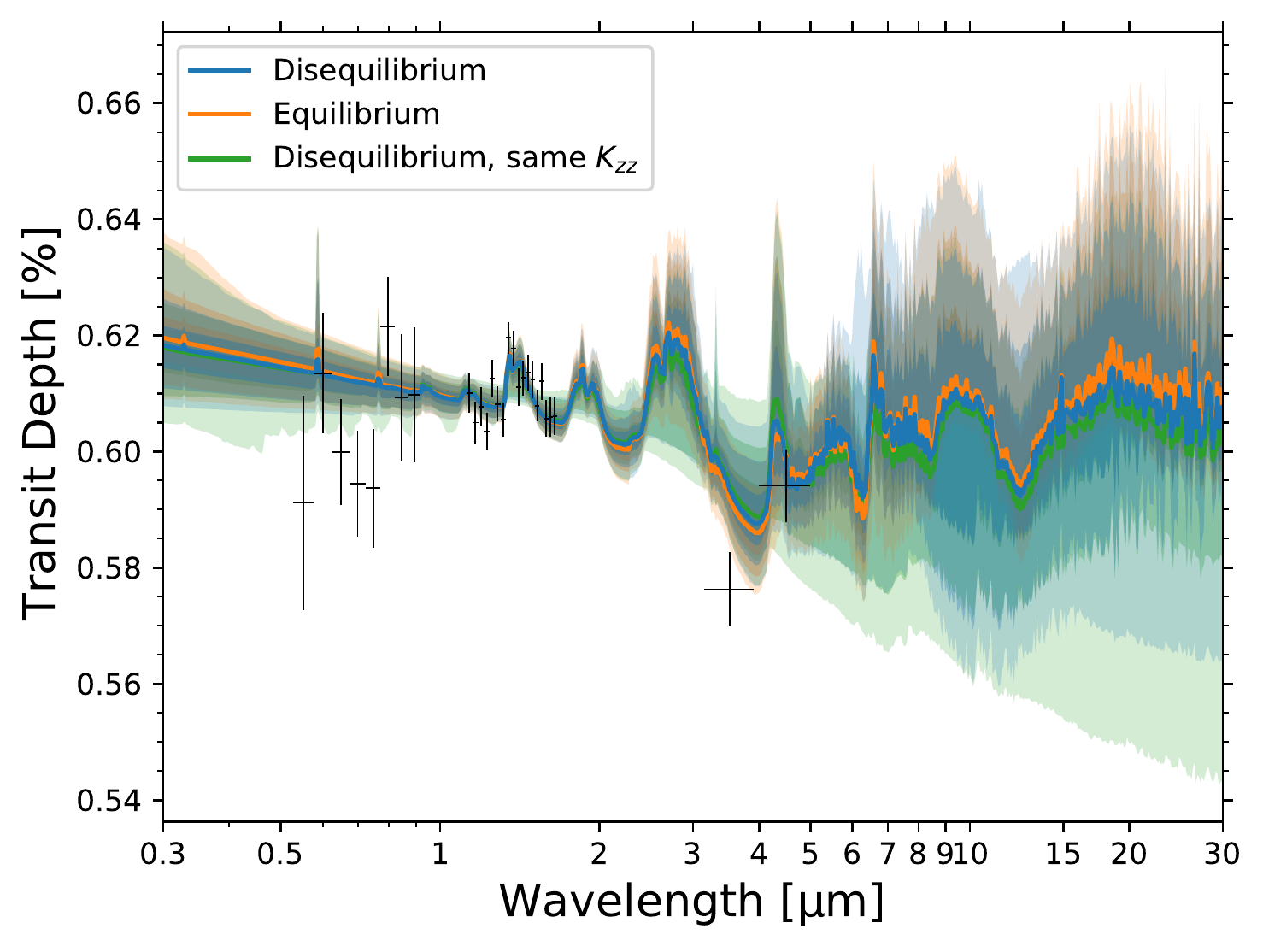}
  \subcaption{GJ~3470b}
 \end{minipage}
 \begin{minipage}{0.5\hsize}
  \centering
  \includegraphics[width=\hsize]{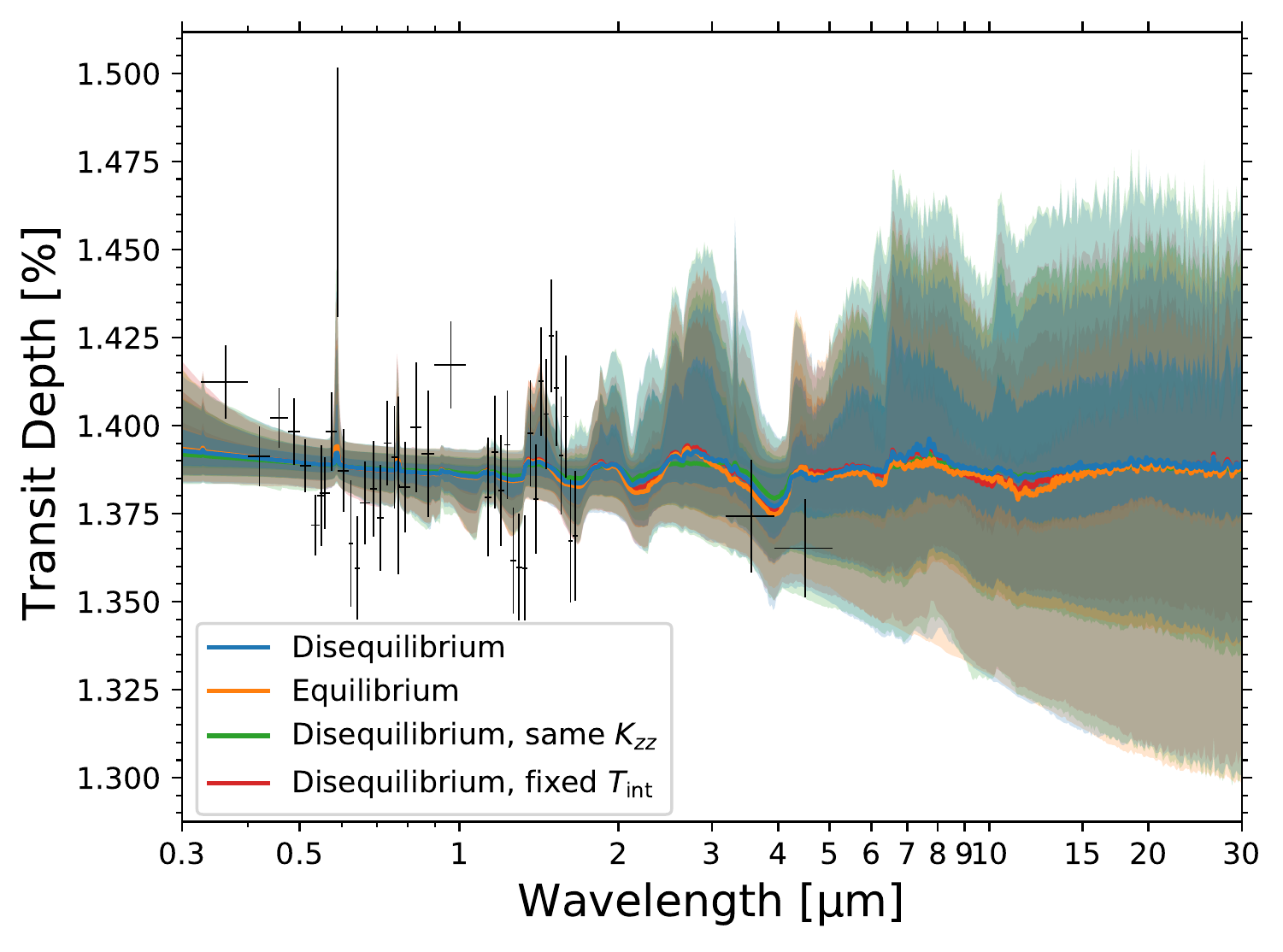}
  \subcaption{HAT-P-1b}
 \end{minipage}
 \begin{minipage}{0.5\hsize}
  \centering
  \includegraphics[width=\hsize]{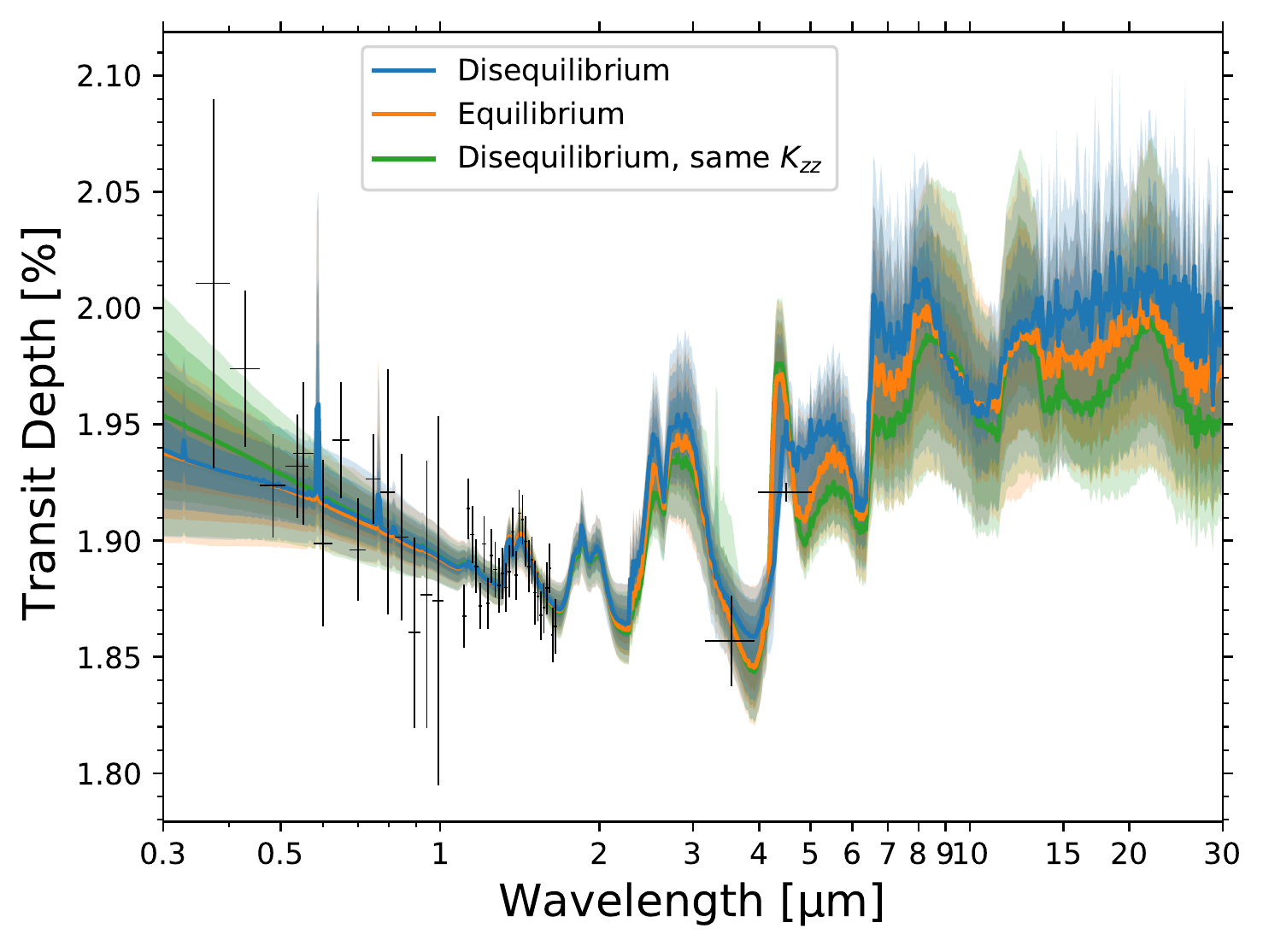}
  \subcaption{HAT-P-12b}
 \end{minipage}
 \begin{minipage}{0.5\hsize}
   \centering
   \includegraphics[width=\hsize]{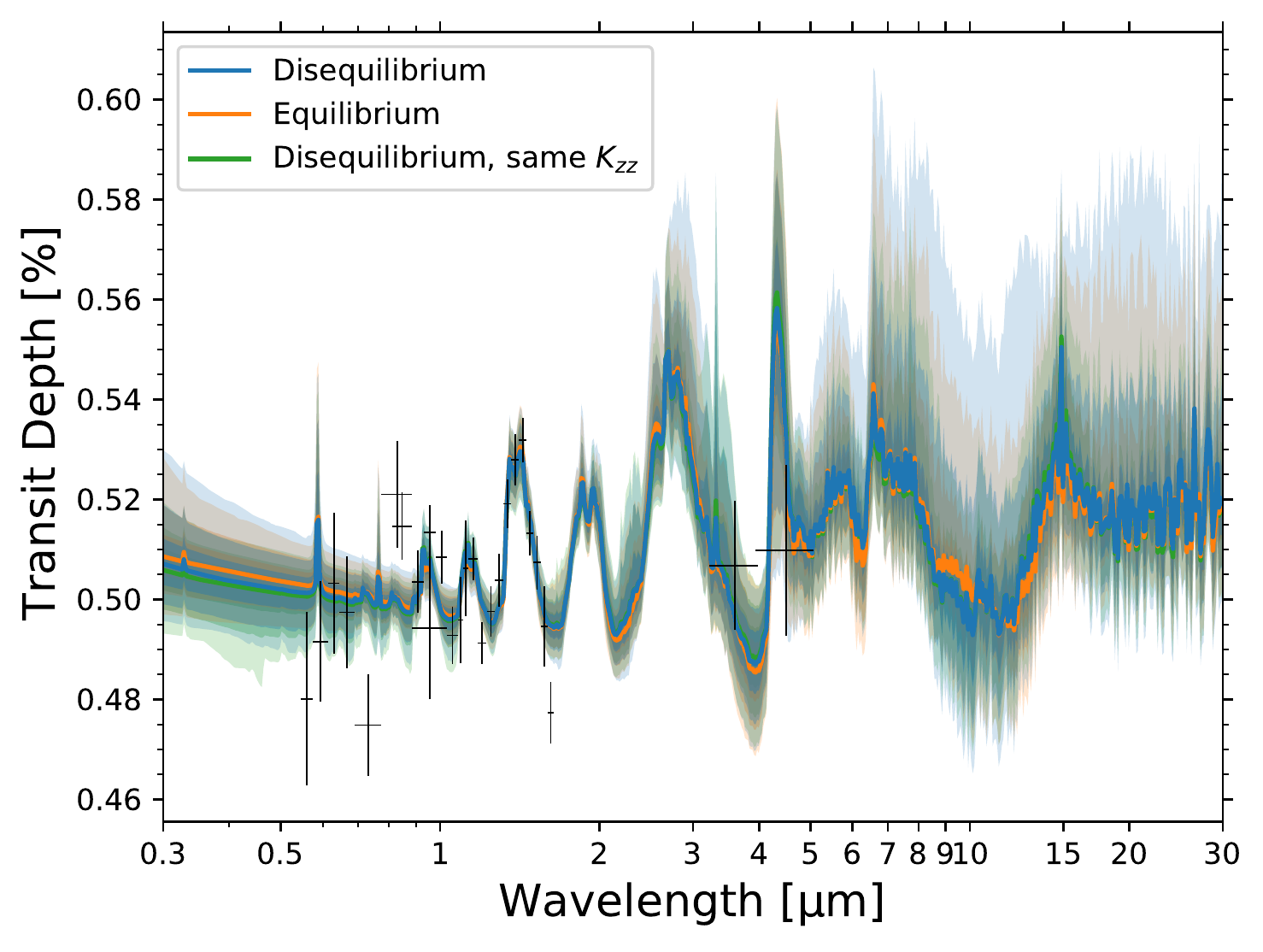}
   \subcaption{HAT-P-26b}
 \end{minipage}
 \begin{minipage}{0.5\hsize}
   \centering
   \includegraphics[width=\hsize]{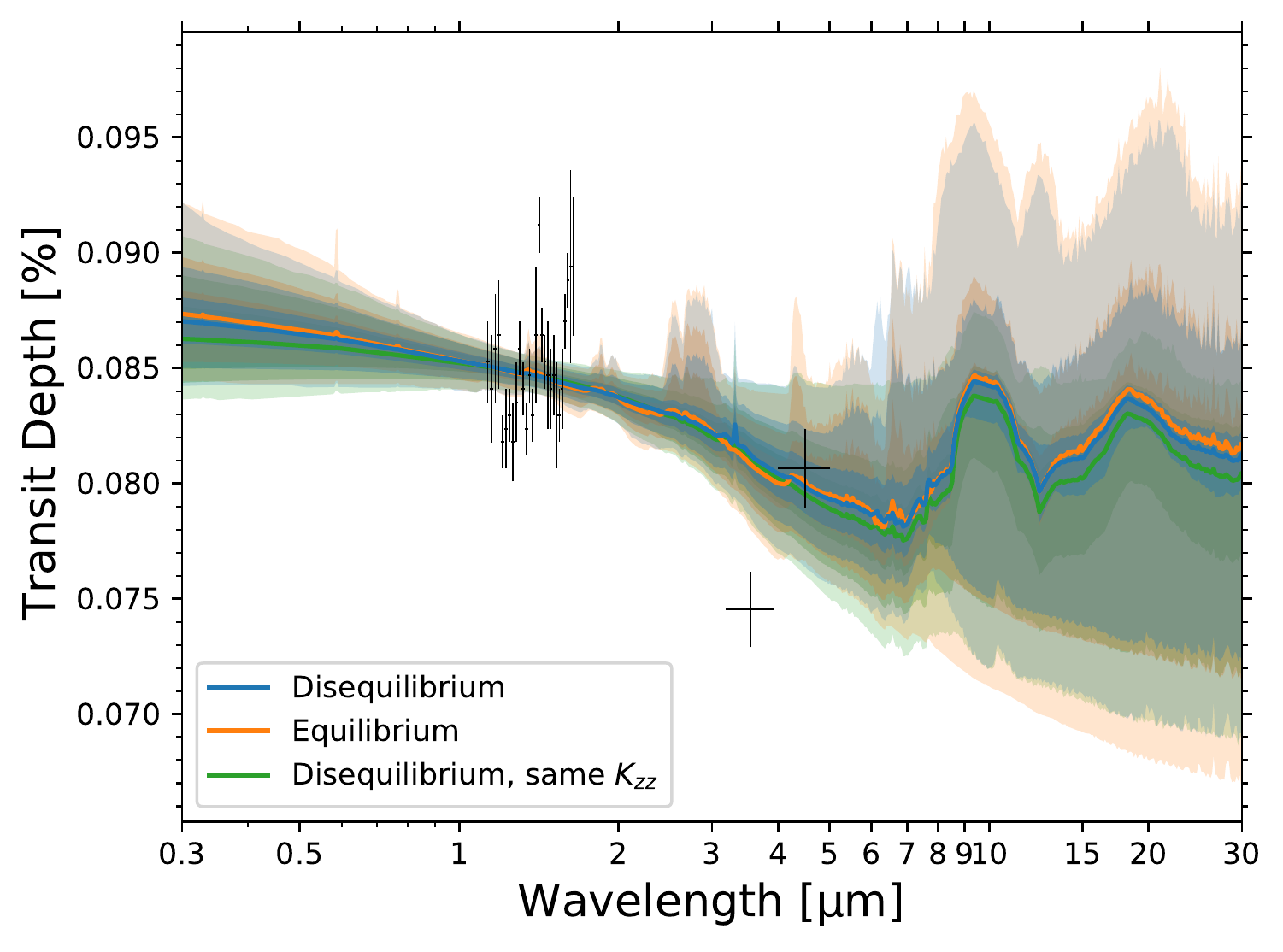}
   \subcaption{HD~97658b}
 \end{minipage}
 \begin{minipage}{0.5\hsize}
  \centering
  \includegraphics[width=\hsize]{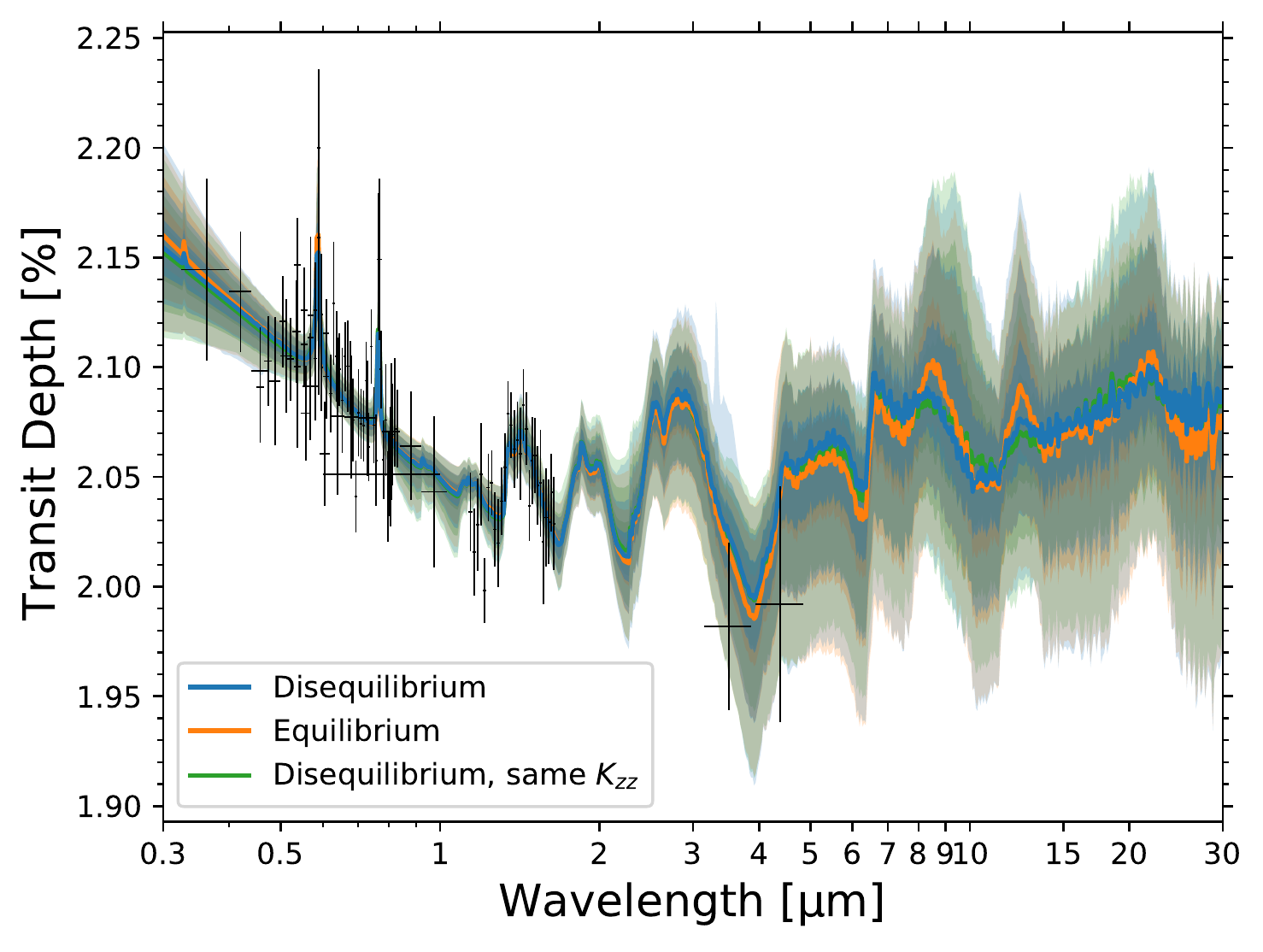}
  \subcaption{WASP-6b}
 \end{minipage}
 \caption{Continued.
 \vspace{5cm}
}
\end{figure*}
\begin{figure*}
\ContinuedFloat
 \begin{minipage}{0.5\hsize}
   \centering
   \includegraphics[width=\hsize]{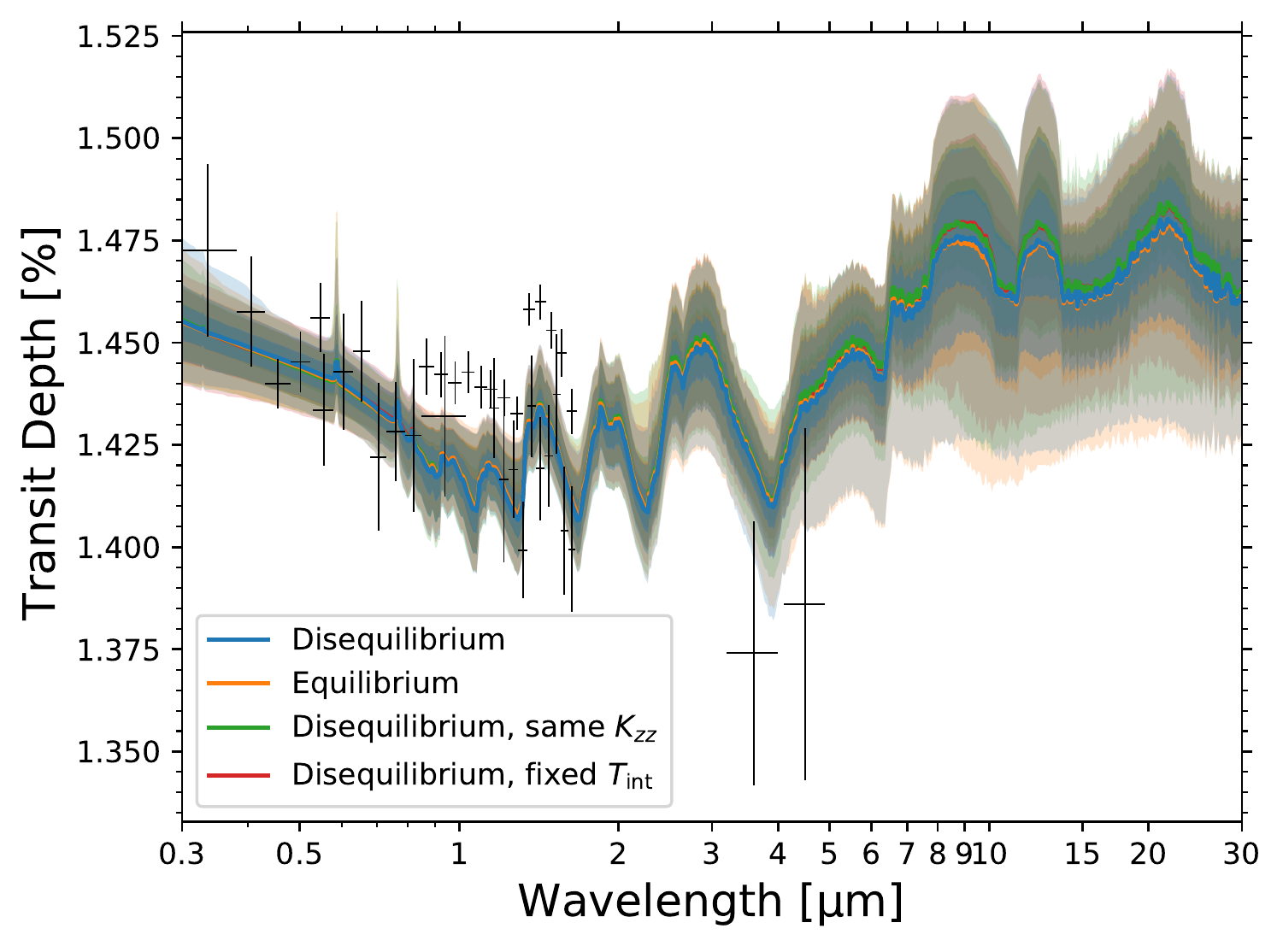}
   \subcaption{WASP-12b}
 \end{minipage}
 \begin{minipage}{0.5\hsize}
   \centering
   \includegraphics[width=\hsize]{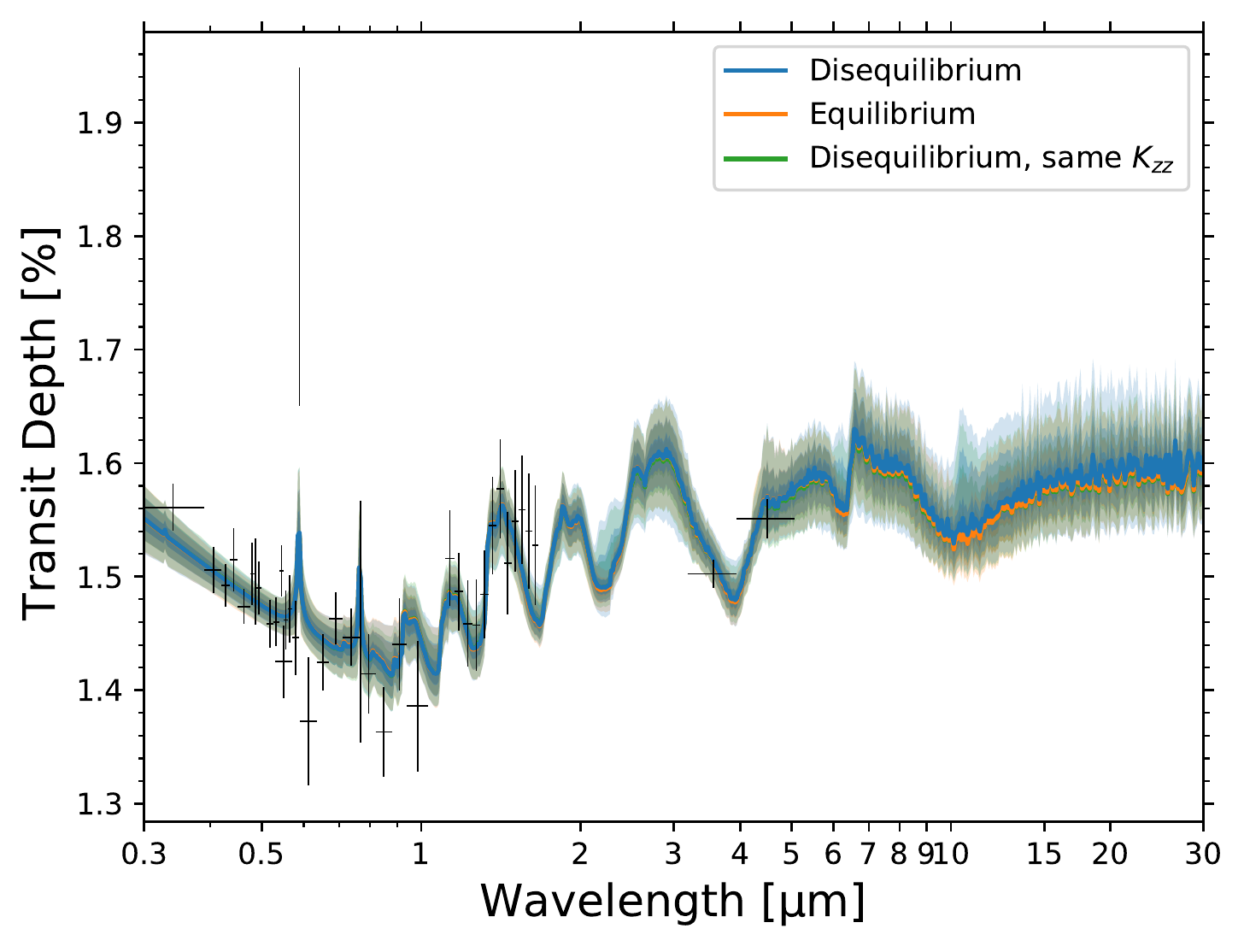}
   \subcaption{WASP-17b}
 \end{minipage}
 \begin{minipage}{0.5\hsize}
   \centering
   \includegraphics[width=\hsize]{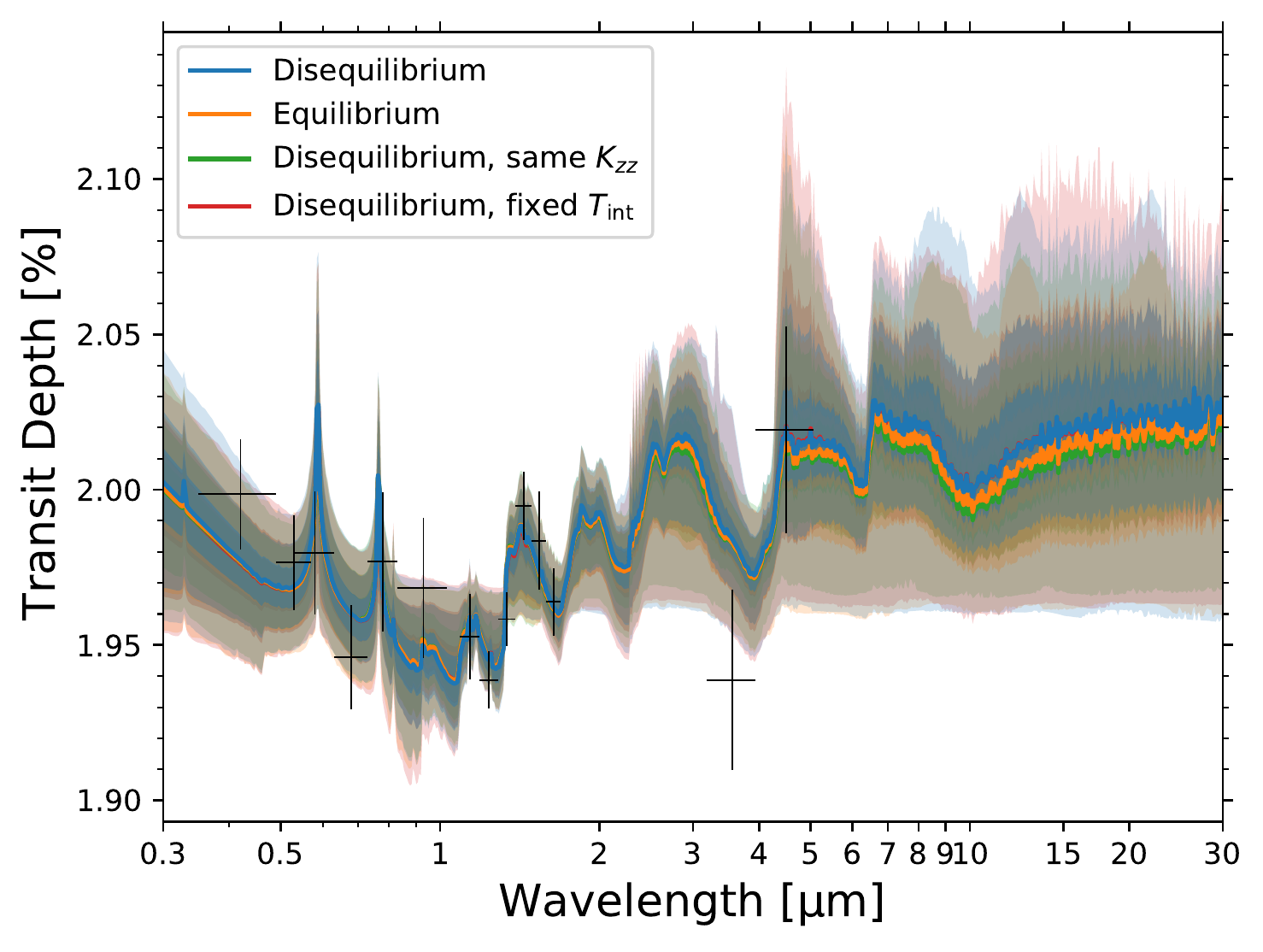}
   \subcaption{WASP-19b}
 \end{minipage}
 \begin{minipage}{0.5\hsize}
   \centering
   \includegraphics[width=\hsize]{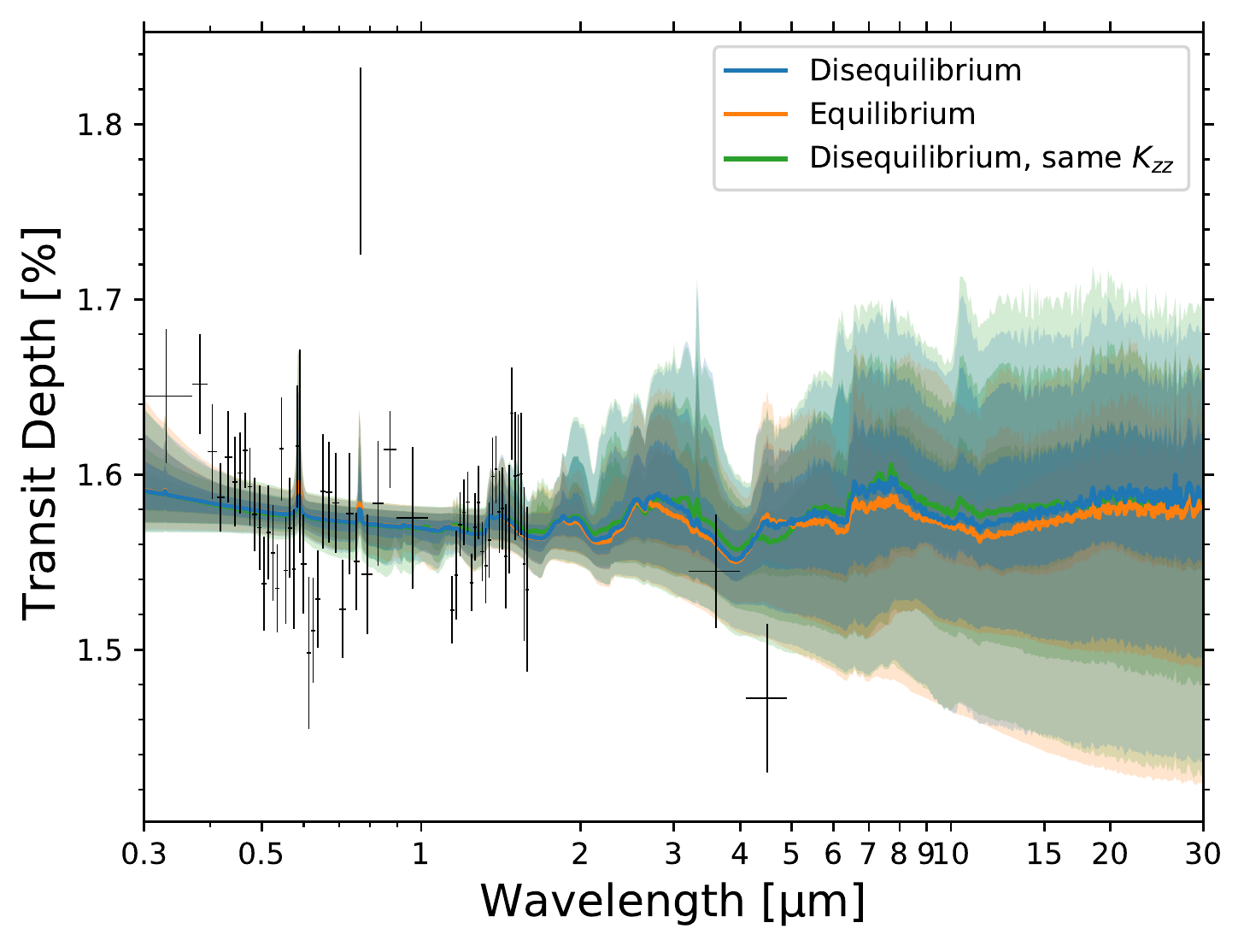}
   \subcaption{WASP-31b}
 \end{minipage}
 \caption{Continued. Note that for (m)~WASP-12b, even though the observed data of \citet{2015ApJ...814...66K} are allowed to vertically shift in each retrieval simulation (see \S~\ref{sec:application} for details), their original values are plotted.}
\end{figure*}

\begin{figure*}
 \begin{subfigure}{\textwidth}
   \centering
   \includegraphics[height=4.6cm]{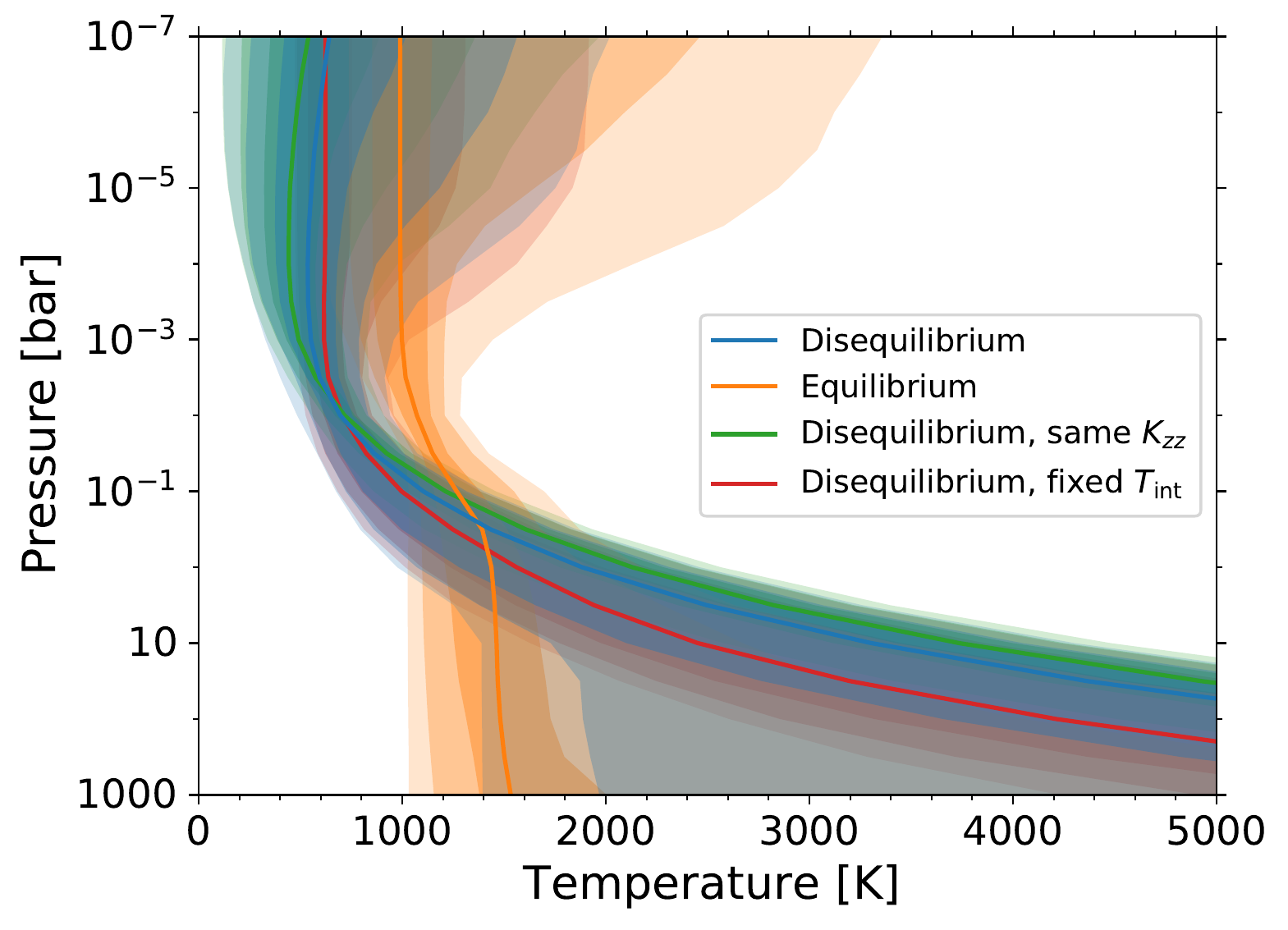}
   \centering
   \includegraphics[height=4.6cm]{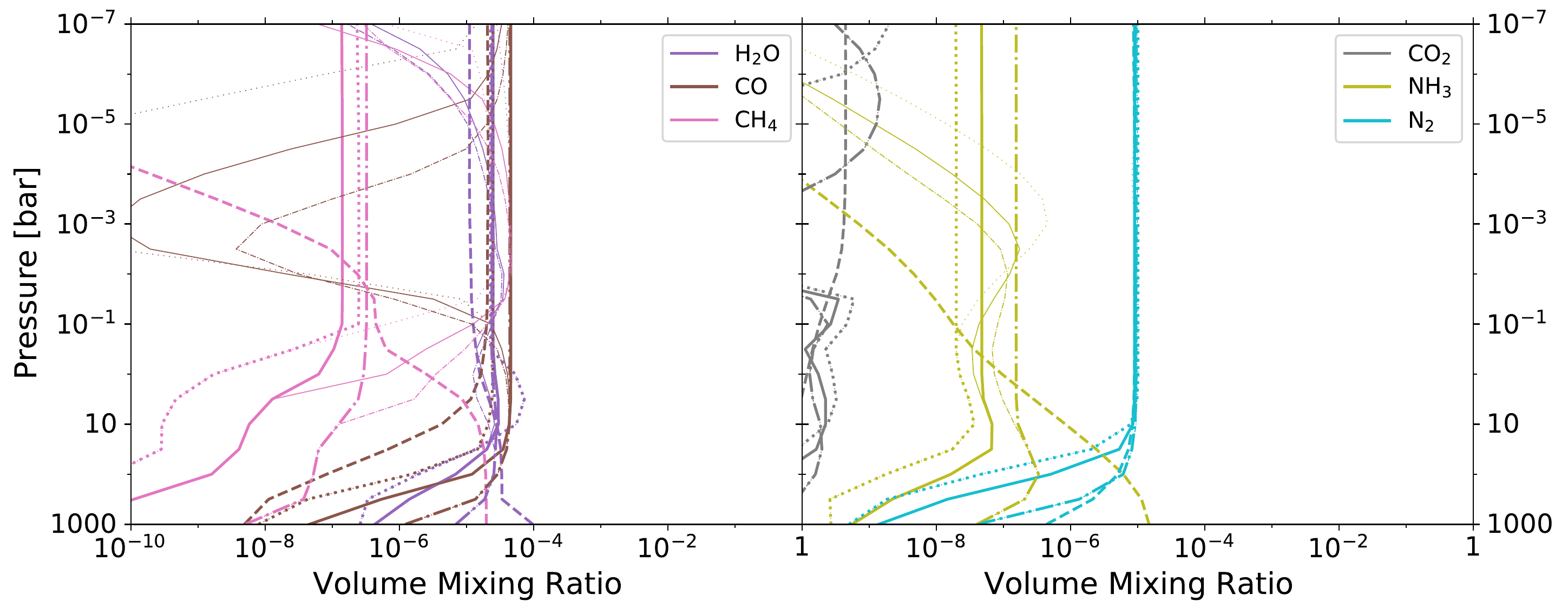}
   \caption{HD~209458b}
 \end{subfigure}
 \begin{subfigure}{\textwidth}
   \centering
   \includegraphics[height=4.6cm]{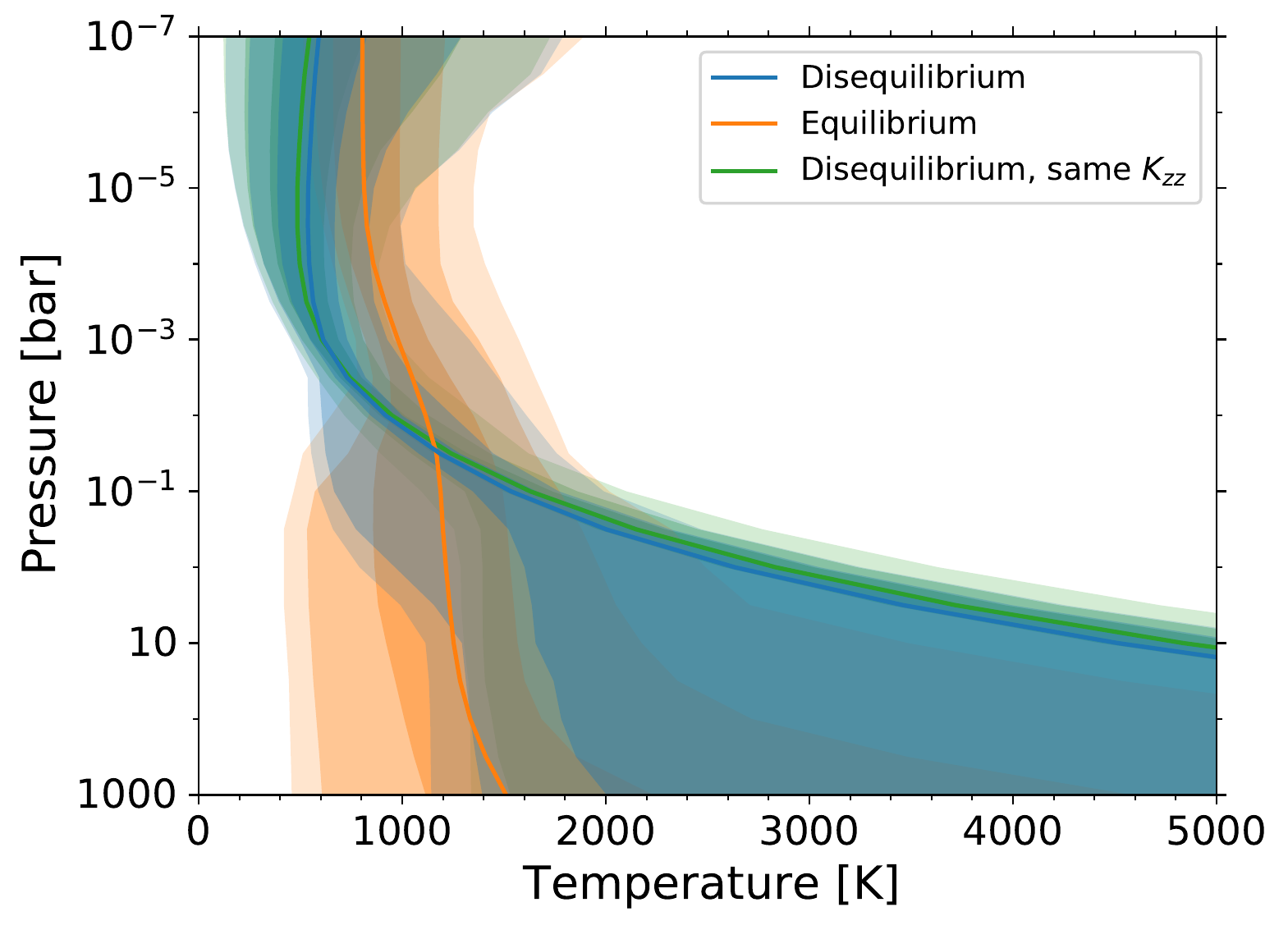}
   \centering
   \includegraphics[height=4.6cm]{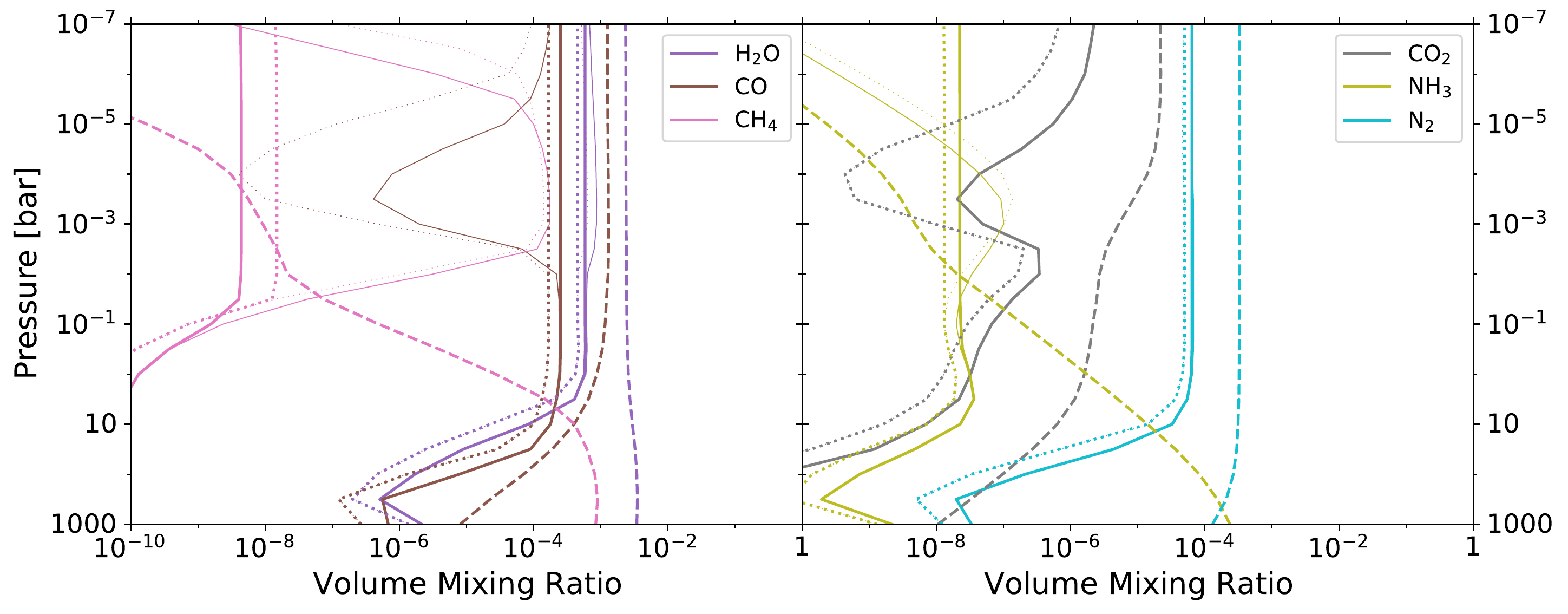}
   \caption{WASP-39b}
 \end{subfigure}
 \begin{subfigure}{\textwidth}
   \centering
   \includegraphics[height=4.6cm]{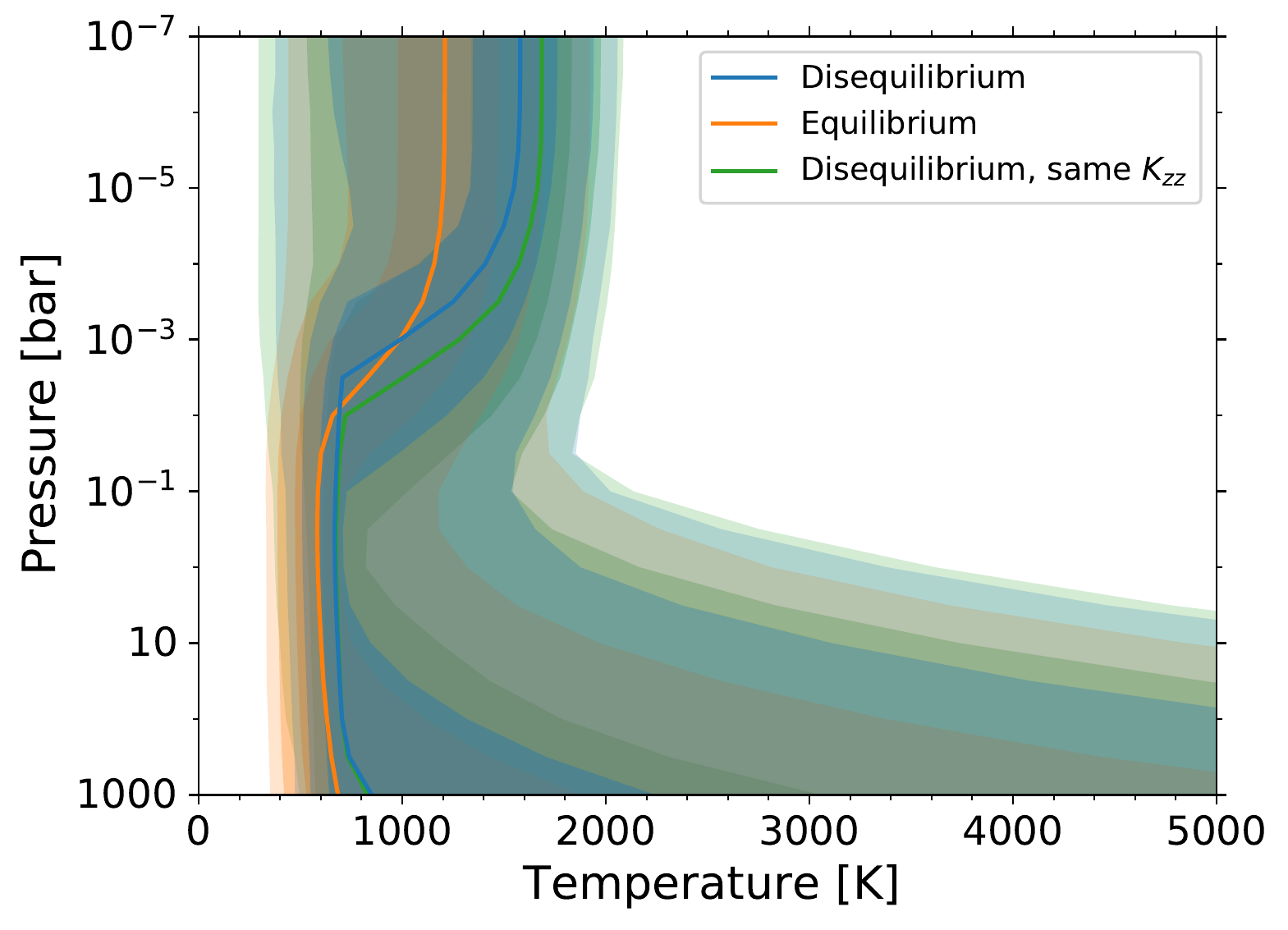}
   \centering
   \includegraphics[height=4.6cm]{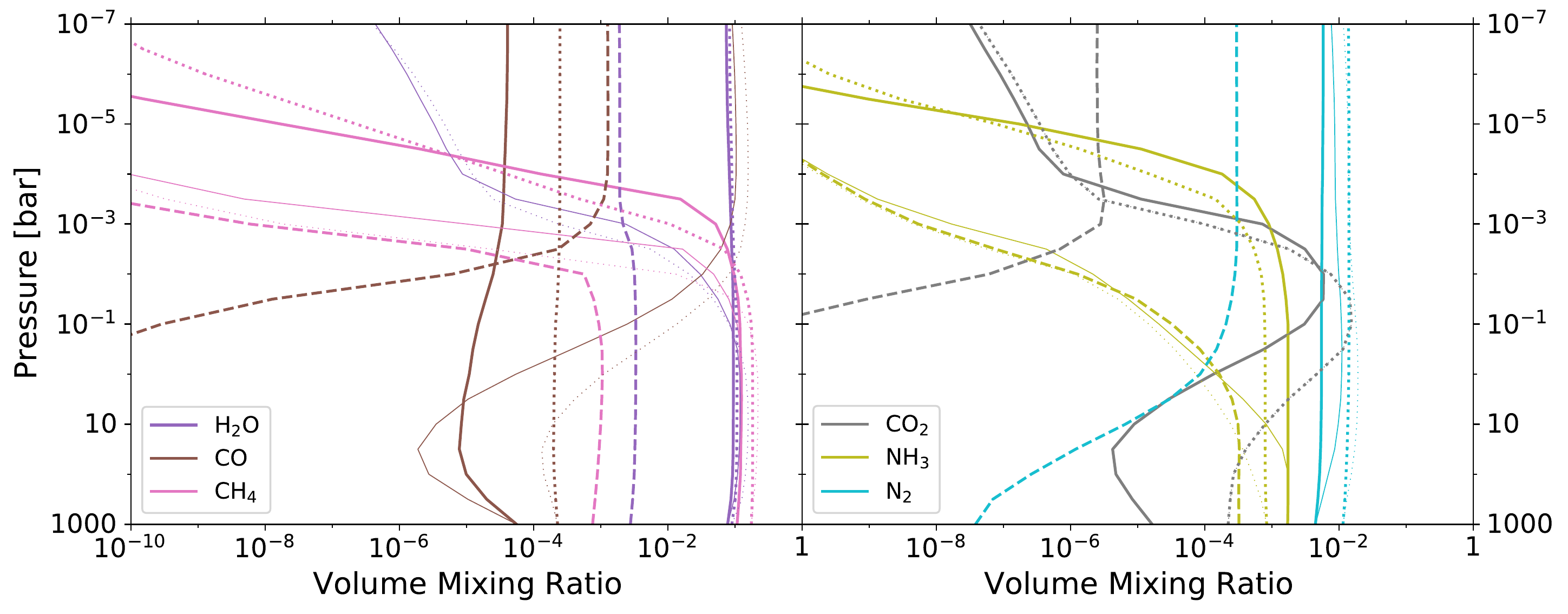}
   \caption{HAT-P-11b}
 \end{subfigure}
 \begin{subfigure}{\textwidth}
   \centering
   \includegraphics[height=4.6cm]{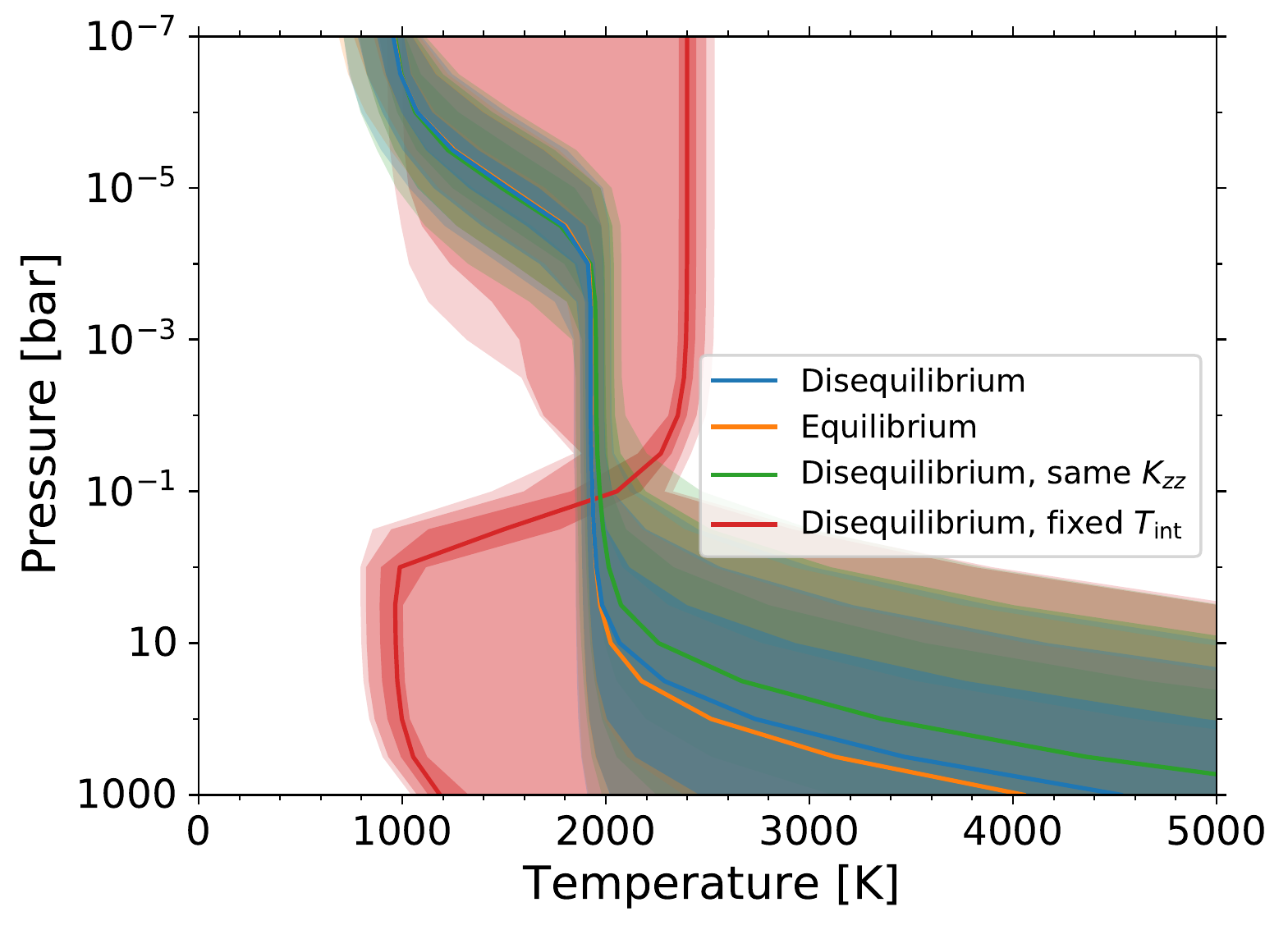}
   \centering
   \includegraphics[height=4.6cm]{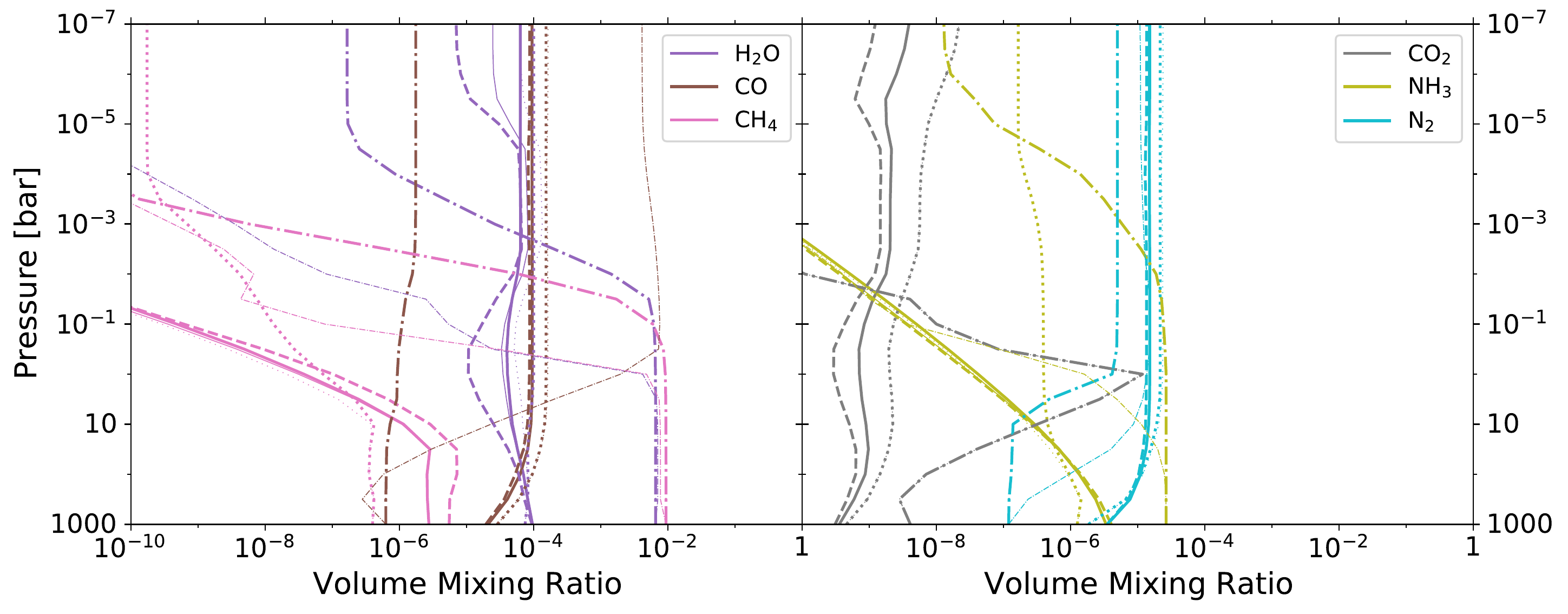}
   \caption{HD~189733b}
 \end{subfigure}
 \caption{\edt{Atmospheric profiles for four representative planets, while those for the other planets are presented in Fig.~\ref{fig:profile2} of Appendix~\ref{sec:profiles}.}
 (Left) Pressure--temperature profiles retrieved from the standard disequilibrium retrieval (blue), equilibrium retrieval (orange), and disequilibrium retrievals with the same $K_{zz}$ (green) and with fixed $T_\mathrm{int}$ (red; only shown for planets with masses larger than half Jupiter mass). The solid lines show the \hl{median} models while the shaded regions indicate the 1, 2, and 3$\sigma$ confidence intervals.
 (Right) \edt{Median} abundance profiles of $\mathrm{H_2O}$, $\mathrm{CO}$, and $\mathrm{CH_4}$ (left) and those of $\mathrm{CO_2}$, $\mathrm{NH_3}$, and $\mathrm{N_2}$ (right) %\hl{at the best-fit case, which is defined as the minimum $\chi^2$,} 
 from the standard disequilibrium retrieval (thick solid lines), equilibrium retrieval (thick dashed lines), and disequilibrium retrievals with the same $K_{zz}$ (thick dotted lines) and with fixed $T_\mathrm{int}$ (thick dash-dotted lines; only shown for planets with masses larger than half Jupiter mass). For reference, \edt{the median abundance profiles for the case where the equilibrium chemistry is imposed}
 %the equilibrium abundance profiles for the best-fit cases of the disequilibrium retrievals 
 are shown with the corresponding thin lines.
 %\edt{Note that those abundance profiles are calculated with the pressure-temperature profiles at the best-fit case, not the median pressure-temperature profiles shown in the right panel.}
 }
 \label{fig:profile}
\end{figure*}

The second column of Table~\ref{table:evidence} presents the natural logs of the Bayes factors between the standard disequilibrium and equilibrium retrievals.
It can be seen that except for HD~209458b, and tentatively \edt{GJ~3470b and} WASP-39b, the differences are almost negligible \citep[cf.][]{2008ConPh..49...71T}, which means that the retrieved parameters from either of the two retrievals are barely favored over the parameters from the other retrieval.
\edt{For HD~209458b and WASP-39b, the natural logs of the Bayes factors indicate that the disequilibrium scenario is favored over the equilibrium scenario by $\geq 4.1\sigma$ and $\geq 2.7\sigma$, respectively, calculated with Eq.~(27) of \citet{2008ConPh..49...71T}.}
\edt{For GJ~3470b, the value of the Bayes factor implies the preferability of the equilibrium scenario with $\geq 3.0\sigma$.
However, as mentioned in \S~\ref{sec:application}, for this planet, a few species still deviate from their equilibrium abundances even for the lower bound of the adopted $K_{zz}^\mathrm{Ch}$ range, namely $10^0$~$\mathrm{cm}^2$~$\mathrm{s}^{-1}$.
Thus, it is uncertain whether the ``true'' equilibrium scenario is indeed favored.
Retrieval simulation extending the lower bound of $K_{zz}^\mathrm{Ch}$ is needed to confirm this.
}
For the retrieved spectra, thermal structures, abundance profiles, and parameters (see Figs.~\ref{fig:spectra}, \ref{fig:profile}, \ref{fig:profile2}, and \ref{fig:corner}), HD~209458b and WASP-39b exhibit relatively large differences between the disequilibrium and equilibrium retrievals while they are largely similar within the uncertainty range for the other planets.
This is consistent with the relatively large eddy diffusion coefficients retrieved for those two planets, such as $\log_{10}{\left( K_{zz}^\mathrm{Ch}~[\mathrm{cm}^2~\mathrm{s}^{-1}] \right)} = 13.12^{+1.39}_{-4.61}$ for HD~209458b and $8.61^{+3.68}_{-4.22}$ for WASP-39b, inferring that disequilibrium chemistry is playing an essential role in their atmospheres.

\begin{table*}
\caption{\hl{N}atural logs of the Bayes factors between the different retrievals}             % title of Table
\label{table:evidence}      % is used to refer this table in the text
\centering                          % used for centering table
\begin{threeparttable}
\begin{tabular}{l c c c}        % centered columns (4 columns)
\hline\hline                 % inserts double horizontal lines
Planet 
& \hl{$\ln{{{E_\mathrm{diseq}}}/{{E_\mathrm{eq}}}}$} 
& \hl{$\ln{{{E_{\mathrm{diseq, same} K_{zz}}}}/{{E_\mathrm{diseq}}}}$}
& \hl{$\ln{{{E_\mathrm{diseq, T_\mathrm{int}}}}/{{E_\mathrm{diseq}}}}$} \\    % table heading 
\hline                        % inserts single horizontal line
GJ~436b & \edt{1.5} & \edt{1.8} & \\
GJ~1214b & $-$1.5 & 0.15 & \\
GJ~3470b & \edt{$-$3.0} & \edt{$-$2.2} & \\
HAT-P-1b & $-$0.26 & $-$0.62 & $-$0.16 \\
HAT-P-11b & $-$0.80 & $-$0.57 & \\
HAT-P-12b & $-$0.12 & $-$2.1 & \\
HAT-P-26b & $-$1.1 & $-$1.7 & \\
HD~97658b & $-$1.3 & 0.28 & \\
HD~189733b & 0.025 & $-$0.89 & $-$28 \\
HD~209458b & 6.9 & $-$1.0 & 0.29 \\
WASP-6b & 0.073 & $-$1.3 & \\
WASP-12b & 0.79 & $-$0.94 & $-$0.37 \\
WASP-17b & $-$0.32 & 1.2 & \\
WASP-19b & $-$0.014 & 0.38 & $-$0.47 \\
WASP-31b & $-$0.16 & $-$0.51 & \\
WASP-39b & 2.4 & 0.47 & \\
\hline                                   %inserts single line
\end{tabular}
\edt{$E_\mathrm{diseq}$, $E_\mathrm{eq}$, $E_{\mathrm{diseq, same} K_{zz}}$, and $E_\mathrm{diseq, T_\mathrm{int}}$ are the Bayesian evidences of the standard disequilibrium retrieval, equilibrium retrieval, and disequilibrium retrievals with the same $K_{zz}$ and with fixed $T_\mathrm{int}$, respectively.}
%\item[1] \hl{Bayesian evidence of} the standard disequilibrium retrieval
%\item[2] \hl{Bayesian evidence of} the equilibrium retrieval
%\item[3] \hl{Bayesian evidence of} the disequilibrium retrieval with same $K_{zz}$
%\item[4] \hl{Bayesian evidence of} the disequilibrium retrieval with fixed $T_\mathrm{int}$
\end{threeparttable}
\end{table*}

For HD~209458b, while the retrieved spectra from both retrievals match {relatively} well in the wavelength range below 1.6~$\mu$m (see Fig.~\ref{fig:spectra}{a}), where precise observational data exists, differences are observed around 2.5--4.0, 6.5--9, and 9--13~$\mu$m.
The discrepancies at the former two wavelength regions mainly come from the difference in the $\mathrm{CH_4}$ abundance.
A comparison of the {pink} thick solid line and {pink} thin solid line in Fig.~\ref{fig:profile}({a}) %in the best-fit case of the disequilibrium retrieval, 
shows that the $\mathrm{CH_4}$ abundance is quenched in the deeper atmosphere with a pressure around 1~bar, resulting in a smaller abundance %in the upper atmosphere 
compared to the equilibrium retrieval {case} ({pink} thick dashed line).
%In the equilibrium retrieval case (orange line), the broad feature at 3.1--4.0~$\mu$m comes from $\mathrm{CH_4}$, whereas in the disequilibrium retrieval case (blue line), $\mathrm{CH_4}$ contributes only to the distinct feature at 3.3~$\mu$m and instead $\mathrm{H_2O}$ is the main absorber over the broad range of 2.5--4.0~$\mu$m.
%This is because in the disequilibrium retrieval case, 
%
%As for the difference between 6.5 and 9~$\mu$m too, the negligible contribution from $\mathrm{CH_4}$ in the disequilibrium retrieval case causes such discrepancy.
%
While quenching reduces the $\mathrm{CH_4}$ abundance in the observable pressure region, it works to maintain a high $\mathrm{NH_3}$ abundance (compare {yellow} thick solid line and {yellow} thin solid line in Fig.~\ref{fig:profile}{a}).
This results in the prominent $\mathrm{NH_3}$ absorption feature at 10.5~$\mu$m for the disequilibrium retrieval case (blue line in Fig.~\ref{fig:spectra}{a}), while that feature hardly exists in the equilibrium retrieval case (orange line).
%Regarding the feature around 10.5~$\mu$m, 
%in the equilibrium retrieval case, it comes from $\mathrm{H_2O}$ on top of the broad collision-induced absorption (CIA) by $\mathrm{H_2}$--$\mathrm{H_2}$ while $\mathrm{NH_3}$ hardly contributes to that feature.
%On the other hand, in the disequilibrium retrieval case, $\mathrm{H_2O}$ and $\mathrm{NH_3}$ equally contribute to the feature at the same wavelength in addition to the $\mathrm{H_2}$--$\mathrm{H_2}$ CIA.
%This is because quenching .
%than its abundance of the best-fit case from the equilibrium retrieval (purple thick dashed line in the same figure).

\edt{To understand which observed features require the condition of disequilibrium chemistry for HD~209458b, we have additionally performed standard disequilibrium and equilibrium retrievals systematically excluding each observed data. Natural logs of the Bayes factors for those additional retrievals are presented in Table~\ref{table:evidence_wo}.
It can be seen that the data of {HST}/WFC3/G141 ($1.1$--$1.6$~$\mu$m) requires the disequilibrium chemistry condition most for the case of HD~209458b while it is also important for the {HST}/STIS data  ($0.3$--$0.9$~$\mu$m).
We have confirmed that the best-fit parameter set, which yields the minimum $\chi^2$ when compared to the observed data, from the equilibrium retrieval explains the observed $1.3$--$1.6$~$\mu$m feature by $\mathrm{H_2O}$ alone.
On the other hand, in the disequilibrium retrieval case, $\mathrm{NH_3}$ also partially contributes to reproducing the feature, yielding a better match to the observed data and thus demonstrating the preferability of the disequilibrium retrieval. 
Here we mention that this slight evidence regarding the $\mathrm{NH_3}$ feature was already noted by \citet{2017MNRAS.469.1979M}, who used the same data analyzed by \citet{Sing:2016hi} for their retrieval simulations and raised the possibility of disequilibrium chemistry playing a role in the atmosphere of this planet.
}

\begin{table}
\caption{\edt{Natural logs of the Bayes factors between the disequilibrium and equilibrium retrievals for the results without a specific data set for HD~209458b}}             % title of Table
\label{table:evidence_wo}      % is used to refer this table in the text
\centering                          % used for centering table
%\begin{threeparttable}
\begin{tabular}{l c}        % centered columns (4 columns)
\hline\hline                 % inserts double horizontal lines
\edt{Data}
& \edt{$\ln{{{E_\mathrm{diseq}}}/{{E_\mathrm{eq}}}}$} 
\\    % table heading 
\hline                        % inserts single horizontal line
\edt{all} & \edt{6.9} \\
\edt{without {HST}/STIS ($0.3$--$0.9$~$\mu$m)} & \edt{0.35} \\
\edt{without {HST}/WFC3/G141 ($1.1$--$1.6$~$\mu$m)} & \edt{$-$0.12} \\
\edt{without {\textit{Spitzer}}/IRAC/Ch1 (3.6~$\mu$m)} & \edt{6.8} \\
\edt{without {\textit{Spitzer}}/IRAC/Ch2 (4.5~$\mu$m)} & \edt{6.9} \\
% \edt{WASP-39b} & \edt{all} & \edt{2.4} \\
%  & \edt{without {HST}/STIS ($\mu$m)} & \edt{-1.1} \\
%  & \edt{without {HST}/WFC3/G102 ($\mu$m)} &  \\
%  & \edt{without {HST}/WFC3/G141 ($\mu$m)} & \edt{-0.41} \\
%  & \edt{without {Spitzer}/IRAC/Ch1 (3.6~$\mu$m)} & \edt{1.0} \\
%  & \edt{without {Spitzer}/IRAC/Ch2 (4.5~$\mu$m)} & \edt{2.3} \\
\hline                                   %inserts single line
\end{tabular}
%\item[1] \hl{Bayesian evidence of} the standard disequilibrium retrieval
%\item[2] \hl{Bayesian evidence of} the equilibrium retrieval
%\item[3] \hl{Bayesian evidence of} the disequilibrium retrieval with same $K_{zz}$
%\item[4] \hl{Bayesian evidence of} the disequilibrium retrieval with fixed $T_\mathrm{int}$
%\end{threeparttable}
\end{table}

\hl{Recently, \citet{2021Natur.592..205G} {performed} high-resolution transmission spectroscopy of HD~209458b and reported the detection of six species, $\mathrm{H_2O}$, $\mathrm{CO}$, $\mathrm{HCN}$, $\mathrm{CH_4}$, $\mathrm{NH_3}$, and $\mathrm{C_2H_2}$.
In their analysis, the disequilibrium scenario was strongly disfavored, though they did not deny the possibility that disequilibrium processes were in effect to some extent.
This discrepancy may partly be because they assumed specific thermal and abundance profiles for their disequilibrium scenario while we allow those profiles to vary within our retrieval.}

Next, regarding WASP-39b, noticeable differences are found around 0.3--0.5, 2.7, 4.3, and 15~$\mu$m
%and the longward of 17~$\mu$m 
(see Fig.~\ref{fig:spectra}b).
%exclude 10.5 and longward of 17
In the case of the equilibrium retrieval (orange line), the favored cloud \edt{distributions contribute to} %opacity 
\edt{flattening} the Rayleigh scattering slope and fail to reproduce the observed steepness of the optical slope.
The small difference at 2.7~$\mu$m arises from the smaller $\mathrm{H_2O}$ abundance in the disequilibrium retrieval case (compare the purple thick solid line and purple thick dashed line in Fig.~\ref{fig:profile}b).
Due to the absence of $\mathrm{CO_2}$ in the upper atmosphere (gray thick solid line in the same figure), its strong features at 4.3 and 15~$\mu$m hardly exist in the disequilibrium retrieval case (blue line in Fig.~\ref{fig:spectra}b).
%10 from higher temperature despite larger $\mathrm{H_2O}$ absorption in the eq. in this case cooler temperature in the lower atmosphere dakara?
%\yuic{check longward of 17 and 10. 10 not seem to be NH3}.
%17: H2o contribution larger in the eq.
%same as 10
%10.5 and 17 baseline difficulr.

%\edt{For WASP-39b, when judging from the values of natural log of the Bayes factors between the standard disequilibrium and equilibrium retrievals for the data without a specific data set (Table~\ref{table:evidence_wo}), the data of {HST}/STIS requires the disequilibrium chemistry most.}

% \edt{As for the retrieved thermal profiles, although the number of the samples is limited, higher temperatures in the lower atmospheres and lower temperatures in the upper atmospheres are favoured for the disequilibrium retrievals when compared to the equilibrium retrieval cases besides the case of HD~209458b, such as those of WASP-39b, HAT-P-26b, HD~97658b, GJ~1214b, and GJ~3470b. We consider that this is also due to the similar reason for the case of HD~209458b, namely behaviours of the abundances of $\mathrm{NH_3}$ and/or $\mathrm{CH_4}$ under disequilibrium chemistry as we have explained above.}
% \edt{Here we note that given the stronger features of $\mathrm{NH_3}$ and $\mathrm{CH_4}$ at longer wavelengths such as the $\mathrm{NH_3}$ feature at 10.5~$\mu$m and $\mathrm{CH_4}$ feature at 3.3~$\mu$m, the absence of their features at the wavelength range of {HST}/WFC3/G141 and the existence of their features at longer wavelengths would enable us to tightly constrain the eddy diffusivity in the atmosphere.}

\edt{For the retrieved thermal profiles, although the number of samples is limited, a higher temperature in the lower atmosphere is favored for the disequilibrium retrieval, while the equilibrium retrieval prefers an isothermal-like profile for several planet samples such as HD~209458b, WASP-39b, GJ~1214b, GJ~3470b, HAT-P-26b, and HD~97658b (see Figs.~\ref{fig:profile} and ~\ref{fig:profile2}).
We speculate that this is due to the absence of $\mathrm{NH_3}$ and/or $\mathrm{CH_4}$ absorption features for some of the above planets.
This is because if moderate vertical diffusion is imposed for the cool lower atmosphere, such as the equilibrium retrieval cases of the above planets, their abundances would be quenched to the larger values because of their stability at low temperatures and high pressures.
Thus, the temperature in the lower atmosphere needs to be increased to reduce their abundances.}

The planets for which at least one of the retrieved parameters differs by more than $1\sigma$ between the standard disequilibrium and equilibrium retrievals are GJ~436b, GJ~1214b, \edt{GJ~3470b}, HAT-P-11b, HAT-P-12b, HAT-P-26b, HD~209458b, WASP-6b, and WASP-39b, and their differences are summarized in Table~\ref{table:param_diff}.
Aside from HD~209458b and WASP-39b, some differences are also found for HAT-P-11b \edt{and GJ~436b}. 
For \edt{HAT-P-11b}, the median values of C/O ratio and metallicity differ by 2.9 and $1.4\sigma$, respectively.
While the retrieved C/O ratio from the disequilibrium retrieval is $0.82^{+0.26}_{-0.42}$, that from the equilibrium retrieval is $0.27^{+0.19}_{-0.12}$.
For the metallicity, the retrieved values from the disequilibrium and equilibrium retrievals are
$\log_{10}{\left( Z~[\mathrm{dex}] \right)} = 2.39^{+0.27}_{-0.34}$ and $0.67^{+1.24}_{-0.68}$, respectively.
For this planet, even though the retrieved eddy diffusion coefficient is relatively small ($\log_{10}{\left( K_{zz}^\mathrm{Ch}~[\mathrm{cm}^2~\mathrm{s}^{-1}] \right)} = 3.74^{+2.67}_{-2.29}$), because of the somewhat low temperature, disequilibrium chemistry plays an important role %in the best-fit case from the disequilibrium retrieval 
(compare the thick and thin solid lines in Fig.~\ref{fig:profile}c).
However, \hl{the} quite small value of the natural log of the Bayes factor in Table~\ref{table:evidence} indicates that the retrieved parameters from the disequilibrium retrieval are never favored over those from the equilibrium case and vice versa. A further observational constraint is needed to determine the atmospheric properties of this planet.
For this purpose, the search for the $\mathrm{CO_2}$ feature at 4.3~$\mu$m is promising because of its absence in the disequilibrium retrieval case (Fig.~\ref{fig:spectra}c).
\edt{For GJ~436b, while the irradiation parameter from the equilibrium retrieval is ${0.01}^{+0.01}_{-0.01}$, that from the disequilibrium retrieval is ${0.06}^{+0.06}_{-0.04}$, resulting in a $4.5\sigma$ difference of this parameter and different retrieved thermal structure (Fig.~\ref{fig:profile2}e).
%This increase of the irradiation parameter leads to the different thermal structure shown in Figure~\ref{fig:profile2}(i).
However, as for the case of HAT-P-11b, the natural log of the Bayes factor is too small to draw a firm conclusion on the parameter.
High-precision observations in the optical wavelength range, where the spectra from the two retrievals show a difference, would be promising to distinguish this.}
These examples raise the possibility that ignoring the effect of disequilibrium chemistry can lead to an incorrect constraint on the atmospheric properties.
%, especially for the metallicity and elemental abundance ratios because of the large impact of the quenching effect on those parameters.

\begingroup
\renewcommand{\arraystretch}{1.3}
\begin{table*}
\caption{Retrieved parameters with more than 1$\sigma$ difference between the standard disequilibrium (diseq.) and equilibrium (eq.) retrievals}             % title of Table
\label{table:param_diff}      % is used to refer this table in the text
\centering                          % used for centering table
\begin{tabular}{l c r r c}        % centered columns (4 columns)
\hline\hline                 % inserts double horizontal lines
Planet & Parameter & Value from diseq. & Value from eq. & Difference \\    % table heading 
\hline
\hline
GJ~436b & $f_\mathrm{irr}$ & ${0.06}^{+0.06}_{-0.04}$ & ${0.01}^{+0.01}_{-0.01}$ & $4.5\sigma$ \\
& $\log_{10}{\left( \kappa_\mathrm{IR}~[\mathrm{cm}^2~\mathrm{g}^{-1}] \right)}$ & ${-1.51}^{+1.79}_{-1.63}$ & ${0.28}^{+2.39}_{-1.78}$ & $1.0\sigma$ \\
& $Z$~[dex] & ${1.63}^{+0.26}_{-0.34}$ & ${1.25}^{+0.32}_{-0.36}$ & $1.2\sigma$ \\
\hline
GJ~1214b & $\log_{10}{\left( T_\mathrm{int}~[K] \right)}$ & ${2.15}^{+0.36}_{-0.43}$ & ${1.54}^{+0.48}_{-0.36}$ & $1.3\sigma$ \\
& $R_\mathrm{ref}$~[$R_\mathrm{J}$] & ${0.24}^{+0.00}_{-0.00}$ & ${0.25}^{+0.00}_{-0.00}$ & $1.3\sigma$ \\
\hline
GJ~3470b & $\log_{10}{\left( \kappa_\mathrm{IR}~[\mathrm{cm}^2~\mathrm{g}^{-1}] \right)}$ & ${-1.45}^{+1.67}_{-1.01}$ & ${-2.34}^{+0.83}_{-0.85}$ & $1.1\sigma$ \\
& $\log_{10}{\left( T_\mathrm{int}~[K] \right)}$ & ${2.53}^{+0.47}_{-0.76}$ & ${1.72}^{+0.51}_{-0.46}$ & $1.6\sigma$ \\
& $R_\mathrm{ref}$~[$R_\mathrm{J}$] & ${0.32}^{+0.01}_{-0.01}$ & ${0.33}^{+0.01}_{-0.01}$ & $2.0\sigma$ \\
\hline
HAT-P-11b & C/O & ${0.82}^{+0.26}_{-0.42}$ & ${0.27}^{+0.19}_{-0.12}$ & $2.9\sigma$ \\
& $Z$~[dex] & ${2.39}^{+0.27}_{-0.34}$ & ${0.67}^{+1.24}_{-0.68}$ & $1.4\sigma$ \\
& $R_\mathrm{ref}$~[$R_\mathrm{J}$] & ${0.39}^{+0.00}_{-0.00}$ & ${0.39}^{+0.00}_{-0.01}$ & $1.2\sigma$ \\
\hline
HAT-P-12b & $Z$~[dex] & ${-0.19}^{+1.64}_{-0.61}$ & ${1.55}^{+0.28}_{-1.52}$ & $1.1\sigma$ \\
\hline
HAT-P-26b & $R_\mathrm{ref}$~[$R_\mathrm{J}$] & ${0.46}^{+0.02}_{-0.02}$ & ${0.49}^{+0.01}_{-0.01}$ & $1.5\sigma$ \\
\hline
HD~209458b & $\log_{10}{\gamma}$ & ${0.77}^{+0.80}_{-1.15}$ & ${-0.77}^{+0.45}_{-0.37}$ & $3.4\sigma$ \\
& $f_\mathrm{irr}$ & ${0.01}^{+0.01}_{-0.00}$ & ${0.09}^{+0.05}_{-0.04}$ & $2.3\sigma$ \\
& $\log_{10}{\left( \kappa_\mathrm{IR}~[\mathrm{cm}^2~\mathrm{g}^{-1}] \right)}$ & ${1.51}^{+1.59}_{-1.91}$ & ${-1.49}^{+0.80}_{-0.77}$ & $3.7\sigma$ \\
& $\log_{10}{\left( T_\mathrm{int}~[K] \right)}$ & ${2.11}^{+0.42}_{-0.39}$ & ${1.59}^{+0.46}_{-0.39}$ & $1.1\sigma$ \\
& C/O & ${0.50}^{+0.21}_{-0.28}$ & ${0.19}^{+0.14}_{-0.07}$ & $2.3\sigma$ \\
& $R_\mathrm{ref}$~[$R_\mathrm{J}$] & ${1.35}^{+0.02}_{-0.01}$ & ${1.36}^{+0.01}_{-0.01}$ & $2.1\sigma$ \\
& $\log_{10}{\left( K_{zz}^\mathrm{Cl}~[\mathrm{cm}^2~\mathrm{s}^{-1}] \right)}$ & ${5.46}^{+0.26}_{-0.30}$ & ${5.89}^{+0.13}_{-0.15}$ & $2.9\sigma$ \\
\hline
WASP-6b & $\log_{10}{\left( T_\mathrm{int}~[K] \right)}$ & ${1.91}^{+0.41}_{-0.58}$ & ${1.49}^{+0.34}_{-0.30}$ & $1.3\sigma$ \\
& $\log_{10}{\left( K_{zz}^\mathrm{Cl}~[\mathrm{cm}^2~\mathrm{s}^{-1}] \right)}$ & ${7.27}^{+1.12}_{-1.11}$ & ${8.15}^{+1.99}_{-0.82}$ & $1.1\sigma$ \\
\hline
WASP-39b & $\log_{10}{\gamma}$ & ${0.50}^{+1.03}_{-2.01}$ & ${-0.90}^{+1.30}_{-0.49}$ & $1.1\sigma$ \\
& $f_\mathrm{irr}$ & ${0.02}^{+0.03}_{-0.01}$ & ${0.07}^{+0.07}_{-0.04}$ & $1.6\sigma$ \\
& $\log_{10}{\left( \kappa_\mathrm{IR}~[\mathrm{cm}^2~\mathrm{g}^{-1}] \right)}$ & ${1.44}^{+1.66}_{-2.06}$ & ${-0.38}^{+0.98}_{-2.37}$ & $1.9\sigma$ \\
& $\log_{10}{\left( T_\mathrm{int}~[K] \right)}$ & ${2.04}^{+0.43}_{-0.34}$ & ${1.52}^{+0.48}_{-0.35}$ & $1.1\sigma$ \\
& Si/O & ${0.01}^{+0.01}_{-0.01}$ & ${0.06}^{+0.06}_{-0.04}$ & $1.2\sigma$ \\
& $Z$~[dex] & ${-0.04}^{+0.35}_{-0.26}$ & ${0.73}^{+0.75}_{-0.65}$ & $1.2\sigma$ \\
& $R_\mathrm{ref}$~[$R_\mathrm{J}$] & ${1.20}^{+0.05}_{-0.02}$ & ${1.27}^{+0.03}_{-0.03}$ & $2.5\sigma$ \\
& $\log_{10}{\left( g~[\mathrm{cm}~\mathrm{s}^{-2}] \right)}$ & ${2.78}^{+0.03}_{-0.05}$ & ${2.73}^{+0.04}_{-0.06}$ & $1.1\sigma$ \\
& $\log_{10}{\left( K_{zz}^\mathrm{Cl}~[\mathrm{cm}^2~\mathrm{s}^{-1}] \right)}$ & ${5.59}^{+0.48}_{-0.38}$ & ${6.44}^{+0.31}_{-0.47}$ & $1.8\sigma$ \\
\hline                                %inserts single line
\end{tabular}
\end{table*}
\endgroup

\subsection{Comparison of disequilibrium retrievals with and without imposing the same $K_{zz}$ for chemistry and cloud calculations} \label{sec:same}
In this subsection, we compare the results from the standard disequilibrium retrieval and a retrieval where the condition $K_{zz}^\mathrm{Ch} = K_{zz}^\mathrm{Cl}$ is imposed to examine how this affects the retrieval of parameters.
As shown in the third column of Table~\ref{table:evidence}, the absolute values of the natural logs of the Bayes factors between the two retrievals are almost negligible for all of our samples \citep[cf.][]{2008ConPh..49...71T}, indicating that the retrieved parameters from either of the two retrievals are barely favored over those from the other.

The planets for which at least one of the retrieved parameters differs by more than $1\sigma$ between the same-$K_\mathrm{zz}$ and standard disequilibrium retrievals, and their differences are summarized in Table~\ref{table:param_diff_same}.
As expected, employing the condition $K_{zz}^\mathrm{Ch} = K_{zz}^\mathrm{Cl}$ strongly affects the retrieval of these two parameters and the cloud nucleation rate.
Large ($\gtrsim 3\sigma$) differences are found for \edt{GJ~436b}, GJ~1214b, HAT-P-12b, HD~189733b, HD~209458b, WASP-17b, WASP-19b, and WASP-39b. For all the planets, those differences are for $K_{zz}^\mathrm{Cl}$ and $K_{zz}^\mathrm{Ch}$ as well as $\dot{\Sigma}$ for HD~189733b.
%This is because of the relatively strong constraint on $K_{zz}^\mathrm{Cl}$ due to its strong effect on the cloud structure \citep{2018ApJ...863..165G, 2019A&A...622A.121O}.

For HD~209458b, for which we find %the tentative evidence 
\edt{an indication} of disequilibrium chemistry from the comparison between the standard disequilibrium and equilibrium retrievals in \S~\ref{sec:diseq}, the retrieved value of $\log_{10}{\left( K_{zz}^\mathrm{Ch}~[\mathrm{cm}^2~\mathrm{s}^{-1}] \right)}$ decreases from ${13.12}^{+1.39}_{-4.61}$ to ${4.17}^{+1.15}_{-1.05}$ with a $1.9\sigma$ difference.
%Despite this decline, however, t
Despite this decrease, the spectrum from the same-$K_{zz}$ retrieval (green line in Fig.~\ref{fig:spectra}a) 
%is quite similar to that from the standard disequilibrium retrieval (blue line in the same figure) and it 
still exhibits certain disequilibrium features, as discussed in \S~\ref{sec:diseq}, such as those at 2.5--4.0, 6.5--9, and 9--13~$\mu$m.
%\hl{The values of $\log_{10}{K_{zz}^\mathrm{Ch}}$ in the best-fit cases of the two retrievals are somewhat comparable (5.47 for the same-$K_{zz}$ retrieval while 10.91 for the standard one),}
%the slightly cooler thermal structure retrieved for the same-$K_{zz}$ retrieval (Figure~\ref{fig:profile}s) allows the smaller $K_{zz}^\mathrm{Ch}$ to cause a similar degree of disequilibrium effect, 
%resulting in similar molecular abundance profiles between the two retrievals (compare thick solid and thick dotted lines in Figure~\ref{fig:profile}n).

Also, for WASP-39b, for which we also find %the tentative evidence 
\edt{an indication} of disequilibrium chemistry in \S~\ref{sec:diseq}, the retrieved spectra, thermal structures, abundance profiles, and parameters (see Figs.~\ref{fig:spectra}b, \ref{fig:profile}b, and \ref{fig:corner}b) are quite similar to the standard disequilibrium retrieval case.
Thus, our findings of %tentative evidence 
\edt{indications} of disequilibrium chemistry for HD~209458b and WAS-39b still hold.

\begingroup
\renewcommand{\arraystretch}{1.3}
\begin{table*}
\caption{Retrieved parameters with more than 1$\sigma$ difference between the same-$K_{zz}$ and standard disequilibrium retrievals}             % title of Table
\label{table:param_diff_same}      % is used to refer this table in the text
\centering                          % used for centering table
\begin{threeparttable}
\begin{tabular}{l c r r c}        % centered columns (4 columns)
\hline\hline                 % inserts double horizontal lines
Planet & Parameter & Value from same-$K_{zz}$ & Value from standard & Difference \\    % table heading 
\hline
\hline
GJ~436b & $\log_{10}{\left( K_{zz}^\mathrm{Cl}~[\mathrm{cm}^2~\mathrm{s}^{-1}] \right)}$ & ${2.37}^{+1.73}_{-1.54}$ & ${5.33}^{+0.40}_{-0.23}$ & $12.9\sigma$ \\
& $\log_{10}{\left( K_{zz}^\mathrm{Ch}~[\mathrm{cm}^2~\mathrm{s}^{-1}] \right)}$ & ${2.37}^{+1.73}_{-1.54}$ & ${7.77}^{+4.77}_{-5.05}$ & $1.1\sigma$ \\
\hline
GJ~1214b & $\log_{10}{\left( K_{zz}^\mathrm{Cl}~[\mathrm{cm}^2~\mathrm{s}^{-1}] \right)}$ & ${14.07}^{+0.62}_{-6.55}$ & ${6.72}^{+2.06}_{-0.88}$ & $3.6\sigma$ \\
& $\log_{10}{\left( K_{zz}^\mathrm{Ch}~[\mathrm{cm}^2~\mathrm{s}^{-1}] \right)}$ & ${14.07}^{+0.62}_{-6.55}$ & ${8.83}^{+3.98}_{-5.05}$ & $1.3\sigma$ \\
\hline
HAT-P-11b & $\log_{10}{\left( K_{zz}^\mathrm{Cl}~[\mathrm{cm}^2~\mathrm{s}^{-1}] \right)}$ & ${5.61}^{+0.96}_{-2.58}$ & ${6.28}^{+0.77}_{-0.64}$ & $1.0\sigma$ \\
\hline
HAT-P-12b & $\log_{10}{\left( \kappa_\mathrm{IR}~[\mathrm{cm}^2~\mathrm{g}^{-1}] \right)}$ & ${-2.27}^{+0.85}_{-0.89}$ & ${-3.09}^{+0.72}_{-0.61}$ & $1.2\sigma$ \\
& $\log_{10}{\left( T_\mathrm{int}~[K] \right)}$ & ${2.86}^{+0.20}_{-0.26}$ & ${1.97}^{+0.66}_{-0.59}$ & $1.4\sigma$ \\
& Si/O & ${0.03}^{+0.04}_{-0.02}$ & ${0.14}^{+0.09}_{-0.09}$ & $1.2\sigma$ \\
& $Z$~[dex] & ${1.85}^{+0.23}_{-0.26}$ & ${-0.19}^{+1.64}_{-0.61}$ & $1.2\sigma$ \\
& $R_\mathrm{ref}$~[$R_\mathrm{J}$] & ${0.84}^{+0.02}_{-0.02}$ & ${0.86}^{+0.01}_{-0.02}$ & $1.1\sigma$ \\
& $\log_{10}{\left( K_{zz}^\mathrm{Cl}~[\mathrm{cm}^2~\mathrm{s}^{-1}] \right)}$ & ${13.87}^{+0.84}_{-2.48}$ & ${10.18}^{+1.14}_{-1.45}$ & $3.2\sigma$ \\
& $\log_{10}{\left( K_{zz}^\mathrm{Ch}~[\mathrm{cm}^2~\mathrm{s}^{-1}] \right)}$ & ${13.87}^{+0.84}_{-2.48}$ & ${2.66}^{+1.80}_{-1.67}$ & $6.2\sigma$ \\
\hline
HD~97658b & $\log_{10}{\left( K_{zz}^\mathrm{Cl}~[\mathrm{cm}^2~\mathrm{s}^{-1}] \right)}$ & ${12.53}^{+1.40}_{-1.83}$ & ${10.53}^{+0.91}_{-1.18}$ & $2.2\sigma$ \\
& $\log_{10}{\left( K_{zz}^\mathrm{Ch}~[\mathrm{cm}^2~\mathrm{s}^{-1}] \right)}$ & ${12.53}^{+1.40}_{-1.83}$ & ${6.92}^{+4.93}_{-4.61}$ & $1.1\sigma$ \\
\hline
HD~189733b & $\log_{10}{\left( K_{zz}^\mathrm{Cl}~[\mathrm{cm}^2~\mathrm{s}^{-1}] \right)}$ & ${13.04}^{+0.72}_{-0.91}$ & ${11.21}^{+0.51}_{-1.21}$ & $3.6\sigma$ \\
& $\log_{10}{\left( \dot{\Sigma}~[\mathrm{cm}^2~\mathrm{s}^{-1}] \right)}$ & ${-10.98}^{+2.78}_{-1.69}$ & ${-13.99}^{+0.84}_{-1.38}$ & $3.6\sigma$ \\
& $\log_{10}{\left( K_{zz}^\mathrm{Ch}~[\mathrm{cm}^2~\mathrm{s}^{-1}] \right)}$ & ${13.04}^{+0.72}_{-0.91}$ & ${5.43}^{+5.90}_{-3.44}$ & $1.3\sigma$ \\
\hline
HD~209458b & C/O & ${0.19}^{+0.12}_{-0.06}$ & ${0.50}^{+0.21}_{-0.28}$ & $1.1\sigma$ \\
& $\log_{10}{\left( K_{zz}^\mathrm{Cl}~[\mathrm{cm}^2~\mathrm{s}^{-1}] \right)}$ & ${4.17}^{+1.15}_{-1.05}$ & ${5.46}^{+0.26}_{-0.30}$ & $4.3\sigma$ \\
& $\log_{10}{\left( K_{zz}^\mathrm{Ch}~[\mathrm{cm}^2~\mathrm{s}^{-1}] \right)}$ & ${4.17}^{+1.15}_{-1.05}$ & ${13.12}^{+1.39}_{-4.61}$ & $1.9\sigma$ \\
\hline
WASP-12b & $\log_{10}{\left( K_{zz}^\mathrm{Cl}~[\mathrm{cm}^2~\mathrm{s}^{-1}] \right)}$ & ${11.35}^{+2.16}_{-2.08}$ & ${9.94}^{+1.28}_{-1.34}$ & $1.1\sigma$ \\
\hline
WASP-17b & $\log_{10}{\left( K_{zz}^\mathrm{Cl}~[\mathrm{cm}^2~\mathrm{s}^{-1}] \right)}$ & ${2.82}^{+2.13}_{-1.88}$ & ${5.70}^{+0.55}_{-0.46}$ & $6.2\sigma$ \\
\hline
WASP-19b & $\log_{10}{\left( K_{zz}^\mathrm{Cl}~[\mathrm{cm}^2~\mathrm{s}^{-1}] \right)}$ & ${3.32}^{+2.29}_{-2.11}$ & ${6.20}^{+1.66}_{-0.78}$ & $3.7\sigma$ \\
\hline
WASP-39b & $\log_{10}{\left( K_{zz}^\mathrm{Cl}~[\mathrm{cm}^2~\mathrm{s}^{-1}] \right)}$ & ${4.27}^{+1.23}_{-1.01}$ & ${5.59}^{+0.48}_{-0.38}$ & $3.5\sigma$ \\
& $\log_{10}{\left( K_{zz}^\mathrm{Ch}~[\mathrm{cm}^2~\mathrm{s}^{-1}] \right)}$ & ${4.27}^{+1.23}_{-1.01}$ & ${8.61}^{+3.68}_{-4.22}$ & $1.0\sigma$ \\
\hline                                %inserts single line
\end{tabular}
\begin{tablenotes}
\item Note that $K_{zz}$ derived from the same-$K_{zz}$ retrieval, which is used in both cloud and chemistry calculations, is compared to $K_{zz}^\mathrm{Cl}$ and $K_{zz}^\mathrm{Ch}$ from the standard disequilibrium retrieval.
\end{tablenotes}
\end{threeparttable}
\end{table*}
\endgroup

\subsection{Comparison of disequilibrium retrievals with and without a fixed intrinsic temperature} \label{sec:tint}
In this subsection, we compare the results from the standard disequilibrium retrieval and that with a fixed intrinsic temperature.
The motivation to fix the intrinsic temperature is based on the expectation that the retrieval of the chemical eddy diffusion coefficient might degenerate with the intrinsic temperature since the quenching often happens in the deeper atmosphere where the intrinsic temperature has a large influence on the thermal structure.
Despite our expectation, however, such a correlation is not found (see Fig.~\ref{fig:corner}), which may be due to the limits of the current observational precision.
\edt{The correlation coefficients between $\log_{10}{\left( T_\mathrm{int}~[K] \right)}$ and $\log_{10}{\left( K_{zz}^\mathrm{Cl}~[\mathrm{cm}^2~\mathrm{s}^{-1}] \right)}$ are 0.12, 0.071, 0.17, 0.0037, and 0.10 for HAT-P-1b, HD~189733b, HD~209458b, WASP-12b, and WASP-19b, respectively.}
\hl{We note that for all the five targets to which we applied $T_\mathrm{int}$-fixed retrieval, the standard retrieval favors a lower value of $T_\mathrm{int}$ than that inferred from the formula derived by \citet{2019ApJ...884L...6T} (see Table~\ref{table:param_diff_int}).}
As shown in the fourth column of Table~\ref{table:evidence}, the absolute values of the natural logs of the Bayes factors between the two retrievals are quite small \citep[cf.][]{2008ConPh..49...71T} except for HD~189733b\hl{.}

\hl{HD~189733b exhibits significantly smaller Bayesian evidence of} disequilibrium retrieval with fixed $T_\mathrm{int}$, \hl{indicating that the retrieved parameters from that retrieval are} strongly \hl{($\gtrsim 7.8\sigma$)} disfavored compared to \hl{those} from the standard disequilibrium retrieval.
The significant difference in the \hl{Bayesian evidence} is partly due to the worse fit to the observed data points in the bluest part (0.3--0.5~$\mu$m) for the $T_\mathrm{int}$-fixed retrieval (red line in Fig.~\ref{fig:spectra}d).
This discrepancy \hl{is due to} the different retrieved parameters for the clouds.
We note, however, that there is room for discussion regarding the quite steep optical slope of this planet \citep{2013MNRAS.432.2917P}.
Several origins aside from the planetary atmosphere, such as an unknown systematic instrument offset between the optical and near-infrared or the presence of starspots on the host star \citep[e.g.,][]{2014A&A...568A..99O, 2020A&A...643A..64O} have been proposed as being capable of artificially producing such a steep slope in the planetary spectrum.
%On the other hand, the cloud opacity can still reproduce the observed steepness as shown in the best-fit spectra from the standard disequilibrium and equilibrium retrievals.

Comparing the retrieved spectra of HD~189733b from the two retrievals, differences can also be found above 8.0~$\mu$m.
In the standard disequilibrium retrieval case (blue line), cloud features 
%of $\mathrm{MgSiO_3}$ and $\mathrm{SiO_2}$ 
exist around 9.4 and 20~$\mu$m while the $\mathrm{C_2H_2}$ absorption feature is visible at 13.6~$\mu$m in the $T_\mathrm{int}$-fixed retrieval case (red line).
A striking difference is also found in the retrieved thermal structure (see Fig.~\ref{fig:profile}d).
A strong thermal inversion is retrieved from the $T_\mathrm{int}$-fixed retrieval (red line), whereas no such inversion is derived from the standard disequilibrium retrieval (blue line).
%Although the Bayes factor is smaller for the $T_\mathrm{int}$-fixed retrieval, from the physical point of view, thermal inversion can be more realistic considering the nature of the cloud opacity, namely stronger absorption in the optical wavelength range \yuic{cite some papers}.
We note that in our retrieval, we retrieved the parameters regarding the thermal structure independently from the atmospheric constituents.
A retrieval employing a thermal structure consistent with the opacity of the atmospheric {composition} will be the subject of future work.

\edt{The retrieved parameters with more than $1\sigma$ difference between the $T_\mathrm{int}$-fixed and standard disequilibrium retrievals} are summarized in Table~\ref{table:param_diff_int}.
Indeed, quite significant differences are found for the parameters of HD~189733b.

\begingroup
\renewcommand{\arraystretch}{1.3}
\begin{table*}
\caption{Retrieved parameters with more than 1$\sigma$ difference between the $T_\mathrm{int}$-fixed and standard disequilibrium retrievals}             % title of Table
\label{table:param_diff_int}      % is used to refer this table in the text
\centering                          % used for centering table
\begin{tabular}{l c r r c}        % centered columns (4 columns)
\hline\hline                 % inserts double horizontal lines
Planet & Parameter & Value from $T_\mathrm{int}$-fixed & Value from standard & Difference \\    % table heading 
\hline
\hline
HAT-P-1b & $\log_{10}{\left( T_\mathrm{int}~[K] \right)}$ & ${2.68}$ & ${1.81}^{+0.66}_{-0.51}$ & $1.3\sigma$ \\
\hline
HD~189733b & $\log_{10}{\gamma}$ & ${1.66}^{+0.15}_{-0.08}$ & ${-1.31}^{+0.10}_{-0.16}$ & $30.1\sigma$ \\
& $\log_{10}{\left( \kappa_\mathrm{IR}~[\mathrm{cm}^2~\mathrm{g}^{-1}] \right)}$ & ${-3.81}^{+0.28}_{-0.14}$ & ${3.05}^{+0.38}_{-0.35}$ & $19.8\sigma$ \\
& $\log_{10}{\left( T_\mathrm{int}~[K] \right)}$ & ${2.58}$ & ${1.51}^{+0.30}_{-0.30}$ & $3.6\sigma$ \\
& C/O & ${1.21}^{+0.06}_{-0.09}$ & ${0.47}^{+0.24}_{-0.22}$ & $3.1\sigma$ \\
& Si/O & ${0.03}^{+0.02}_{-0.01}$ & ${0.12}^{+0.07}_{-0.06}$ & $1.5\sigma$ \\
& $Z$~[dex] & ${1.06}^{+0.18}_{-0.23}$ & ${-0.56}^{+0.46}_{-0.26}$ & $3.5\sigma$ \\
& $\log_{10}{\left( g~[\mathrm{cm}~\mathrm{s}^{-2}] \right)}$ & ${3.30}^{+0.02}_{-0.02}$ & ${3.38}^{+0.02}_{-0.02}$ & $4.5\sigma$ \\
& $\log_{10}{\left( K_{zz}^\mathrm{Cl}~[\mathrm{cm}^2~\mathrm{s}^{-1}] \right)}$ & ${7.08}^{+0.17}_{-0.13}$ & ${11.21}^{+0.51}_{-1.21}$ & $3.4\sigma$ \\
& $\log_{10}{\left( \dot{\Sigma}~[\mathrm{cm}^2~\mathrm{s}^{-1}] \right)}$ & ${-7.06}^{+0.05}_{-0.11}$ & ${-13.99}^{+0.84}_{-1.38}$ & $8.2\sigma$ \\
& $\log_{10}{\left( K_{zz}^\mathrm{Ch}~[\mathrm{cm}^2~\mathrm{s}^{-1}] \right)}$ & ${12.62}^{+1.16}_{-0.68}$ & ${5.43}^{+5.90}_{-3.44}$ & $1.2\sigma$ \\
\hline
HD~209458b & $\log_{10}{\gamma}$ & ${-1.03}^{+1.62}_{-0.71}$ & ${0.77}^{+0.80}_{-1.15}$ & $1.6\sigma$ \\
& $\log_{10}{\left( \kappa_\mathrm{IR}~[\mathrm{cm}^2~\mathrm{g}^{-1}] \right)}$ & ${-1.35}^{+0.64}_{-0.44}$ & ${1.51}^{+1.59}_{-1.91}$ & $1.5\sigma$ \\
& $\log_{10}{\left( T_\mathrm{int}~[K] \right)}$ & ${2.75}$ & ${2.11}^{+0.42}_{-0.39}$ & $1.5\sigma$ \\
\hline
WASP-12b & $\log_{10}{\left( T_\mathrm{int}~[K] \right)}$ & ${2.71}$ & ${2.03}^{+0.58}_{-0.62}$ & $1.2\sigma$ \\
\hline
WASP-19b & $\log_{10}{\left( T_\mathrm{int}~[K] \right)}$ & ${2.81}$ & ${1.81}^{+0.70}_{-0.51}$ & $1.4\sigma$ \\
\hline                                %inserts single line
\end{tabular}
\end{table*}
\endgroup

\subsection{\hl{Trend of the retrieved parameters}}
\hl{In this subsection, we explore the trend of the atmospheric parameters derived from our retrieval calculations.}
\subsubsection{Eddy diffusion coefficient}
   \begin{figure*}[h!]
   \centering
   \includegraphics[width=\hsize]{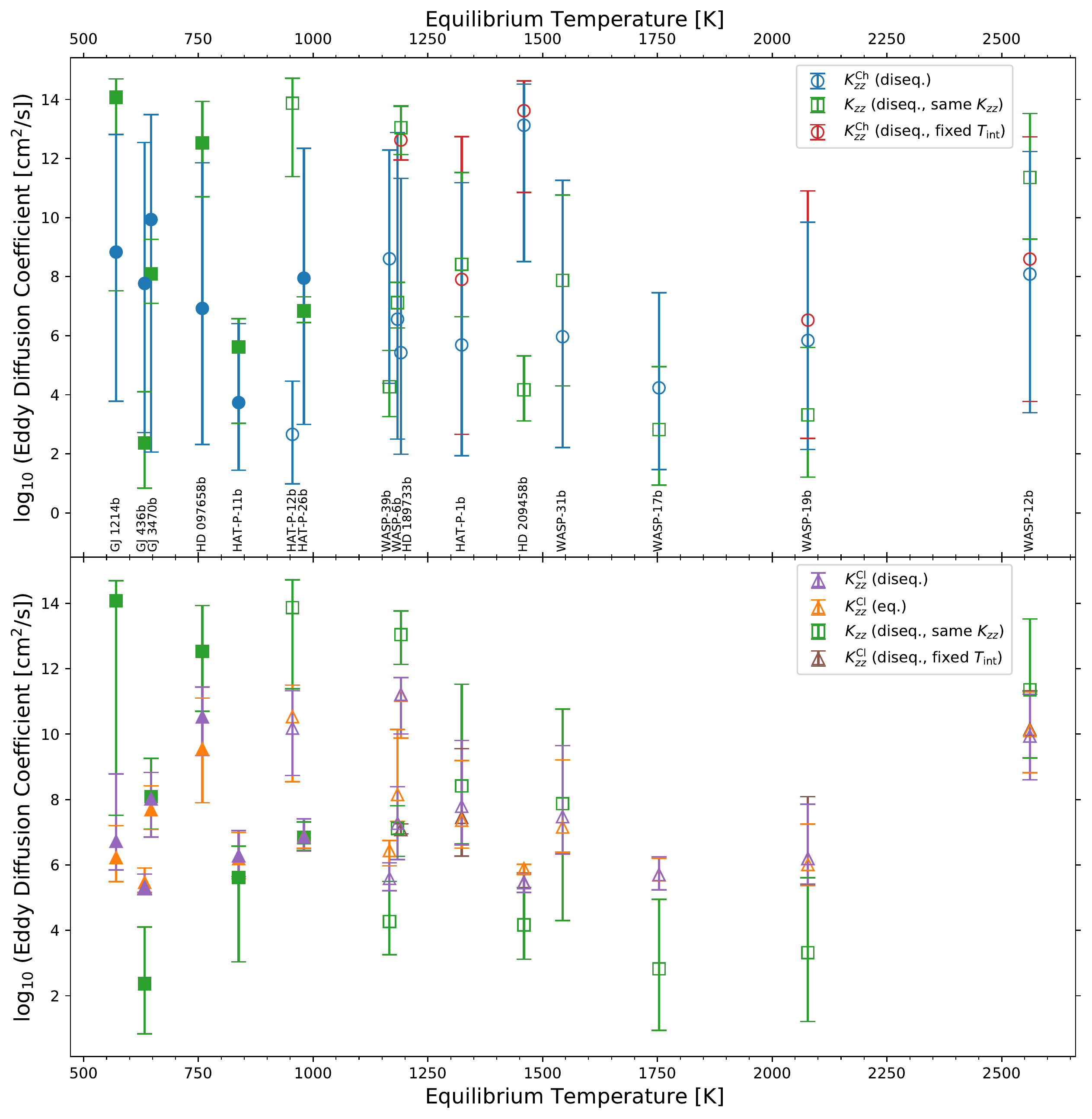}
      \caption{\edt{(Top)} Eddy diffusion coefficients for the chemical structure simulations, $K_{zz}^\mathrm{Ch}$, retrieved from the standard disequilibrium retrieval (blue circle points) and that with fixed $T_\mathrm{int}$ (red circle points; only shown for the planets with masses larger than half Jupiter mass). 
      Also shown are the eddy diffusion coefficients used for both chemistry and cloud simulations in the same-$K_{zz}$ retrieval (green square points).
      \edt{(Bottom) Eddy diffusion coefficients for the cloud structure simulations, $K_{zz}^\mathrm{Cl}$, retrieved from the standard disequilibrium retrieval (purple triangle points), equilibrium retrieval (orange triangle points), and disequilibrium retrieval with fixed $T_\mathrm{int}$ (brown triangle points; again, only shown for the planets with masses larger than half Jupiter mass).
      The eddy diffusion coefficients used for both chemistry and cloud simulations in the same-$K_{zz}$ retrieval are also plotted (green square points, as in the top panel).}
      The unfilled points indicate our hot Jupiter samples, while the filled points are the Neptune samples, which we define as masses smaller than 0.1 times the Jupiter mass.
      \edt{The error bars indicate 1$\sigma$ uncertainty from the retrievals.}}
   \label{Fig:kzz}
   \end{figure*}

Figure~\ref{Fig:kzz} shows the retrieved values of the eddy diffusion coefficients as a function of the equilibrium temperature for our samples \edt{(top panel for the chemical eddy diffusion coefficient $K_{zz}^\mathrm{Ch}$ and bottom panel for the cloud eddy diffusion coefficient $K_{zz}^\mathrm{Cl}$)}.
First, neither $K_{zz}^\mathrm{Ch}$ (blue circle points in the top panel) nor $K_{zz}^\mathrm{Cl}$ (purple triangle points in the bottom panel) from the standard disequilibrium retrieval exhibits any clear trend over the equilibrium temperature, at least for the current observed data.
On the other hand, a larger diffusion coefficient for higher temperature has been predicted from both theory and numerical GCM simulations because of the increasing speed of vertical winds \citep{2019ApJ...881..152K}.
This trend is also tentatively indicated by the recent analysis of the {\textit{Spitzer}} transit depths of about fifty gas giants by \citet{2021A&A...648A.127B}.
The higher observational precision achievable with the \textit{James Webb} Space Telescope \citep[JWST;][]{2006SSRv..123..485G} and the Atmospheric Remote-sensing Infrared Exoplanet Large-survey \citep[Ariel;][]{2018ExA....46..135T} will enable us to further explore such predictions.

Compared to $K_{zz}^\mathrm{Ch}$ (blue circle points), $K_{zz}^\mathrm{Cl}$ (purple triangle points) is retrieved with much smaller uncertainties due to the strong effect of the diffusion coefficient on the cloud structure \citep{2018ApJ...863..165G, 2019A&A...622A.121O}.
Also, the comparison of $K_{zz}^\mathrm{Cl}$ between the standard disequilibrium and equilibrium retrievals (purple and orange triangle points in the bottom panel) shows that they are quite similar.
%This is reasonable considering that the cloud formation usually occurs in the upper atmosphere thus less sensitive to the disequilibrium chemistry.

Next, $K_{zz}$ from the disequilibrium retrieval employing the condition $K_{zz}^\mathrm{Ch} = K_{zz}^\mathrm{Cl}$ (green square points in both panels) does not exhibit any clear trend over the equilibrium temperature either.
In some cases, the 1$\sigma$ uncertainty range of $K_{zz}$ from the same-$K_{zz}$ retrieval lies outside the median values of $K_{zz}^\mathrm{Ch}$ and $K_{zz}^\mathrm{Cl}$ from the standard disequilibrium retrieval.
This is probably because we have imposed the strong condition $K_{zz}^\mathrm{Ch} = K_{zz}^\mathrm{Cl}$ with a constant value throughout the atmosphere even though the location of the quenching and clouds affecting the spectrum are usually different and several orders of magnitude difference of $K_{zz}$ within the atmosphere is expected from GCM simulations \citep{2018ApJ...866....1Z, 2019ApJ...881..152K}.
%The uncertainty of $K_{zz}$ is mostly between those of $K_{zz}^\mathrm{Ch}$ and $K_{zz}^\mathrm{Cl}$ from the standard disequilibrium retrieval (blue circle and purple triangle points) since $K_{zz}$ is used in both the chemistry and cloud simulations in the same-$K_{zz}$ retrieval.

Among our samples, HD~209458b has a markedly large retrieved $K_{zz}^\mathrm{Ch}$ (blue circle point) from the standard disequilibrium retrieval.
%For this planet, $K_{zz}^\mathrm{Ch}$ and $K_{zz}^\mathrm{Cl}$ (purple triangle point) differ by more than several orders of magnitude with almost the same location of the quenching and clouds.
%and this might reflect the difference of the diffusion coefficient throughout the atmosphere.
%Even though we have retrieved those two different eddy diffusion coefficients completely independently in the standard disequilibrium retrieval
%since the locations of the quenching and clouds are usually different and thus to allow those two diffusion coefficients to have different values, 
%For the case of this planet, however, 
%In the best-fit case from the disequilibrium retrieval, while quenching happens in the deeper atmosphere of the pressure around 1~bar  (compare thick and thin solid lines in Figure~\ref{fig:profile}n), clouds also exist around the pressure level of 1~bar.
When the condition $K_{zz}^\mathrm{Ch} = K_{zz}^\mathrm{Cl}$ is imposed, the retrieved value of $K_{zz}$ (green square point) decreases, especially when compared to $K_{zz}^\mathrm{Ch}$.
%In the best-fit case of this retrieval, quenching happens around the pressure level of 0.1~bar (compare thick and thin dotted lines in Figure~\ref{fig:profile}t and \ref{fig:profile}u) while clouds still exist around the pressure level of 1~bar.
%Thus, given these similar locations, for this planet, the value of $K_{zz}$ retrieved from the same-$K_{zz}$ retrieval can be more realistic.
However, as we mentioned in \S~\ref{sec:same}, even with this small $K_{zz}$, the retrieved spectrum still exhibits disequilibrium features.
Thus, we propose that this planet is an ideal target for studying disequilibrium chemistry in exoplanet atmospheres.
%Employing the same value for those two diffusion coefficients is left for future work.
%Thus this might indicate the much higher eddy diffusivity in the cloud-forming region compared to the quenching altitude in the atmosphere of this planet.
%Indeed, higher eddy diffusion coefficient with decreasing pressure has been predicted from both theory and GCM simulations owing to the faster vertical winds \citep{2018ApJ...866....1Z, 2019ApJ...881..152K}.
%

$K_{zz}^\mathrm{Ch}$ retrieved from the $T_\mathrm{int}$-fixed disequilibrium retrieval (red circle points in the top panel) is tightly constrained compared to the standard disequilibrium retrieval for HD~189733b and HD~209458b, \edt{while this is not the case for HAT-P-1b, WASP-12b, and WASP-19b}.
\edt{There are several possible reasons for this.
First, for HD~189733b, the value of $T_\mathrm{int}$ we have fixed with the formula of \citet{2019ApJ...884L...6T} is different from the retrieved value from the standard equilibrium retrieval by more than $3\sigma$ (see Table~\ref{table:param_diff_int}), and the $T_\mathrm{int}$-fixed disequilibrium retrieval results in a different thermal profile with high temperatures in the upper atmosphere (see Fig.~\ref{fig:profile}d).
This requires a larger $K_{zz}^\mathrm{Ch}$ to reproduce the observed amplitude of the 1.4~$\mu$m $\mathrm{H_2O}$ absorption feature by overcoming its thermal dissociation due to high temperatures (compare the purple thick dash-dotted line and purple thin dash-dotted line in Fig.~\ref{fig:profile}d).
On the other hand, in the case of the standard disequilibrium retrieval, \edttwo{the retrieved small value of $K_{zz}^\mathrm{Ch}$ indicates that} disequilibrium chemistry is \edttwo{not significantly preferred over equilibrium chemistry}. \edttwo{This is also seen in the almost consistent median abundance profiles between the standard disequilibrium and equilibrium retrievals (compare the thick solid lines and dashed lines in the right panel of Fig.~\ref{fig:profile}d).} Thus, $K_{zz}^\mathrm{Ch}$ remains unconstrained \edttwo{while it needs to be small enough}.
Next, for HD~209458b, disequilibrium chemistry plays an important role in its atmosphere, unlike the cases of WASP-12b and WASP-19b (see Figs.~\ref{fig:profile} and \ref{fig:profile2}).
Since the quenching happens in the deep atmosphere where the intrinsic temperature has a big influence on the thermal structure, the value of $K_{zz}^\mathrm{Ch}$ could be tightly constrained when $T_\mathrm{int}$ is fixed to a certain value for the atmospheres with disequilibrium chemistry.
Given the somewhat stronger constraint on $K_{zz}^\mathrm{Ch}$ for HD~209458b, we ideally expect a correlation between $T_\mathrm{int}$ and $K_{zz}^\mathrm{Ch}$ in the results of the standard disequilibrium retrieval for this planet, which we do not find, as mentioned in \S~\ref{sec:tint}.
We consider that the complexity of the retrieval we have performed, namely retrieval with more than ten retrieval parameters, could cause this apparent absence of the correlation.
Moreover, we speculate that the precision of the observational data is also related because if the observational uncertainty is large, it would not have an impact on the constraint on the parameters even if we fix any of the parameters.
The observational data for HD~189733b and HD~209458b are relatively precise when compared to those of the other three planets.
Indeed, we have found that when all the observational errors are artificially reduced to half their values, $K_{zz}^\mathrm{Ch}$ becomes somewhat tightly constrained for HAT-P-1b while not for WASP-12b and WASP-19b when $T_\mathrm{int}$ is fixed. In the atmosphere of HAT-P-1b, disequilibrium chemistry has a modest impact (see Fig.~\ref{fig:profile2}h).
Future investigations are needed to draw firm conclusions about the points raised.
}
%This is understandable considering their relatively high-precision observational data and that the quenching happens in the deeper atmosphere of those planets where the intrinsic temperature has a big influence on the thermal structure (see Figures~\ref{fig:profile}l and \ref{fig:profile}n).

\edt{Finally}, $K_{zz}^\mathrm{Cl}$ retrieved from the $T_\mathrm{int}$-fixed disequilibrium retrieval (brown triangle points in the bottom panel) are largely similar to those of the standard disequilibrium retrieval except for HD~189733b, for which the retrieved thermal structure differs significantly (see Fig.~\ref{fig:profile}d).

\subsubsection{\hl{Metallicity and C/O ratio}}
\begin{figure*}
 \begin{minipage}{0.49\hsize}
   \centering
   \includegraphics[height=6.8cm]{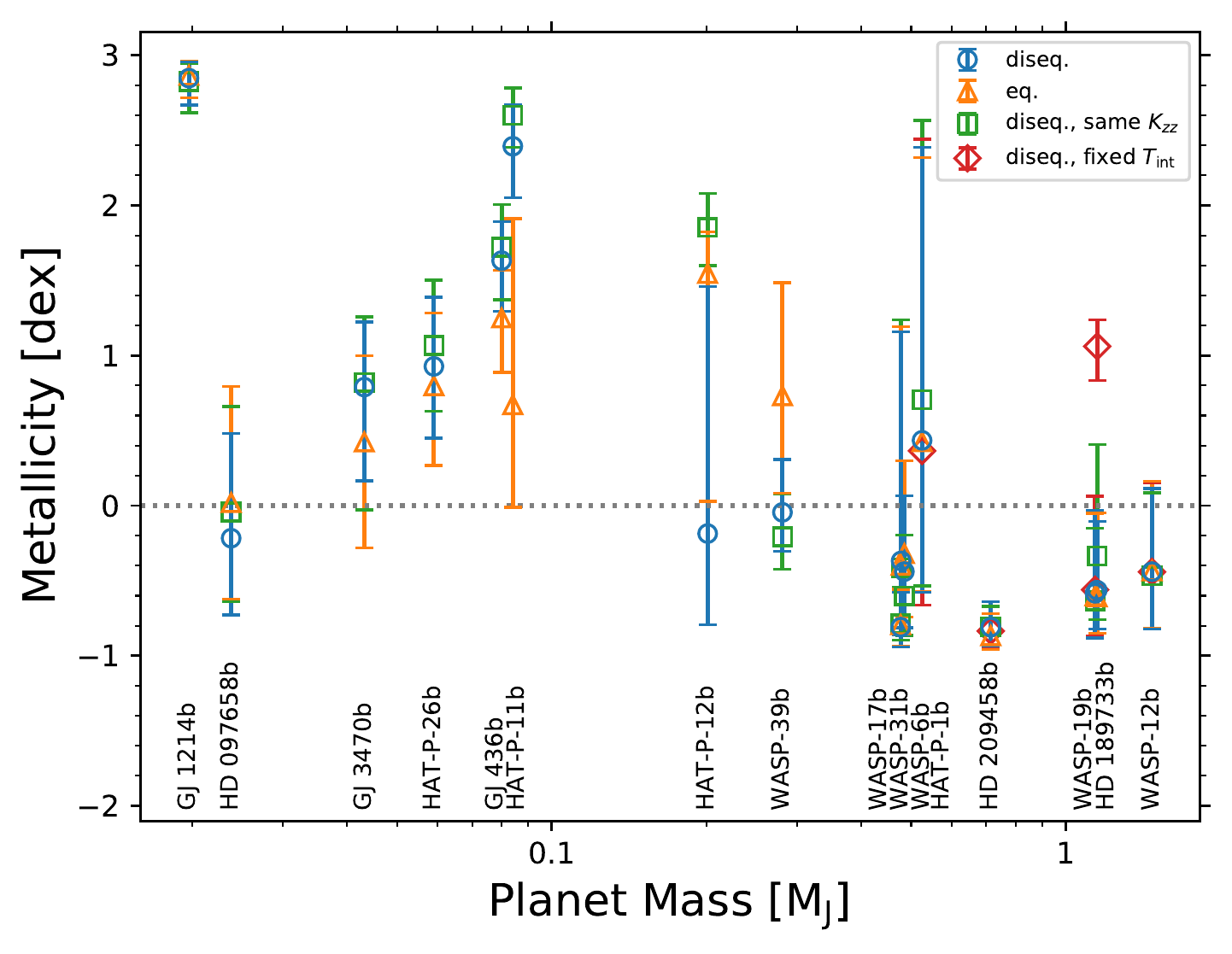}
   \subcaption{Metallicity}
 \end{minipage}
 \begin{minipage}{0.49\hsize}
   \centering
   \includegraphics[height=6.8cm]{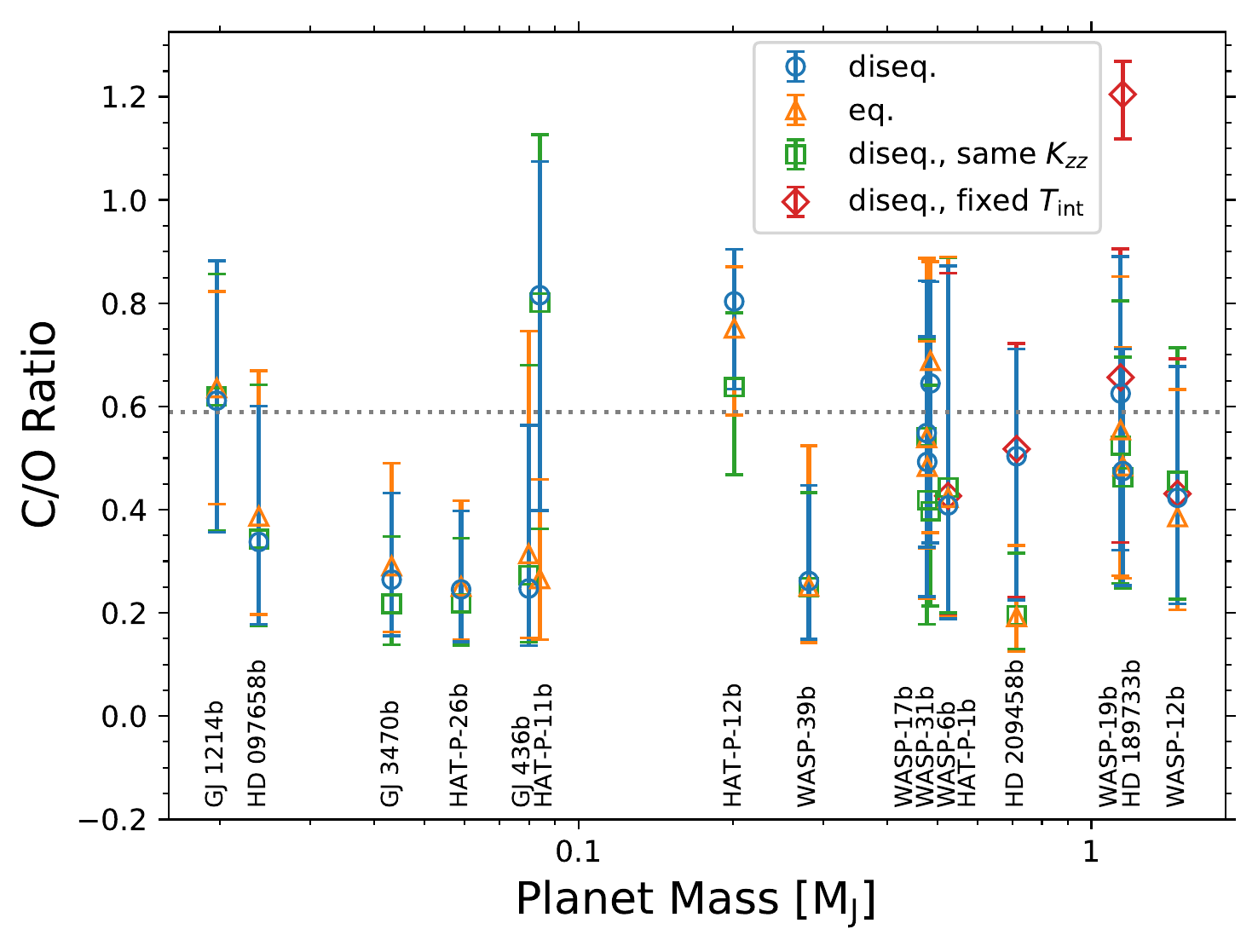}
   \subcaption{C/O ratio}
 \end{minipage}
 \caption{\hl{Atmospheric metallicity (a) and C/O ratio (b) derived from the standard disequilibrium retrieval (blue circle points), equilibrium retrieval (orange triangle points), disequilibrium retrieval with same $K_{zz}$ (green square points), and $T_\mathrm{int}$-fixed disequilibrium retrieval (red diamond points; only shown for the planets with masses larger than half Jupiter mass). The solar metallicity and C/O ratio from the solar photospheric elemental abundance ratios of \citet{2021arXiv210501661A} are marked with dotted lines for reference.}}
 \label{fig:metal}
\end{figure*}

\hl{Figure~\ref{fig:metal} shows the derived values of (a)~metallicity and (b)~C/O ratio from our four different retrievals as a function of the planet mass.
For most planets, the values of the two parameters derived from the different retrievals are consistent within {the} 1$\sigma$ error.
As for the metallicity trend, our results generally agree with the previous studies that found a tentative trend of higher metallicity with decreasing planet mass, such as seen for the solar system planets \citep{2017Sci...356..628W, 2018Natur.557..526N, 2019ApJ...887L..20W}.
On the other hand, we see no clear trend for the C/O ratio, at least for the current observational precision, which is also consistent with previous work \citep{2020A&A...642A..28M}.
Since both metallicity and C/O ratio are regarded as important indicators of planet formation \citep{2011ApJ...743L..16O, 2016A&A...595A..83E, 2016ApJ...832...41M, 2018A&A...613A..14E, 2020MNRAS.499.2229N}, future higher-precision observations are essential to further explore the trends of these parameters toward understanding the origin and formation mechanism of planets.}

\section{Discussion} \label{sec:discussion}
\subsection{Eddy diffusion coefficient}
\hl{
\subsubsection{Range used in retrieval}
As we have mentioned in \S~\ref{sec:application}, for some of the cooler planets of our samples, the lower bound of the $K_{zz}^\mathrm{Ch}$ range adopted in the retrievals, namely $10^{0}$~$\mathrm{cm}^2$~$\mathrm{s}^{-1}$, is not sufficiently small for all the species to be in chemical equilibrium.
To confirm our findings, we have performed additional standard disequilibrium retrievals extending the lower bound of $K_{zz}^\mathrm{Ch}$ to an extremely small value of $10^{-40}$~$\mathrm{cm}^2$~$\mathrm{s}^{-1}$.
We choose this value so that the atmosphere lower than the 0.1~mbar level, which is the lower bound of the valid pressure range for the chemical timescale formula of \citet{2018ApJ...862...31T}, is in chemical equilibrium when calculated with the best-fit parameter set of the standard disequilibrium retrieval for all of our samples except WASP-39b.
%1~mbar is approximately the pressure we probe via transmission spectroscopy, which can be also noticed from the relatively smaller confidence interval ranges in the thermal profile (see Figure~\ref{fig:profile}), while 0.1~$\mu$bar is the upper boundary of the atmosphere calculated inside ARCiS.
We note that while we have found that the best-fit parameter set for WASP-39b requires $K_{zz}^\mathrm{Ch} \lesssim 10^{-60}$~$\mathrm{cm}^2$~$\mathrm{s}^{-1}$, the resultant temperature in its upper atmosphere is lower than the lower bound of the valid temperature range for the chemical timescale of \citet{2018ApJ...862...31T}, namely 500~K.
Thus it is uncertain whether such a small value of $K_{zz}^\mathrm{Ch}$ is needed to achieve the equilibrium condition in its atmosphere.
Regarding our finding of the \edt{indication} %tentative evidence 
of disequilibrium chemistry for WASP-39b, we have confirmed that the temperature in the relatively lower atmosphere where the quenching happens is within the validated range, so this does not affect our conclusion.
}

\hl{Figure~\ref{fig:posterior} shows the posterior probability distribution of $K_{zz}^\mathrm{Ch}$ for (a)~HD~209458b, (b)~WASP-39b, and (c)~HD~189733b.
For HD~209458b and WASP-39b, for which we have found \edt{indicative} %tentative evidence 
evidence of disequilibrium chemistry, the peak of the posterior distribution remains at a similar value regardless of the adopted wider range, ensuring the relatively large values of $K_{zz}^\mathrm{Ch}$ we have retrieved for these two planets.
On the other hand, for all the other planets, as for the case of (c)~HD~189733b, we find that adopting a smaller lower bound for the $K_{zz}^\mathrm{Ch}$ range, the posterior distribution flattens and the median and 1$\sigma$ confidence interval shift toward smaller values, inferring that the retrieval of $K_{zz}^\mathrm{Ch}$ is affected by the choice of the $K_{zz}^\mathrm{Ch}$ range and thus its retrieved value is not reliable.
This is consistent with \edt{our results that the disequilibrium scenarios are never favored over the equilibrium ones}
%almost negligible values for the absolute values of the natural logs of the Bayes %factors between the standard disequilibrium and equilibrium retrievals 
for all of our samples except HD~209458b and WASP-39b in \S~\ref{sec:diseq}.
\begin{figure}
 \begin{minipage}{0.45\hsize}
   \centering
   \includegraphics[width=\hsize]{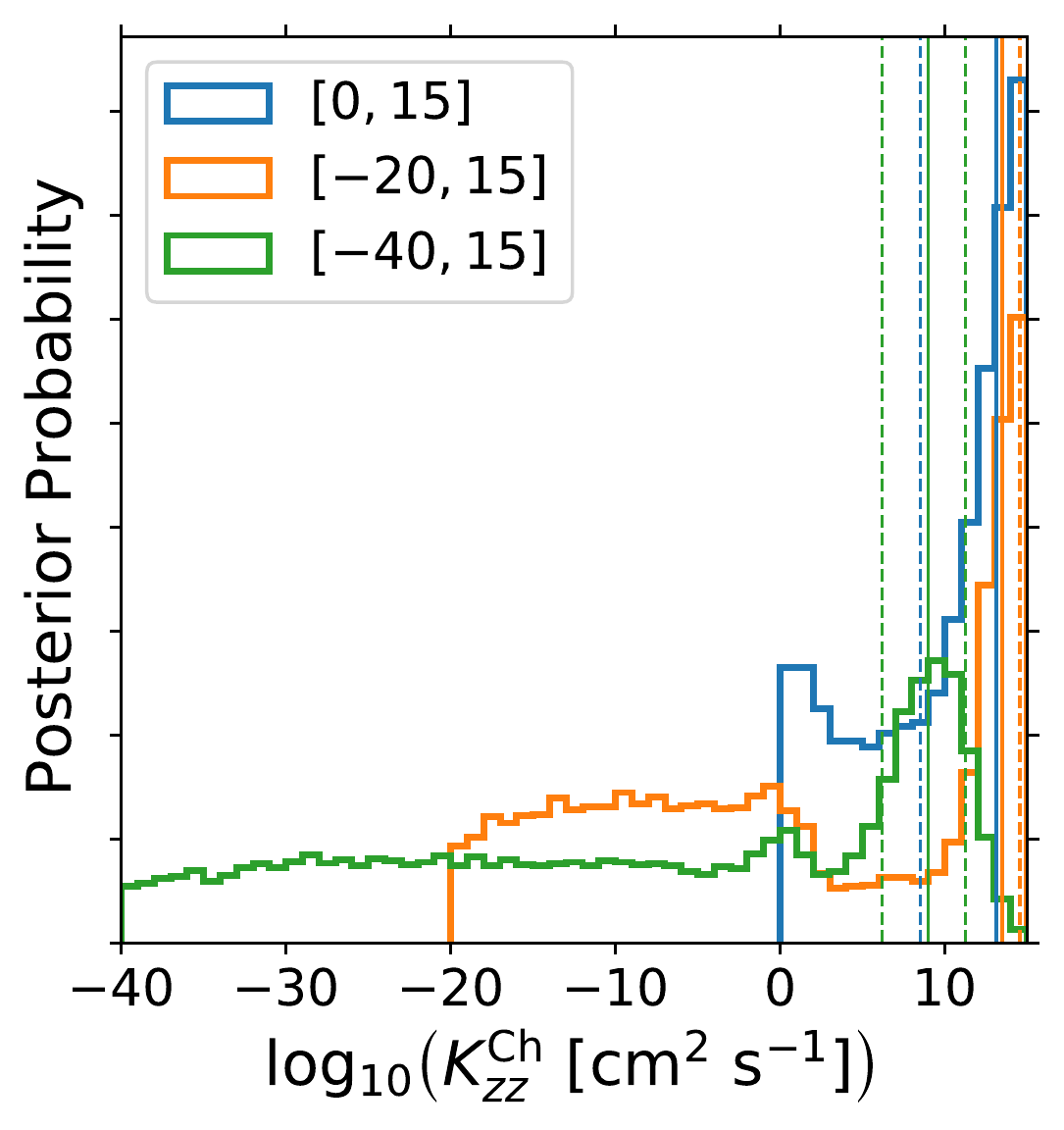}
   \subcaption{HD~209458b}
 \end{minipage}
 \begin{minipage}{0.45\hsize}
   \centering
   \includegraphics[width=\hsize]{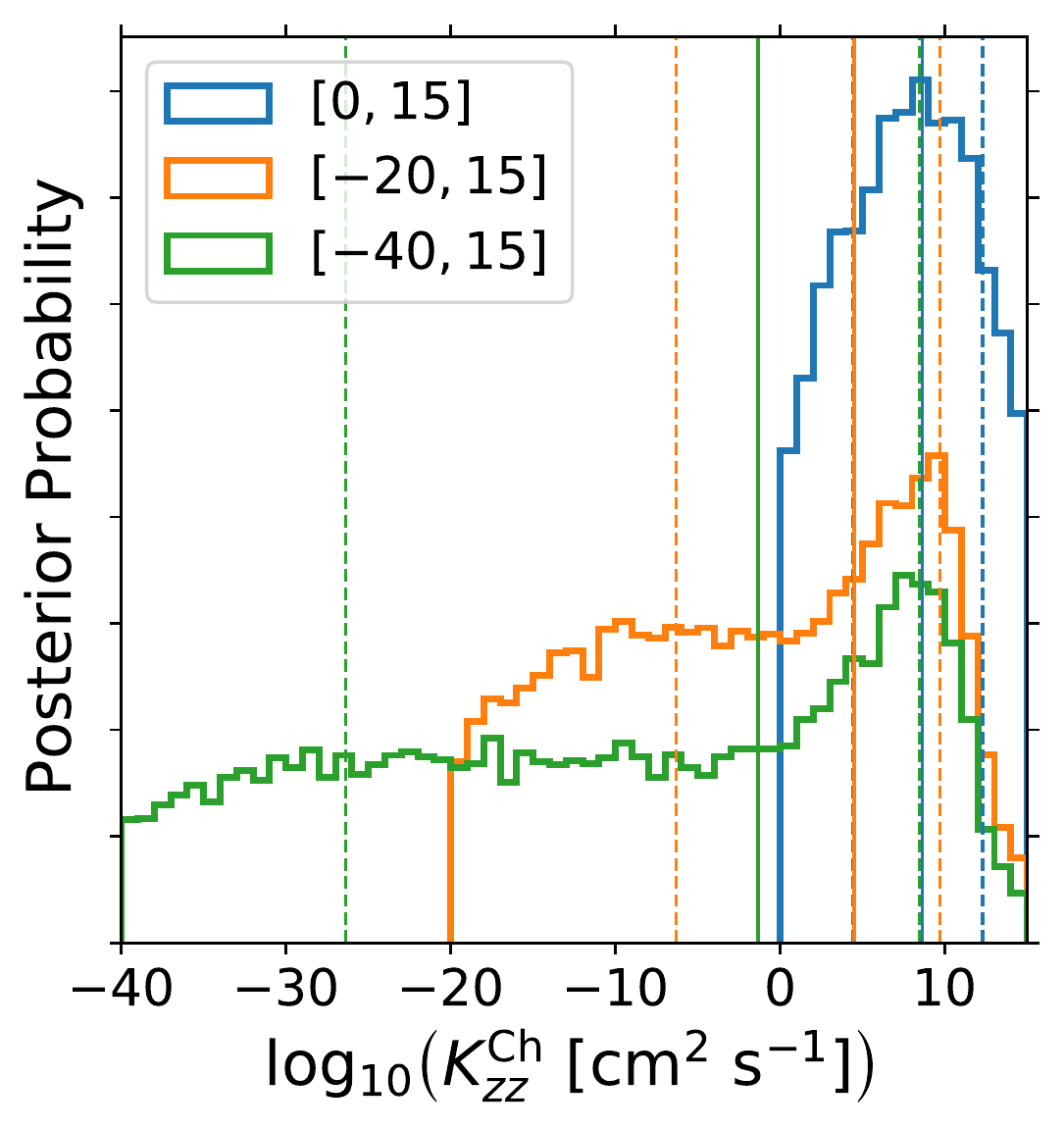}
   \subcaption{WASP-39b}
 \end{minipage}
 \begin{minipage}{1.0\hsize}
   \centering
   \includegraphics[width=0.45\hsize]{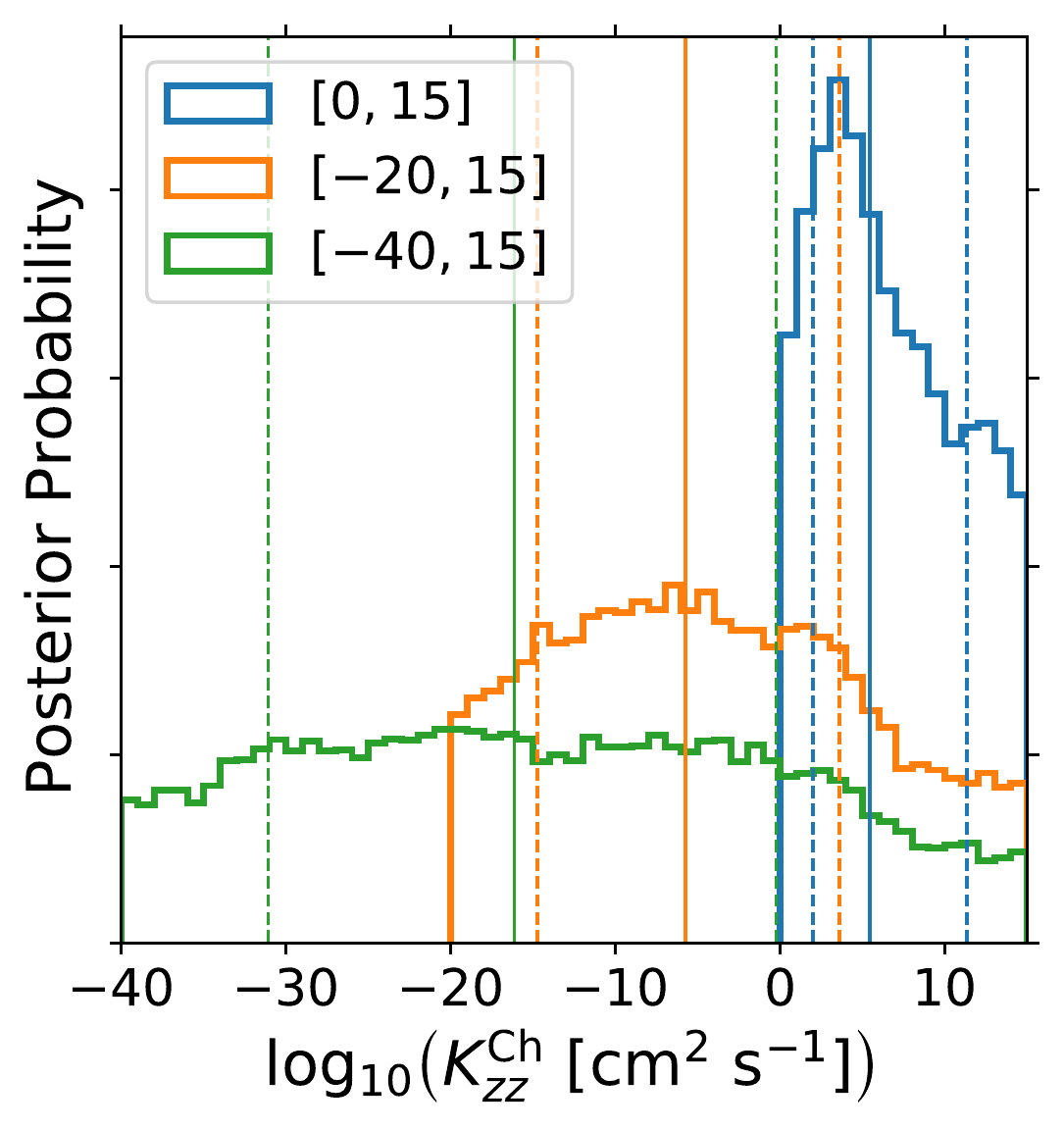}
   \subcaption{HD~189733b}
 \end{minipage}
 \caption{Posterior probability distribution of $\log_{10}{\left( K_{zz}^\mathrm{Ch}~[\mathrm{cm}^2~\mathrm{s}^{-1}] \right)}$ from the standard disequilibrium retrievals with different ranges adopted, $\left[ 0, 15 \right]$ (blue), \edt{$\left[ -20, 15 \right]$ (orange)}, and $\left[ -40, 15 \right]$ (green). The vertical solid and dashed lines indicate the median value and the 1 $\sigma$ confidence interval, respectively.}
 \label{fig:posterior}
\end{figure}
\subsubsection{Profile}}
While we have used a constant eddy diffusion coefficient throughout the atmosphere in this study, $K_{zz}$ is expected to vary over several orders of magnitude within the atmosphere \citep{2013A&A...558A..91P, 2018ApJ...866....1Z, 2019ApJ...881..152K}, which is also observed in the atmospheres of the solar system planets \citep[e.g.,][]{1981JGR....86.3617A, 2005JGRE..110.8001M}.
Employing an eddy diffusion coefficient profile, such as an increasing coefficient with decreasing pressure, will be the subject of future work.

%Also, as the first step to retrieve $K_{zz}$ of exoplanet atmospheres, we have retrieved $K_{zz}$ used in the chemistry calculation and cloud simulation completely independently.
%This is because of the possible different locations of quenching and cloud formation.
%We leave the retrieval simulation imposing the same constant value of $K_{zz}$ throughout the altitude or its vertical profile for both the chemistry and cloud calculations also to future work.

\subsection{Effect of photochemistry}
\edt{As mentioned in the introduction, we have ignored photochemistry.
%, which is of more importance in the upper atmosphere, in our model.
%with pressure of roughly $\lesssim 10^{-4}$~bar, 
Its effect
%on the atmospheric region where we can probe by transmission spectroscopy in the optical and infrared 
is more significant for cooler atmospheres because of the longer timescale of thermochemical reactions, and the atmospheres of the planets close to the host stars.
The UV irradiation from the host star dissociates the molecules \edttwo{such as $\mathrm{H_2O}$} in the upper atmosphere. \edttwo{Thus, if we think naively,} neglecting the photochemistry effect could lead to an underestimate of the chemical eddy diffusion coefficient needed to reproduce the observed absorption features of molecules such as $\mathrm{H_2O}$.\footnote{
\edttwo{We note that reality is more complex. For example, OH molecules produced by the photodissociation of $\mathrm{H_2O}$ react with $\mathrm{H_2}$, restoring $\mathrm{H_2O}$ molecules \citep{2003ApJ...596L.247L}. Hydrogen atoms formed by this process can be crucial for atmospheric chemistry.}}
On the other hand, in hot ($\gtrsim 1000$~K) hydrogen-dominated atmospheres, photodissociation of CO contributes to the formation of additional $\mathrm{CO_2}$ in the upper atmosphere \citep[e.g.,][]{2011ApJ...737...15M, 2012A&A...546A..43V} while in cool ($\lesssim 1000$~K) atmospheres, photodissociation of $\mathrm{CH_4}$ forms both $\mathrm{CO_2}$ and CO at high altitudes \citep[e.g.,][]{2012ApJ...745....3M, 2018ApJ...853....7K}.
Thus, for the spectra with prominent $\mathrm{CO_2}$ and/or CO features, neglecting the photochemistry effect could instead lead to overestimation of the chemical eddy diffusion coefficient.
Among the \textit{Hubble} and {\textit{Spitzer}} data we have used in this study, $\mathrm{CO_2}$ and CO can exhibit their features in {\textit{Spitzer}}'s IRAC/Ch2 (4.5~$\mu$m) band.
%For the observed transmission spectra we have used in this study, however, .......}
%, especially for planets with low temperatures and/or strong UV irradiation from the host stars.}
}

\edt{
To examine the effect of neglecting photochemistry on the observed indications of disequilibrium chemistry, we have performed additional standard disequilibrium and equilibrium retrievals excluding the data of {\textit{Spitzer}}'s IRAC/Ch2 (4.5~$\mu$m) band for WASP-39b, while the results of these retrievals for HD~209458b have already been presented in Table~\ref{table:evidence_wo}.
We find that disequilibrium chemistry remains favored for both planets HD~209458b and WASP-39b, with natural logs of the Bayes factors of
6.9 and 2.3, respectively.
These values are almost the same as those in Table~\ref{table:evidence}, indicating that the features of $\mathrm{CO_2}$ and CO hardly affect the need for disequilibrium chemistry in the atmospheres of those planets.
The retrieved values of the chemical eddy diffusion coefficients are 
$\log_{10}{\left( K_{zz}^\mathrm{Ch}~[\mathrm{cm}^2~\mathrm{s}^{-1}] \right)} = 13.05^{+1.37}_{-2.45}$ and $8.65^{+3.76}_{-4.34}$ for HD~209458b and WASP-39b, respectively, which are also 
similar to the retrieved values for the data including IRAC/Ch2 of ${13.12}^{+1.39}_{-4.61}$ and $8.61^{+3.68}_{-4.22}$.
Thus, the retrieved large value of $K_{zz}^\mathrm{Ch}$ for HD~209458b from the standard disequilibrium retrieval still holds and is not greatly affected by the effect of neglecting photochemistry.}

\subsection{Additional chemical species}
In this study, we have considered the quenching process for the molecules containing the most abundant elements next to hydrogen and helium, namely carbon, oxygen, and nitrogen.
%, that are abundant in hydrogen/helium-dominated atmospheres.
Currently, the reaction rate coefficients are less constrained for the molecules consisting of the other elements, while they are also expected to be subject to the quenching effect.
Derivation of the chemical timescale for molecules such as $\mathrm{H_2S}$ and $\mathrm{PH_3}$ is urgently needed given their relatively good observability and large abundances.

\subsection{Effect of photochemical haze}
While we have considered the effect of clouds in this study using the model of \citet{2019A&A...622A.121O}, we have ignored the effect of photochemical haze that is also expected to form in the atmospheres of exoplanets \citep{2009arXiv0911.0728Z, 2012ApJ...745....3M, 2013ApJ...769....6H, 2013ApJ...775...33M, 2015ApJ...815..110M, 2016ApJ...824..137Z, 2017AJ....153..139G, 2017ApJ...847...32L, 2018ApJ...853....7K, 2020NatAs...4..951G}.
Clouds and haze are expected to have opposite dependences on the eddy diffusivity.
A large $K_{zz}$ can raise clouds, which are usually formed in the relatively lower atmosphere, thus resulting in an optically thick atmosphere \citep{2018ApJ...863..165G, 2019A&A...622A.121O}.
On the other hand, vigorous mixing efficiently removes photochemical haze from the upper atmosphere making the atmosphere optically thinner \citep{2019ApJ...877..109K, 2020ApJ...895L..47O}.
Thus, the inclusion of haze particles, which we will leave for future work, might lead to different results, especially for $K_{zz}$.

\subsection{Toward 2D and 3D retrieval modeling with disequilibrium chemistry}
\edt{In this study, we have considered 1D atmospheres for the sake of simplicity and reducing the computational cost, consistent with most current spectral retrieval models. Recently, \citet{2020MNRAS.493..106I} and \citet{2020AJ....160..137F} explored an extension of the retrieval method beyond 1D modeling since close-in exoplanets are subject to tidal locking and thus significant horizontal and latitudinal variations in the atmospheric properties, which can be explored by phase curve observations, are expected.
The approach we have proposed in this study of including disequilibrium chemistry in spectral retrievals with a physical basis is also applicable to future 2D and 3D retrieval modeling, by adding horizontal or latitudinal transport term to the continuity-transport equation, Eq.~(\ref{eq:cont}) \citep[e.g.,][]{2006ApJ...649.1048C, 2018ApJ...869..107M}.
}

\section{Conclusions} \label{sec:conclusion}
In this study, we have implemented the disequilibrium effect of vertical mixing or quenching to the spectral retrieval code ARCiS \citep{2020A&A...642A..28M} with a physical basis.
Adopting a chemical relaxation method with a chemical timescale derived by \citet{2018ApJ...862...31T}, we have developed a module to compute the profiles of molecular abundances taking the disequilibrium effect into account for the major species in hydrogen/helium-dominated atmospheres, namely $\mathrm{CH_4}$, $\mathrm{CO}$, $\mathrm{H_2O}$, $\mathrm{NH_3}$, $\mathrm{N_2}$, and $\mathrm{CO_2}$. Then using ARCiS updated with this module, we have performed retrievals of the observed transmission spectra of 16 exoplanets with sizes ranging from Jupiter to mini-Neptune.

We have found \edt{indicative} evidence %tentative evidence 
of disequilibrium chemistry for HD~209458b and WASP-39b. For HD~209458b, the retrieved value of the eddy diffusion coefficient, which is used in the chemistry calculation, is as large as $\log_{10}{\left( K_{zz}^\mathrm{Ch}~[\mathrm{cm}^2~\mathrm{s}^{-1}] \right)} = {13.12}^{+1.39}_{-4.61}$ from our standard disequilibrium retrieval\hl{, indicating that} %while $\log_{10}{\left( K_{zz}~[\mathrm{cm}^2~\mathrm{s}^{-1}] \right)} = 4.17^{+1.15}_{-1.05}$ for the disequilibrium retrieval employing the same eddy diffusion coefficient for both cloud and chemistry calculations.
%Despite some difference in the derived $K_{zz}$ values, 
disequilibrium chemistry plays a significant role in determining the molecular abundance profiles in its atmosphere.
%for the best-fit cases of both two retrievals.
Owing to the enhanced abundance of $\mathrm{NH_3}$ due to the quenching effect, its retrieved spectrum exhibits a strong $\mathrm{NH_3}$ absorption feature at 10.5~$\mu$m, which is absent in the retrieved spectrum from the equilibrium retrieval.
This feature is accessible by {JWST}/MIRI. Thus, HD~209458b offers a unique opportunity to study disequilibrium chemistry in exoplanet atmospheres.
Moreover, for HAT-P-11b \edt{and GJ~436b}, we obtained relatively different results between the disequilibrium and equilibrium retrievals \edt{such as a $2.9\sigma$ difference for} the C/O ratio.
This demonstrates the importance of taking disequilibrium chemistry into account for spectral retrieval, where we might otherwise misinterpret results.
%, especially for metallicity and elemental abundance ratios, since the quenching effect has a considerable influence on those parameters.
We have also examined the trend of the retrieved eddy diffusion coefficients over the equilibrium temperature, though no trend was found, possibly due to the limits of the current observational precision.
This study makes clear that including a consideration of disequilibrium chemistry in spectral retrieval is essential in the coming era of {JWST} and Ariel.

\begin{acknowledgements}
     We wish to thank S.-M. Tsai for kindly providing his calculation data for the code validation.
     The insightful advice and valuable comments on the manuscript received from K. Ohno are greatly appreciated.
     Also, we are grateful to P. Woitke, C. Visscher, and J. Moses for their helpful comments.
     {Finally, we wish to thank the anonymous referee for his/her careful reading and constructive comments and the editor E. Lellouch for his helpful comments, both of which significantly helped improve this paper.}
     The numerical computations were carried out on the PC cluster at the Center for Computational Astrophysics, National Astronomical Observatory of Japan.
     This work was supported by JSPS KAKENHI Grant Numbers JP21K13984 and JP21J04998.
    Y.K. acknowledges support from the European Union’s Horizon 2020 Research and Innovation Programme under Grant Agreement 776403 and Special Postdoctoral Researcher Program at RIKEN.
    This work has made use of Numpy \citep{harris2020array}, Matplotlib \citep{Hunter:2007}, and corner.py \citep{corner}, and we are grateful to the developers of those packages.
\end{acknowledgements}

   \bibliographystyle{aa} % style aa.bst
   \bibliography{ads.bib} % your references Yourfile.bib

\begin{appendix}
\onecolumn
\section{Atmospheric Profiles} \label{sec:profiles}
\begin{figure*}[htb!]
 \begin{subfigure}{\textwidth}
  \addtocounter{subfigure}{4}
   \centering
   \includegraphics[height=4.6cm]{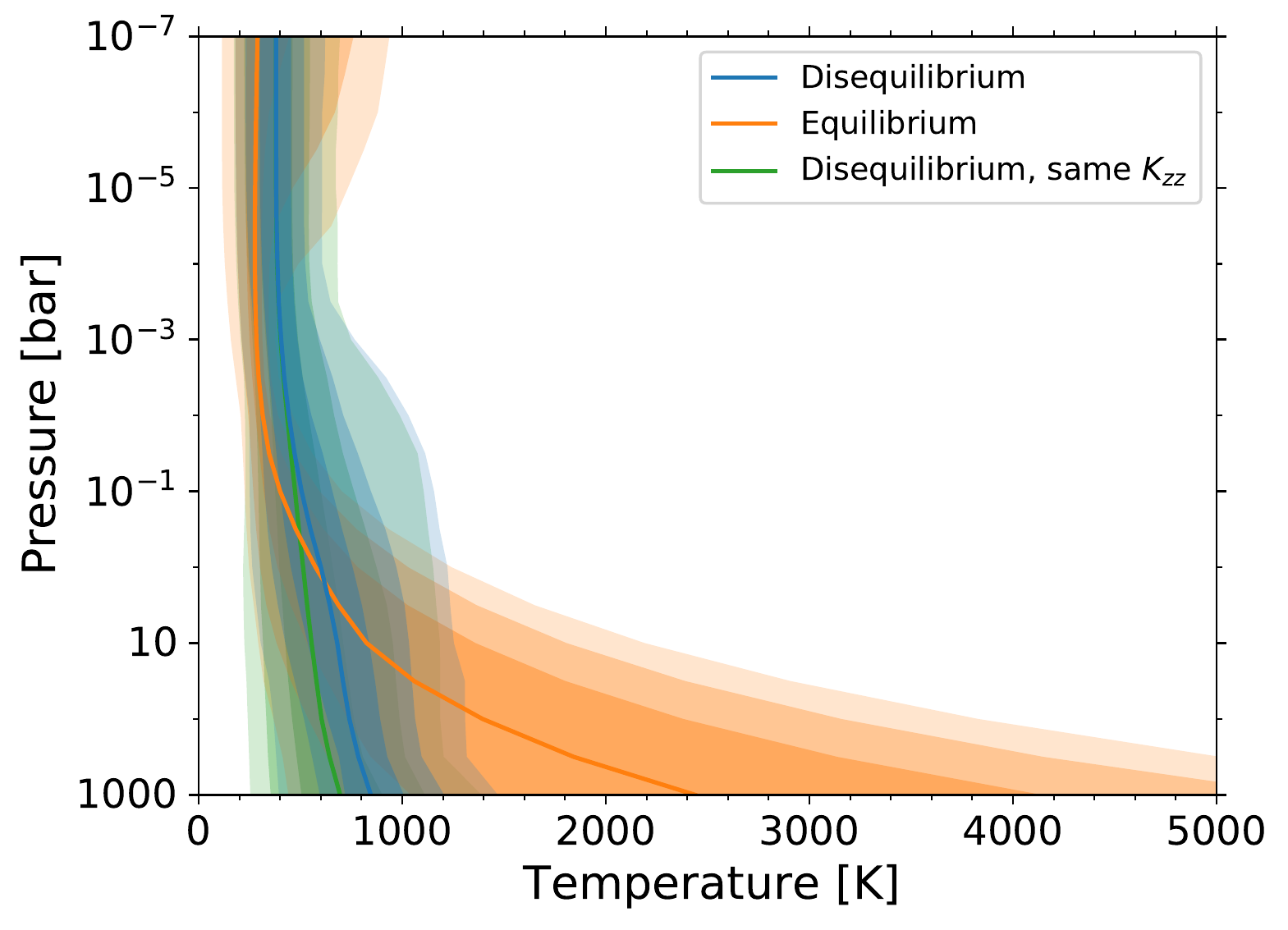}
   \centering
   \includegraphics[height=4.6cm]{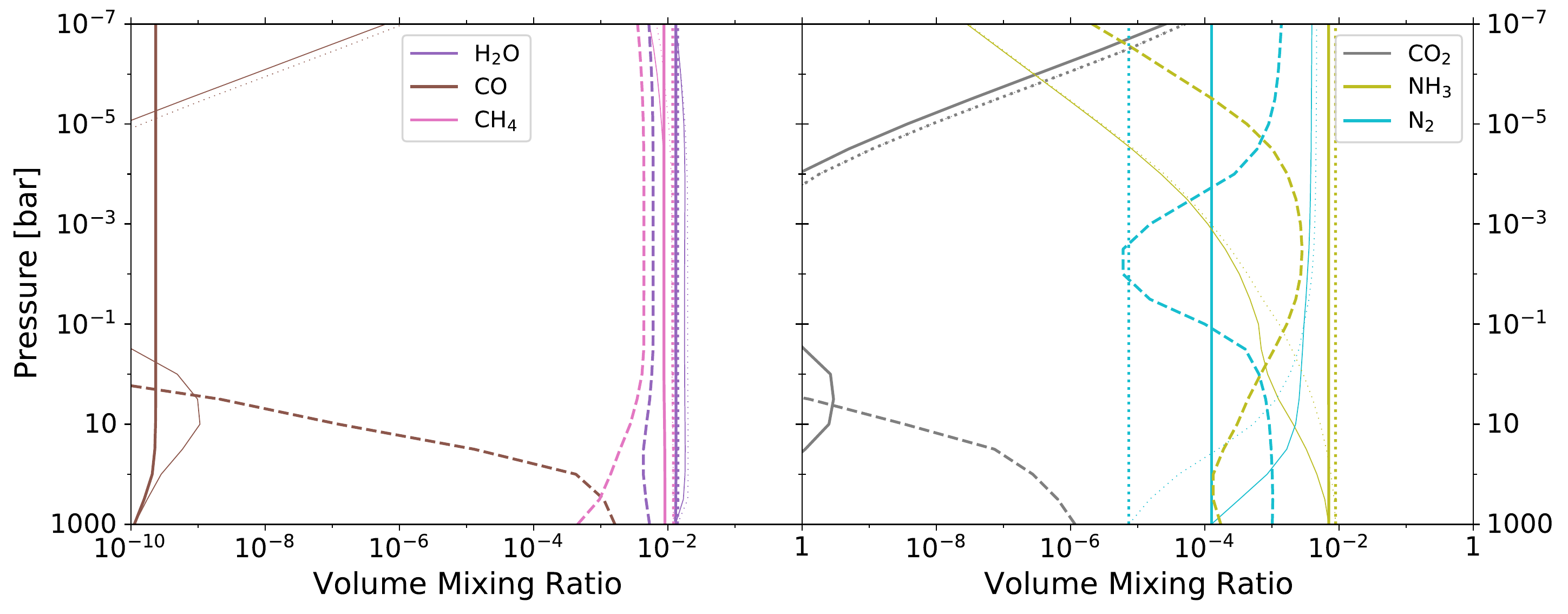}
   \caption{GJ~436b}
 \end{subfigure}
 \begin{subfigure}{\textwidth}
   \centering
   \includegraphics[height=4.6cm]{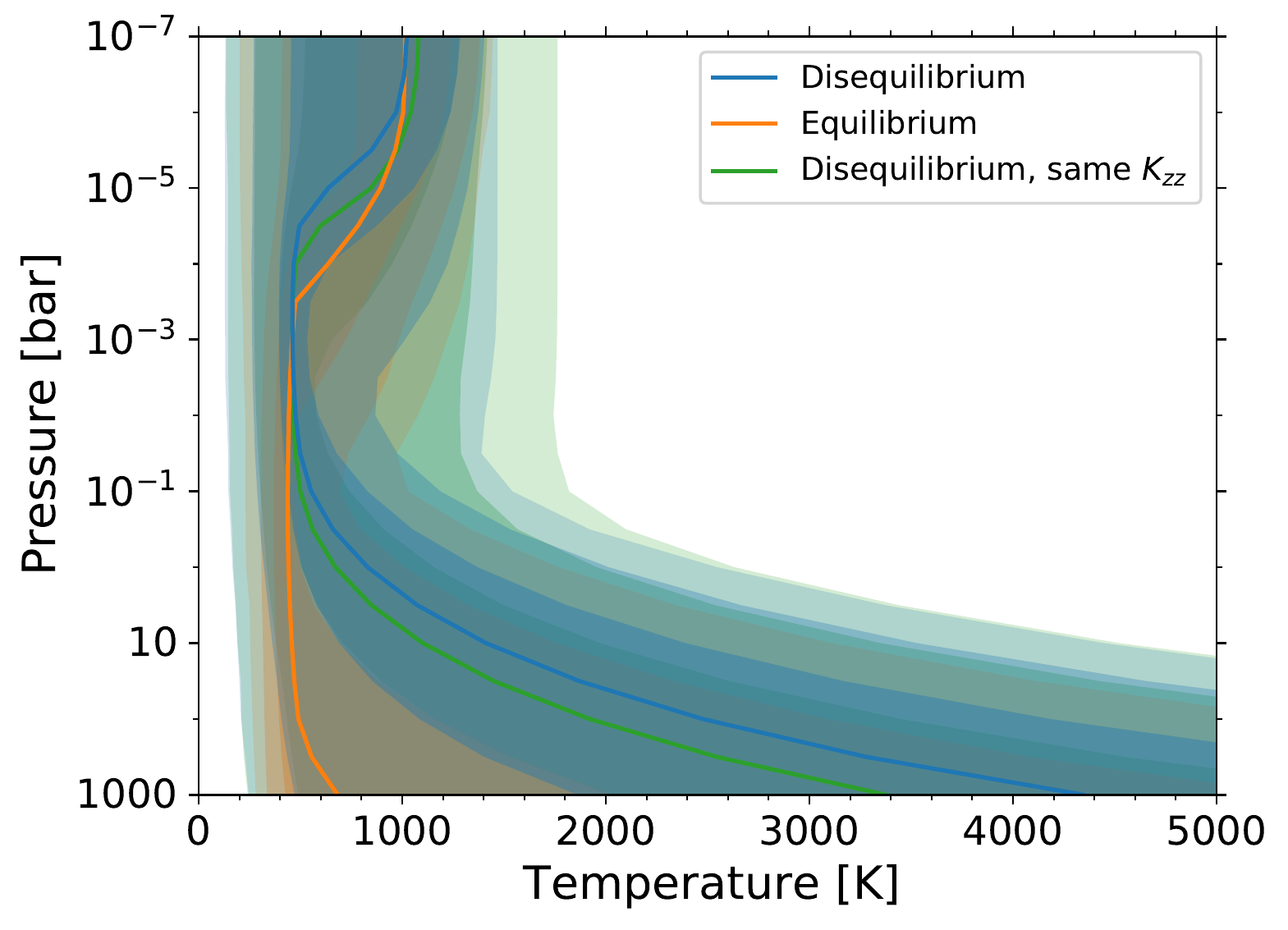}
   \centering
   \includegraphics[height=4.6cm]{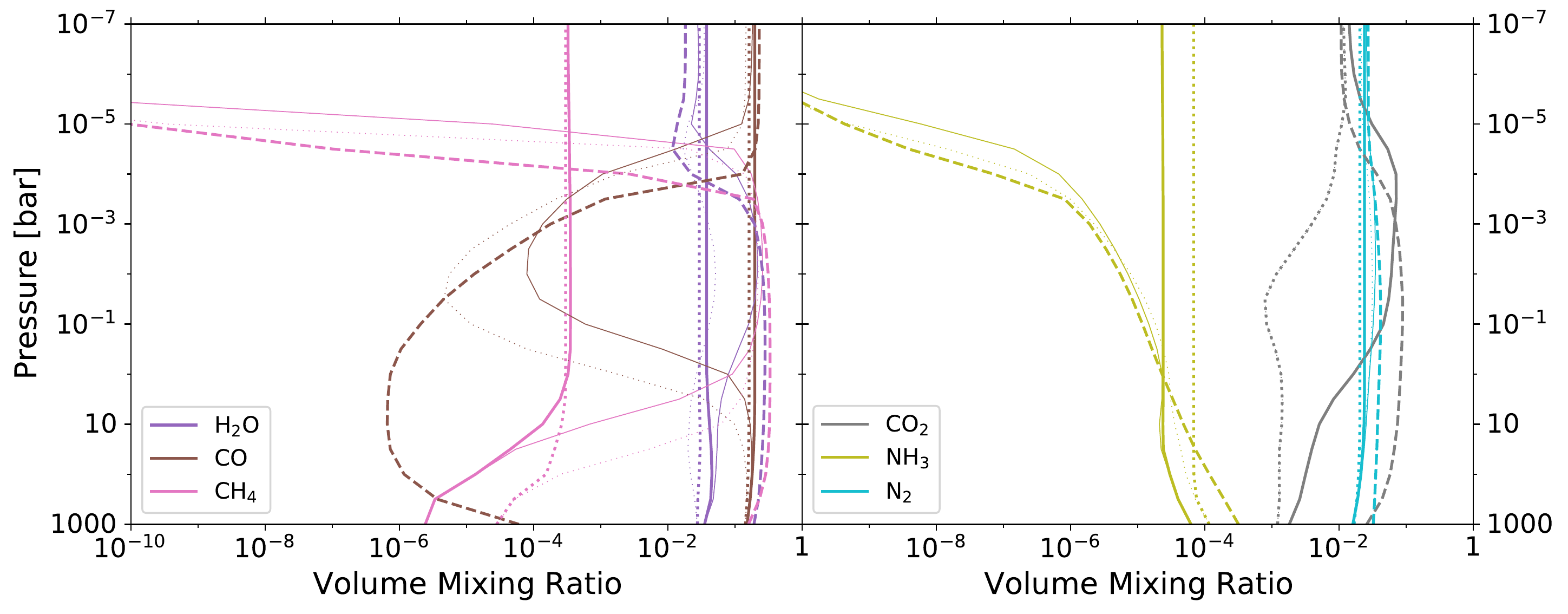}
   \caption{GJ~1214b}
 \end{subfigure}
 \begin{subfigure}{\textwidth}
   \centering
   \includegraphics[height=4.6cm]{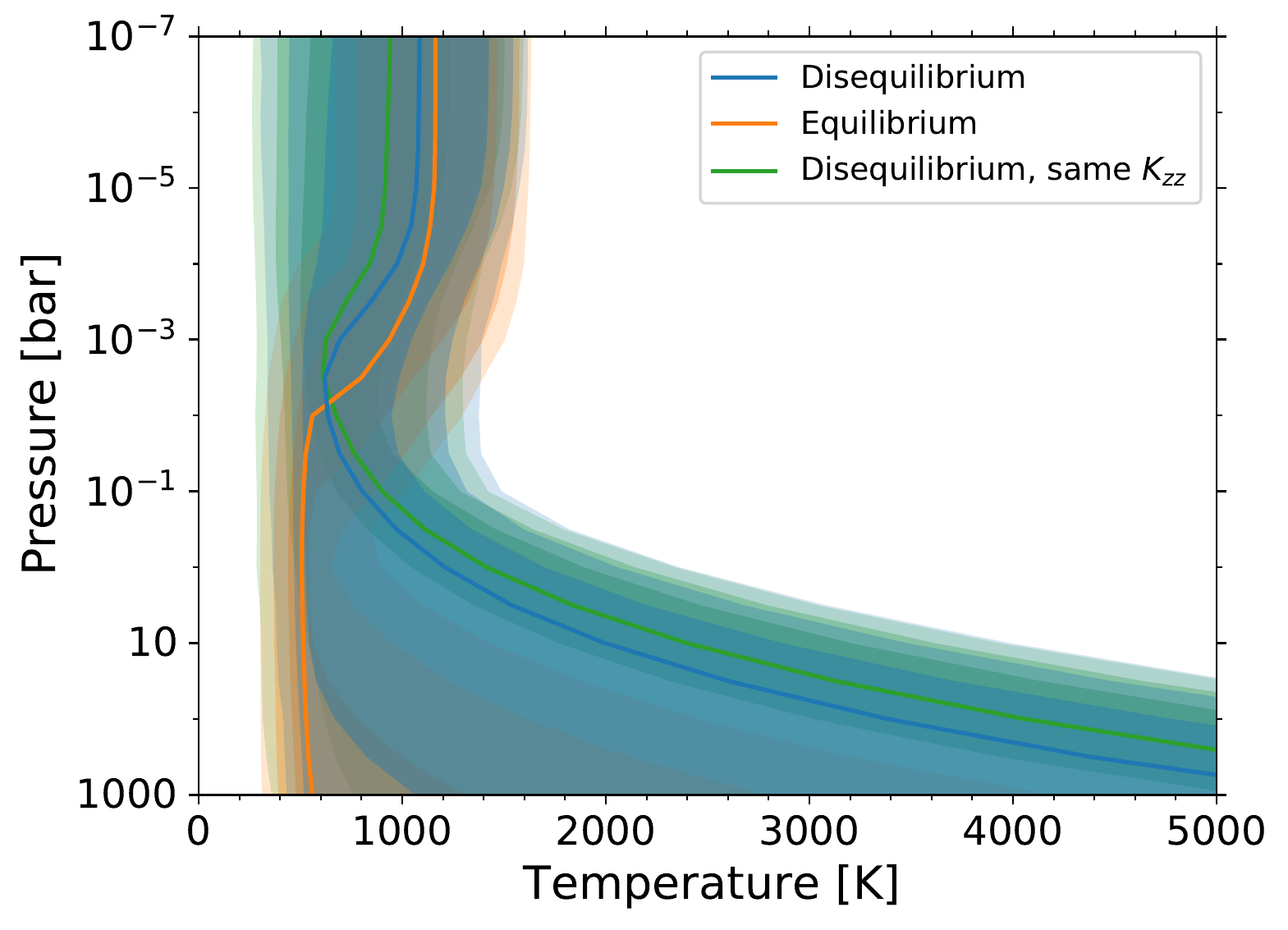}
   \centering
   \includegraphics[height=4.6cm]{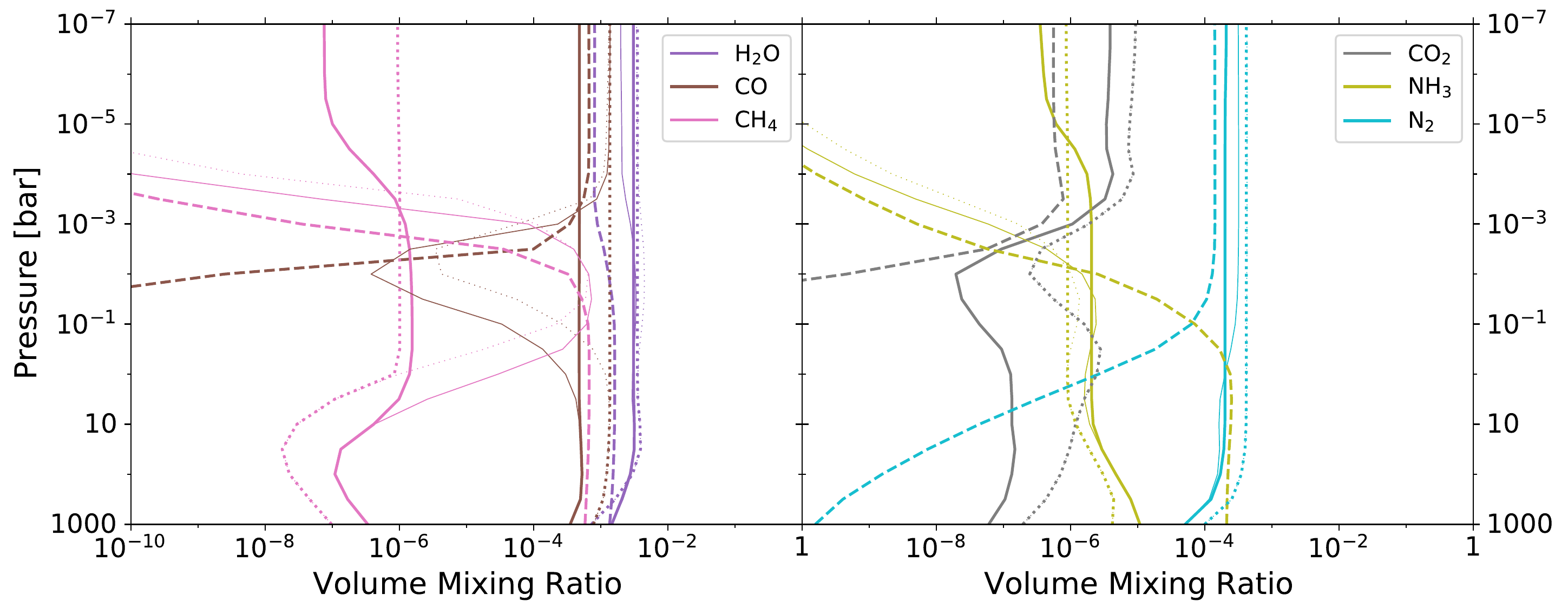}
   \caption{GJ~3470b}
 \end{subfigure}
 \begin{subfigure}{\textwidth}
   \centering
   \includegraphics[height=4.6cm]{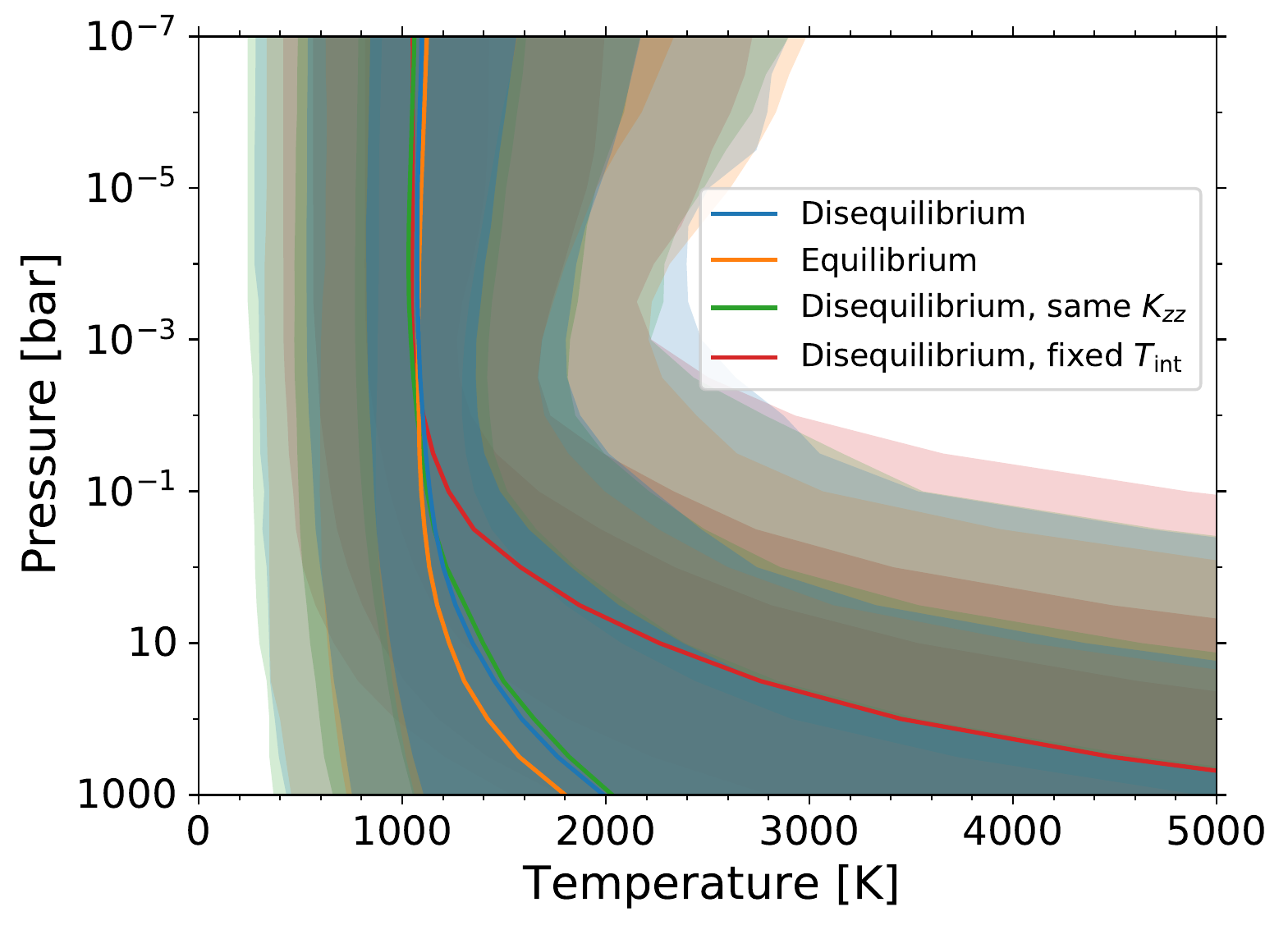}
   \centering
   \includegraphics[height=4.6cm]{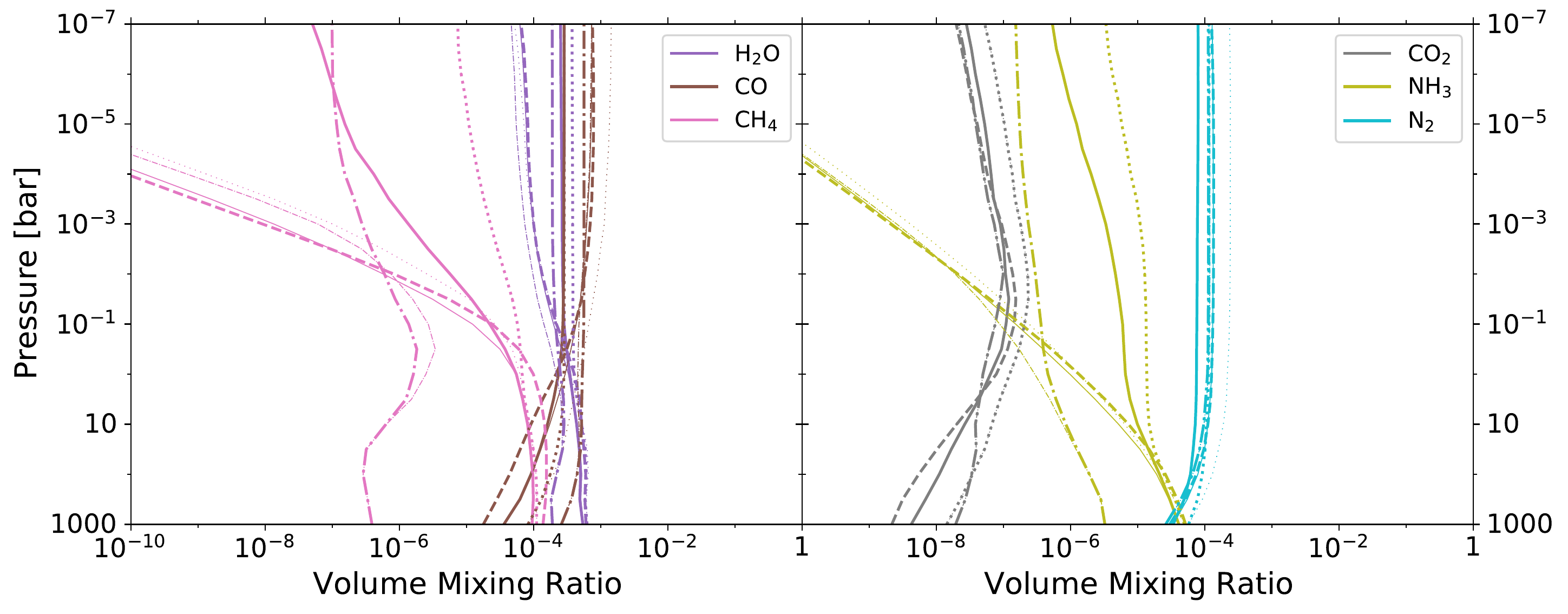}
   \caption{HAT-P-1b}
 \end{subfigure}
 \caption{\edt{As for Fig.~\ref{fig:profile} for the remaining planets.}}
 \label{fig:profile2}
\end{figure*}

\begin{figure*}[htb!]
\ContinuedFloat
 \begin{subfigure}{\textwidth}
   \centering
   \includegraphics[height=4.6cm]{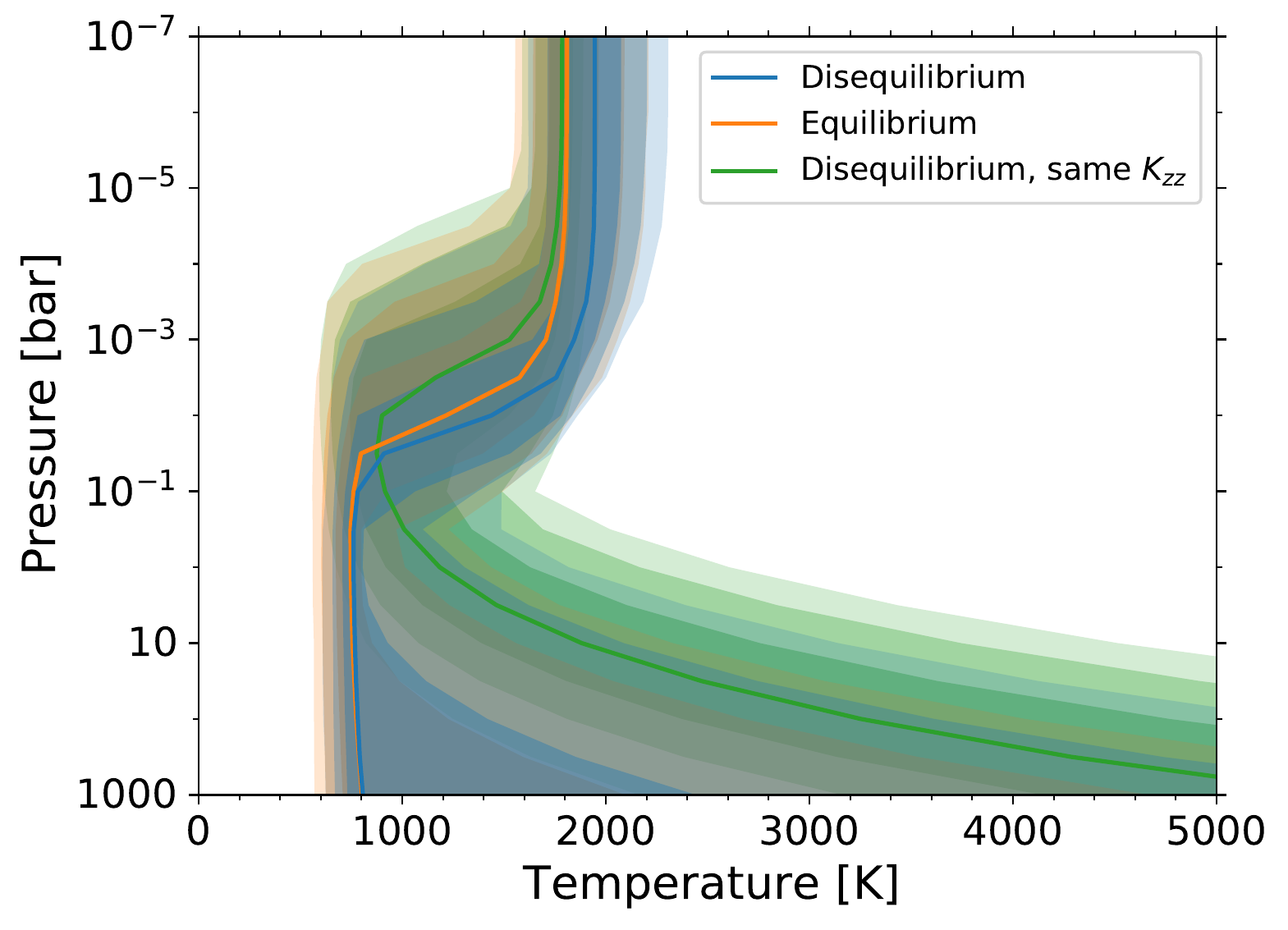}
   \centering
   \includegraphics[height=4.6cm]{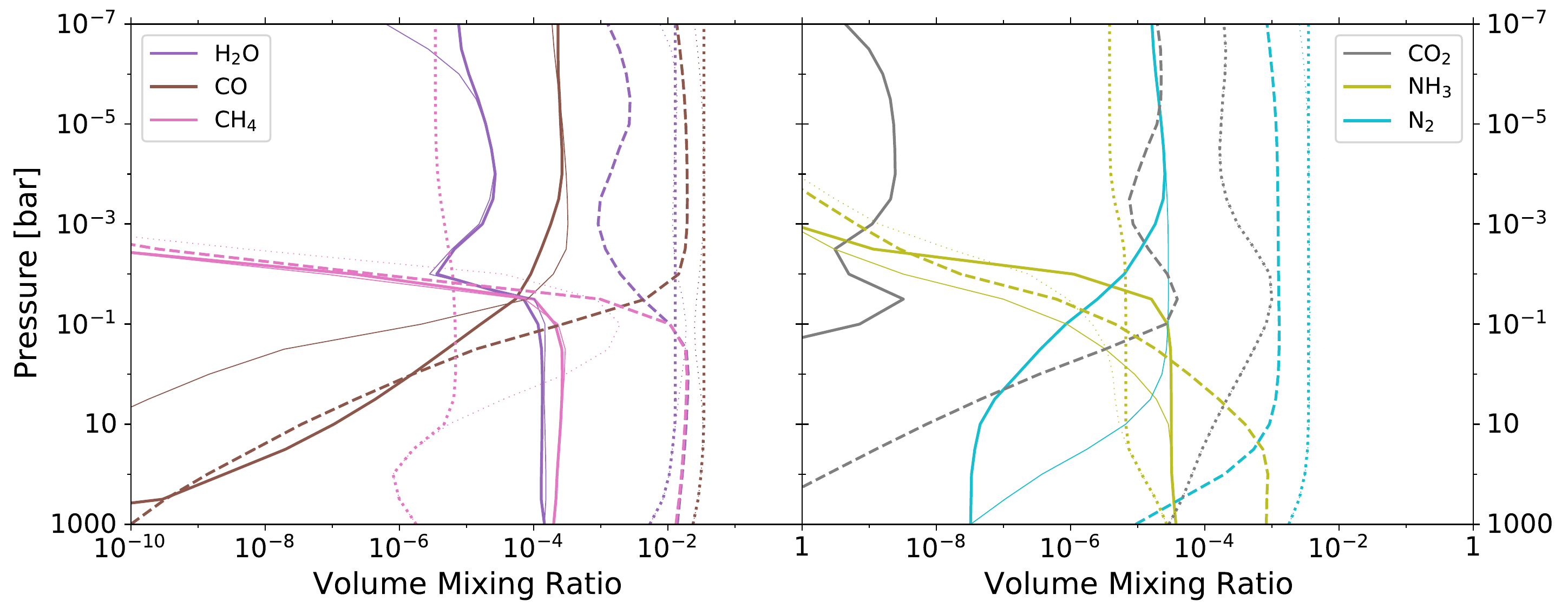}
   \caption{HAT-P-12b}
 \end{subfigure}
 \begin{subfigure}{\textwidth}
   \centering
   \includegraphics[height=4.6cm]{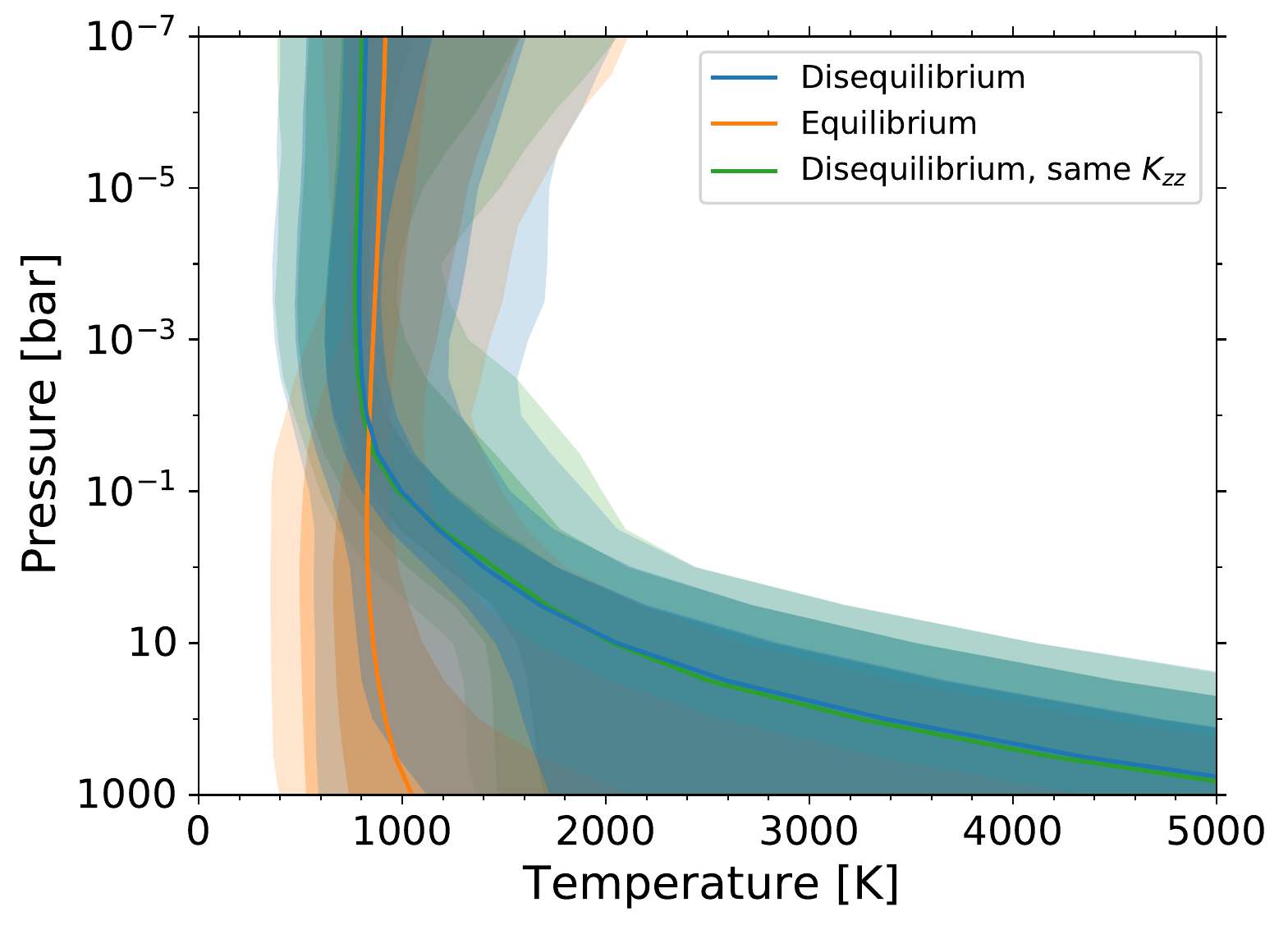}
   \centering
   \includegraphics[height=4.6cm]{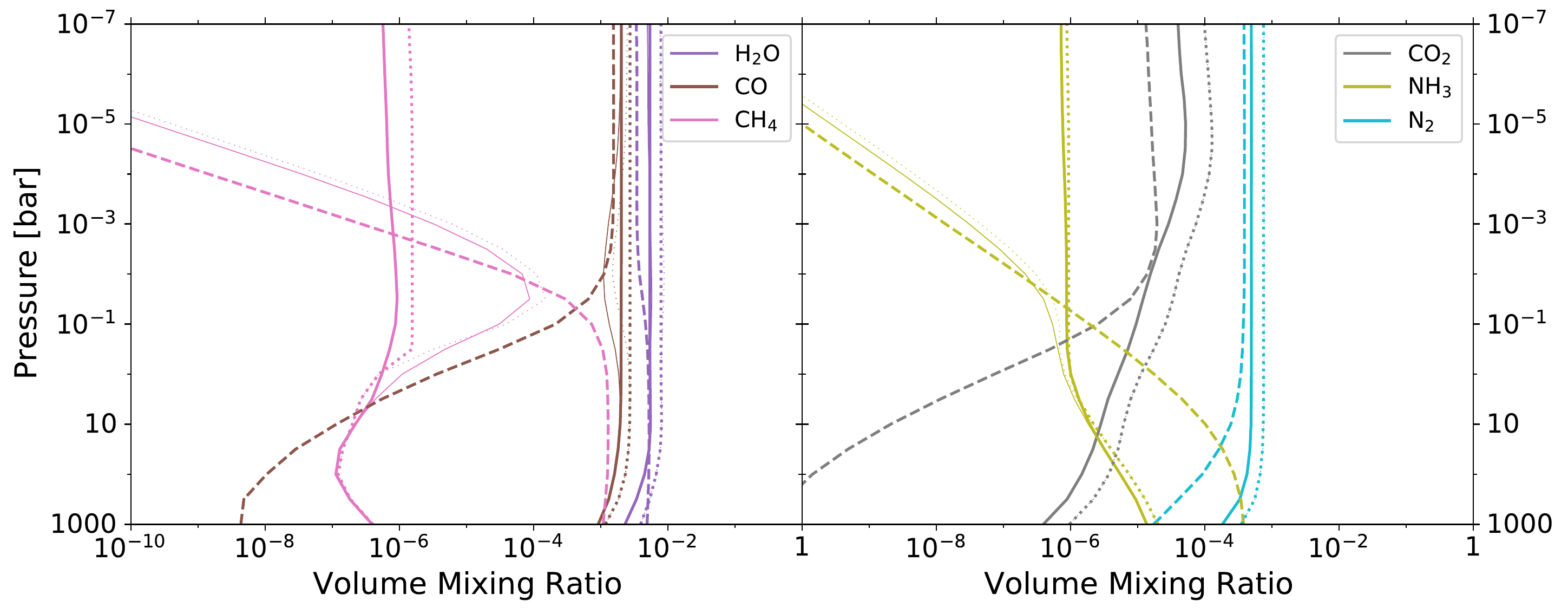}
   \caption{HAT-P-26b}
 \end{subfigure}
 \begin{subfigure}{\textwidth}
   \centering
   \includegraphics[height=4.6cm]{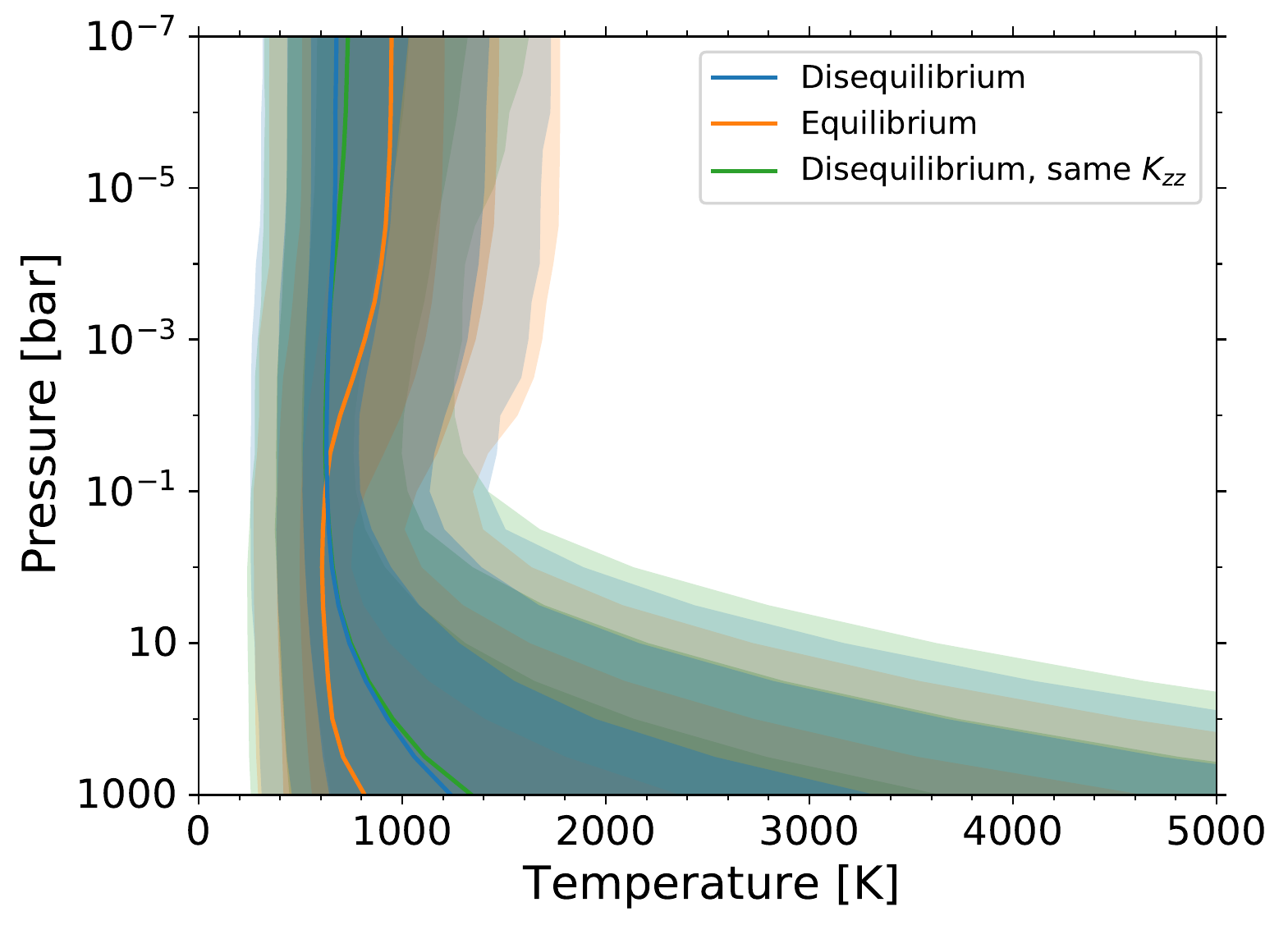}
   \centering
   \includegraphics[height=4.6cm]{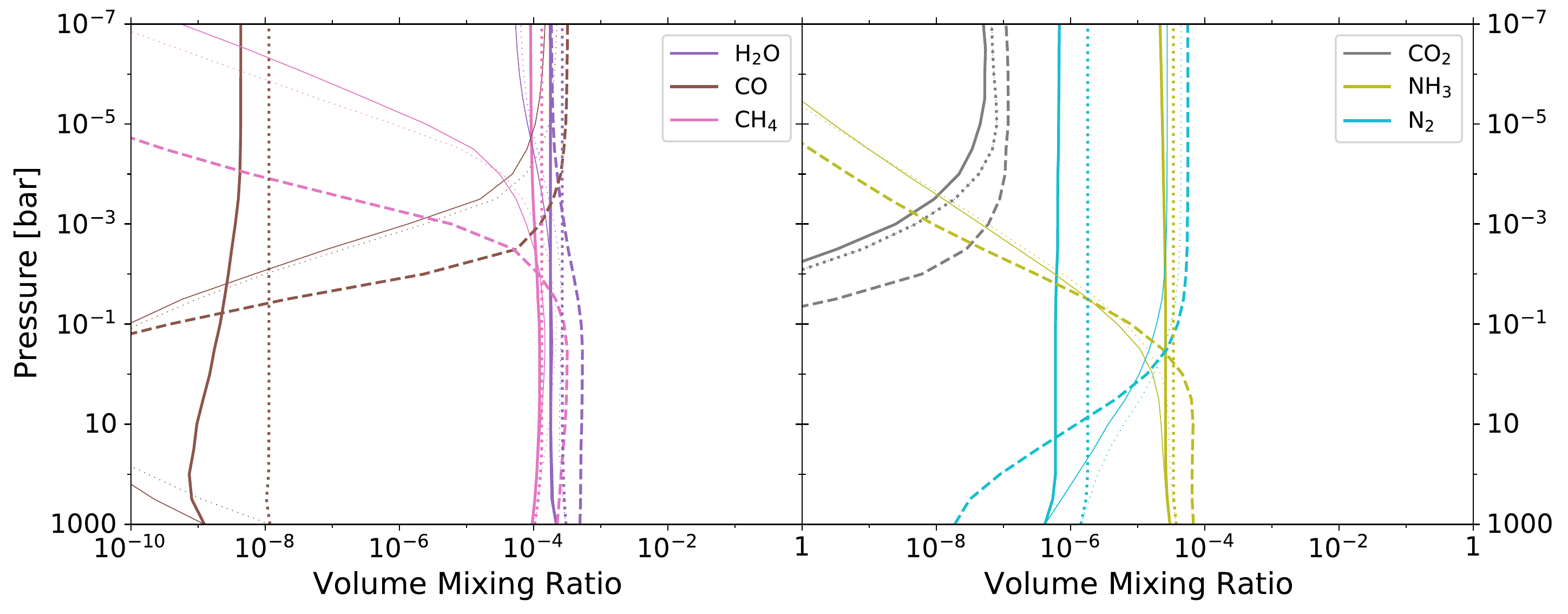}
   \caption{HD~97658b}
 \end{subfigure}
 \begin{subfigure}{\textwidth}
   \centering
   \includegraphics[height=4.6cm]{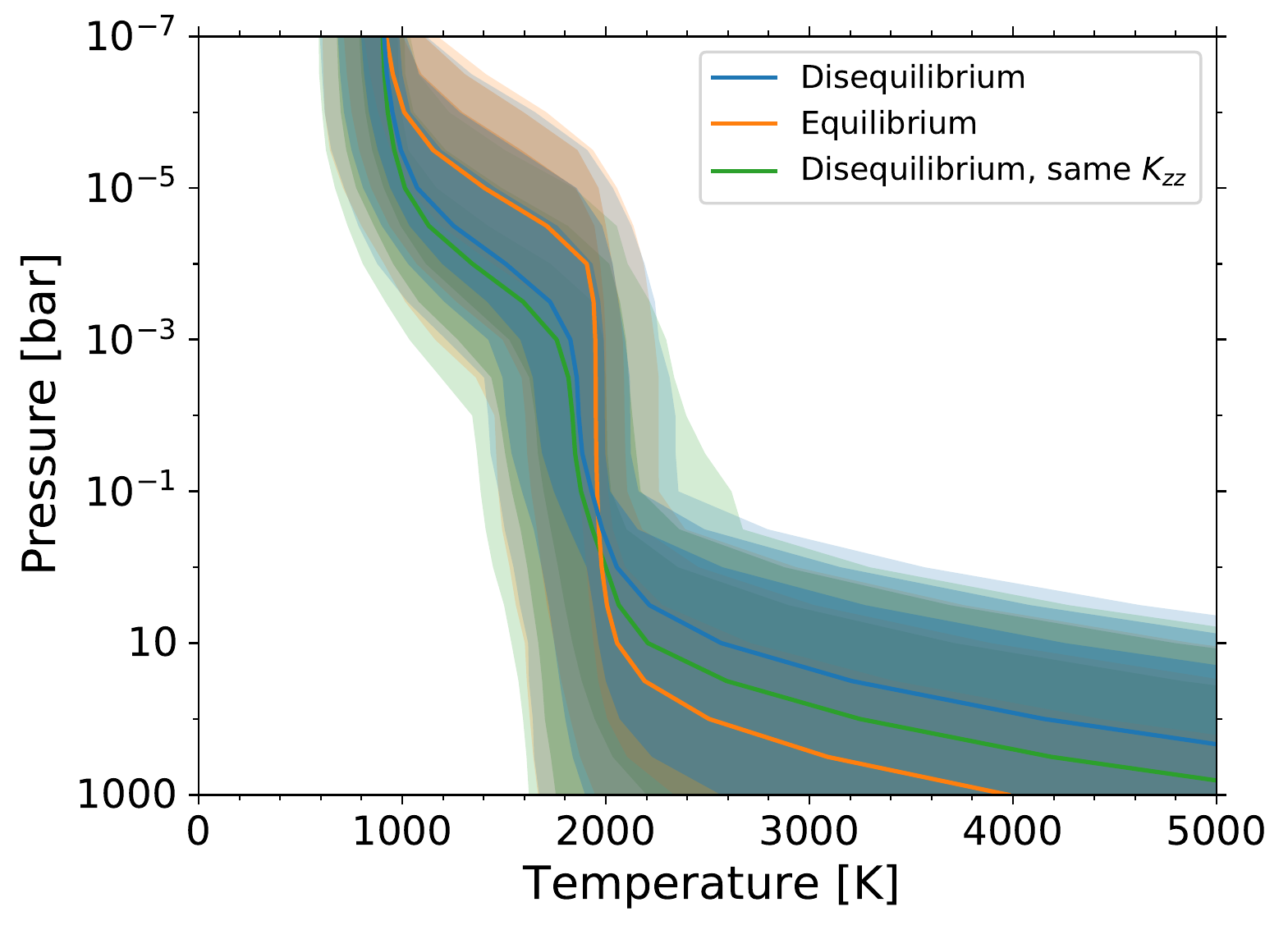}
   \centering
   \includegraphics[height=4.6cm]{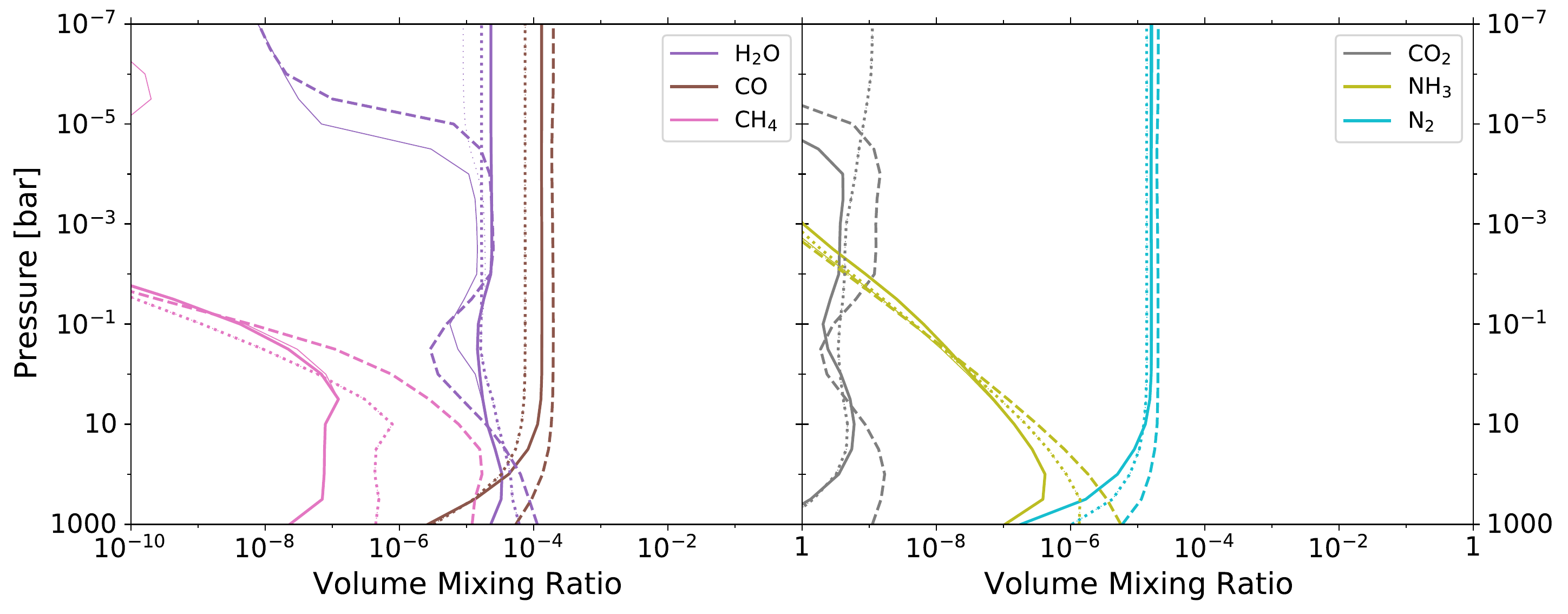}
   \caption{WASP-6b}
 \end{subfigure}
 \caption{Continued.}
\end{figure*}

\begin{figure*}[htb!]
\ContinuedFloat
 \begin{subfigure}{\textwidth}
   \centering
   \includegraphics[height=4.6cm]{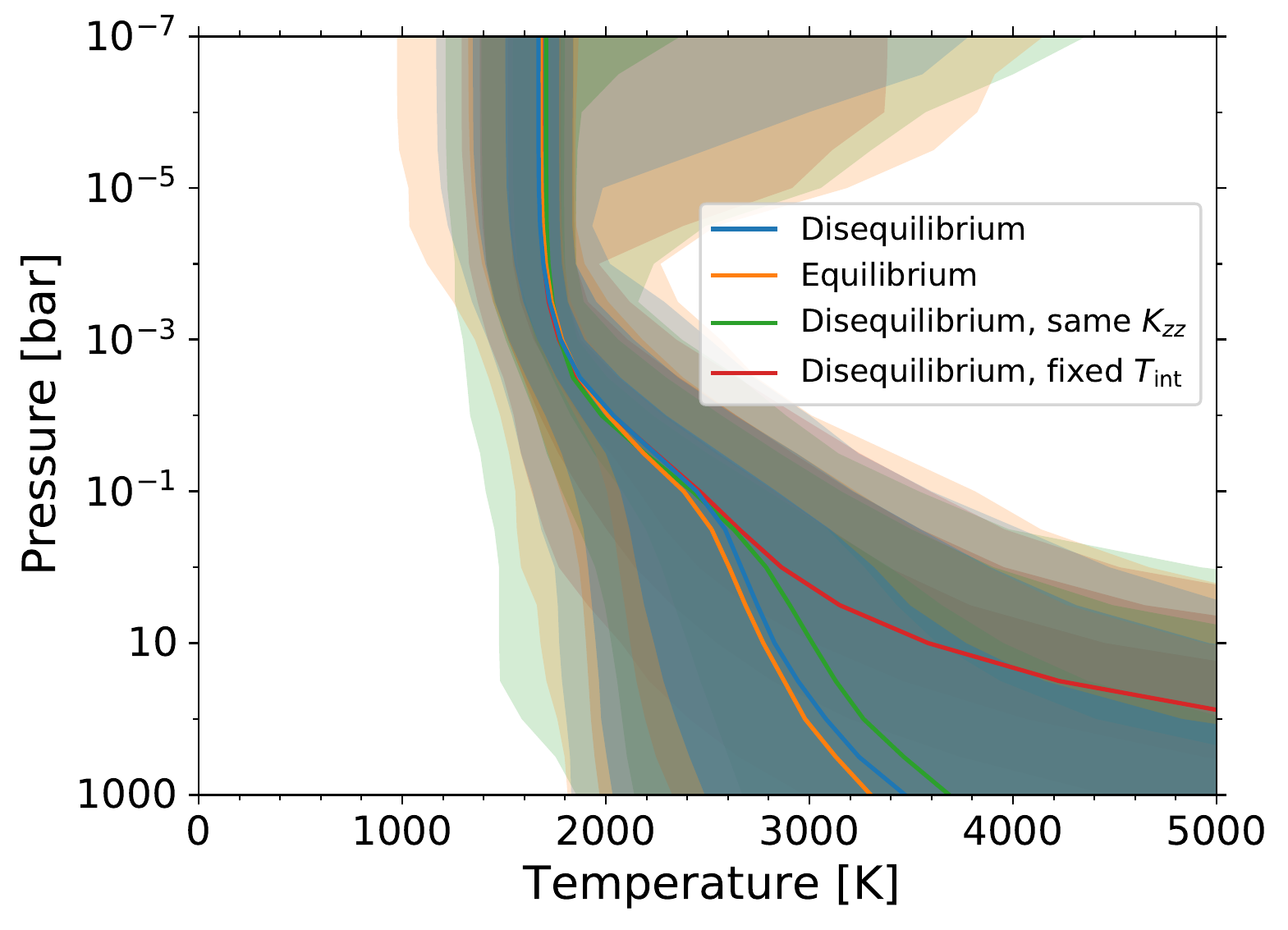}
   \centering
   \includegraphics[height=4.6cm]{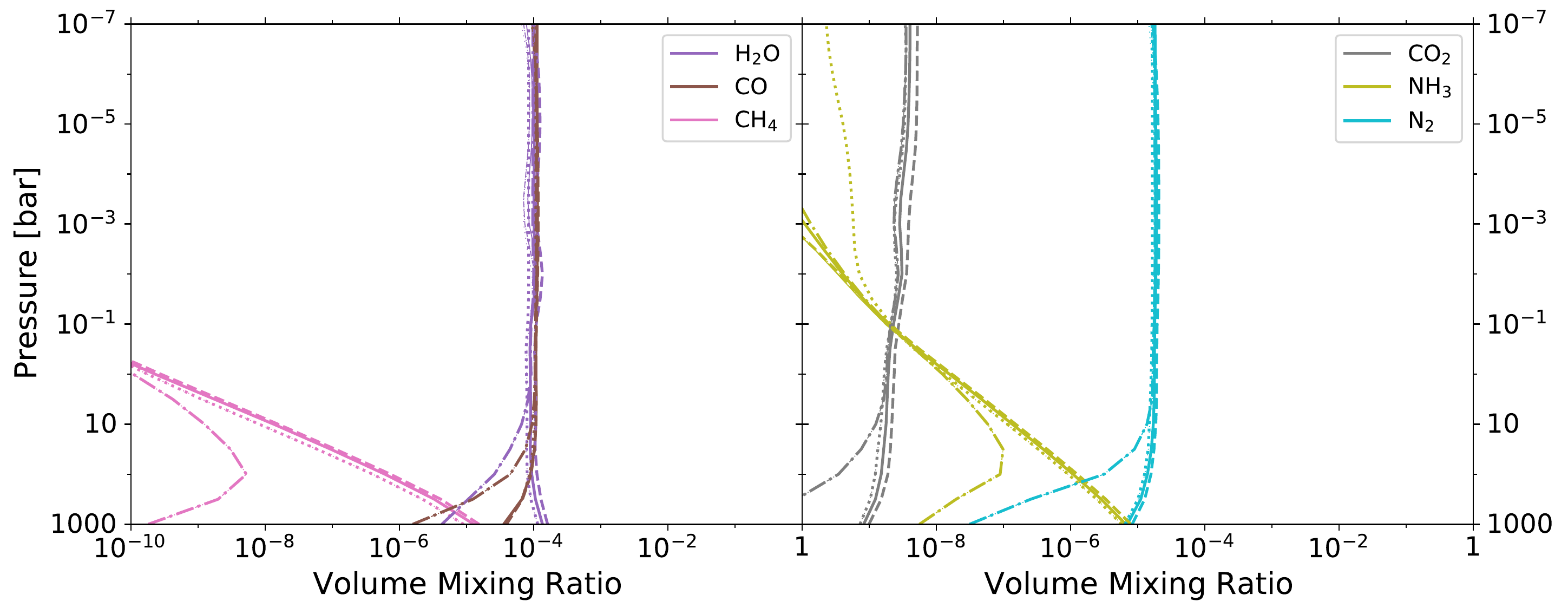}
   \caption{WASP-12b}
 \end{subfigure}
 \begin{subfigure}{\textwidth}
   \centering
   \includegraphics[height=4.6cm]{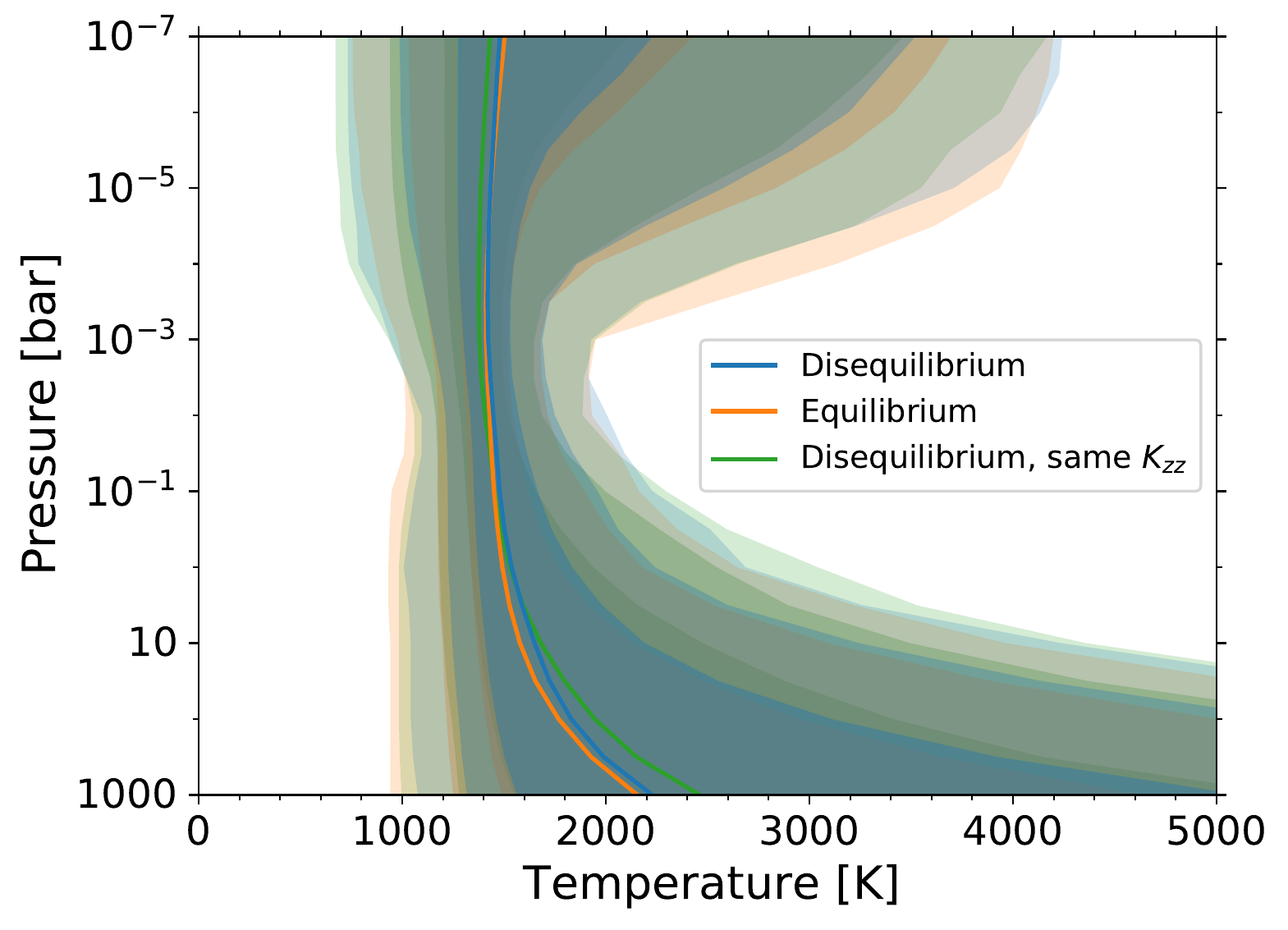}
   \centering
   \includegraphics[height=4.6cm]{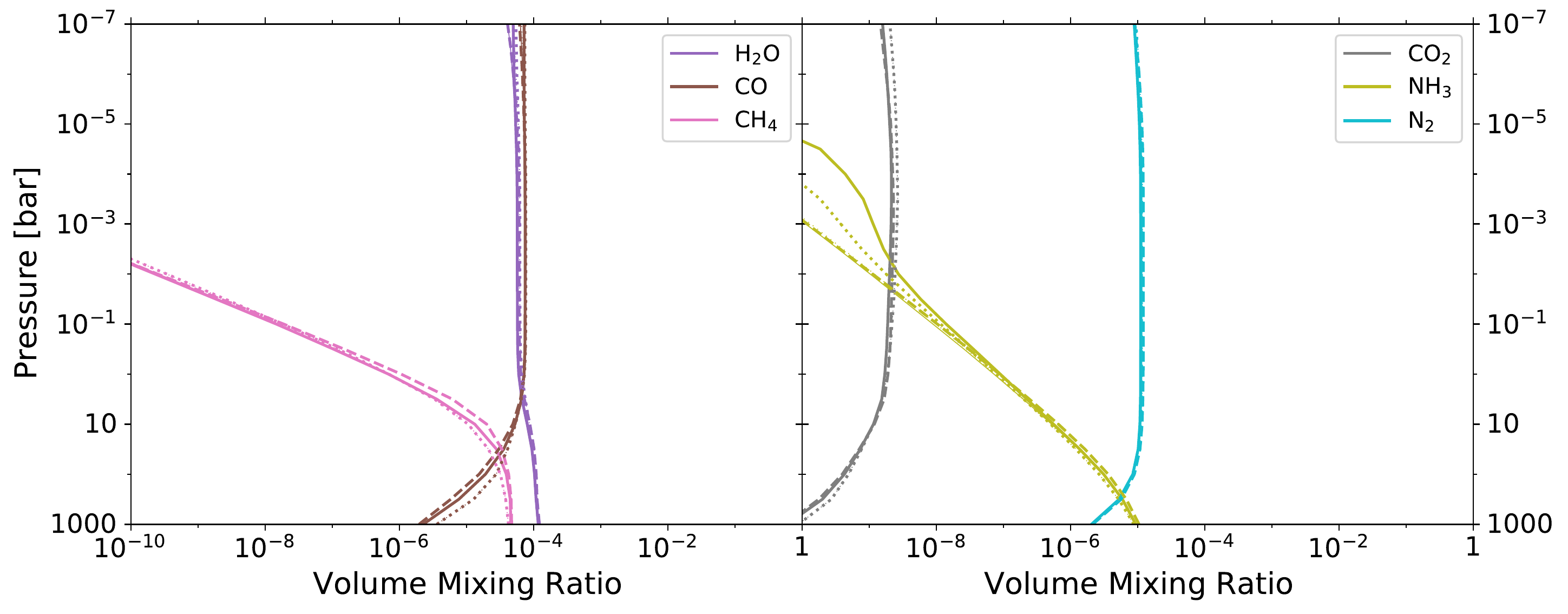}
   \caption{WASP-17b}
 \end{subfigure}
 \begin{subfigure}{\textwidth}
   \centering
   \includegraphics[height=4.6cm]{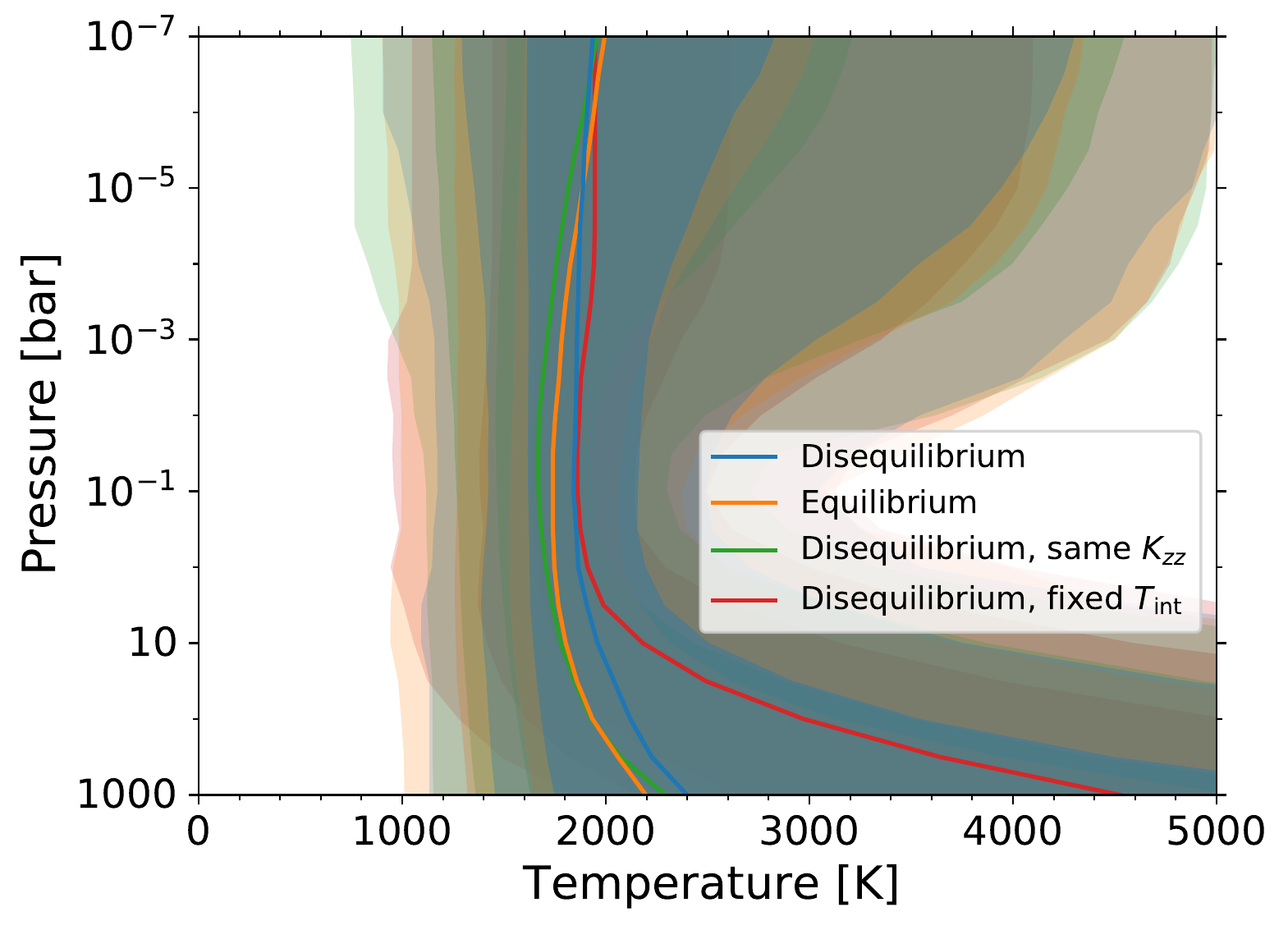}
   \centering
   \includegraphics[height=4.6cm]{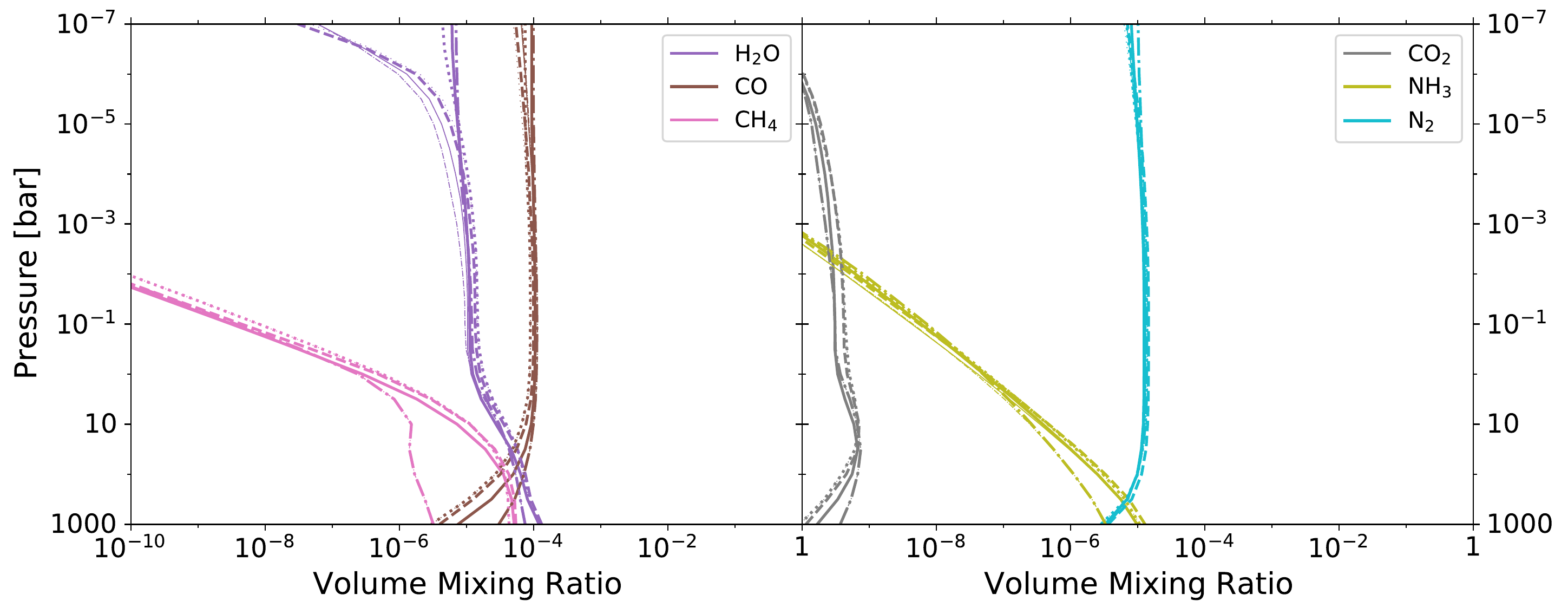}
   \caption{WASP-19b}
 \end{subfigure}
 \begin{subfigure}{\textwidth}
   \centering
   \includegraphics[height=4.6cm]{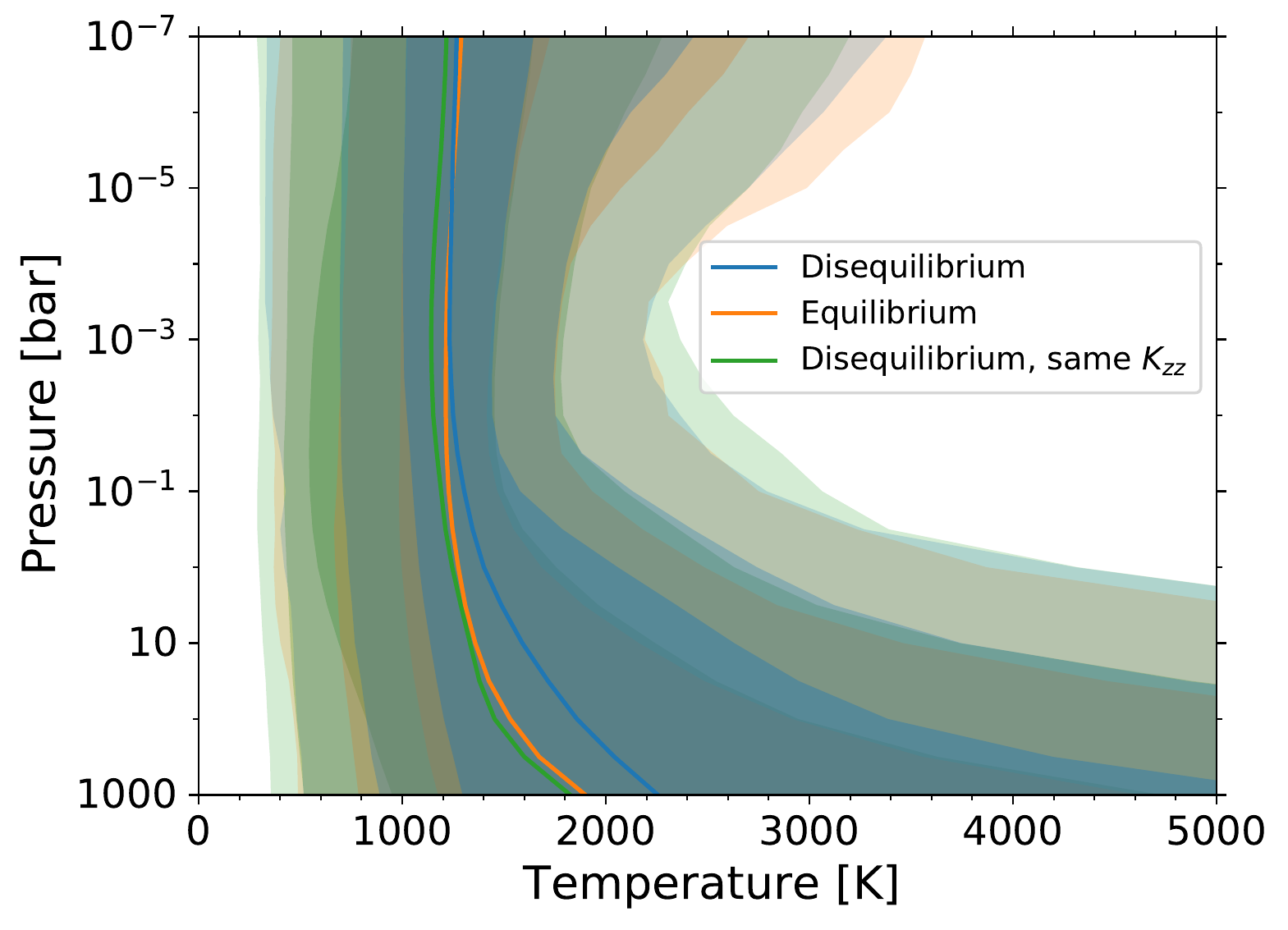}
   \centering
   \includegraphics[height=4.6cm]{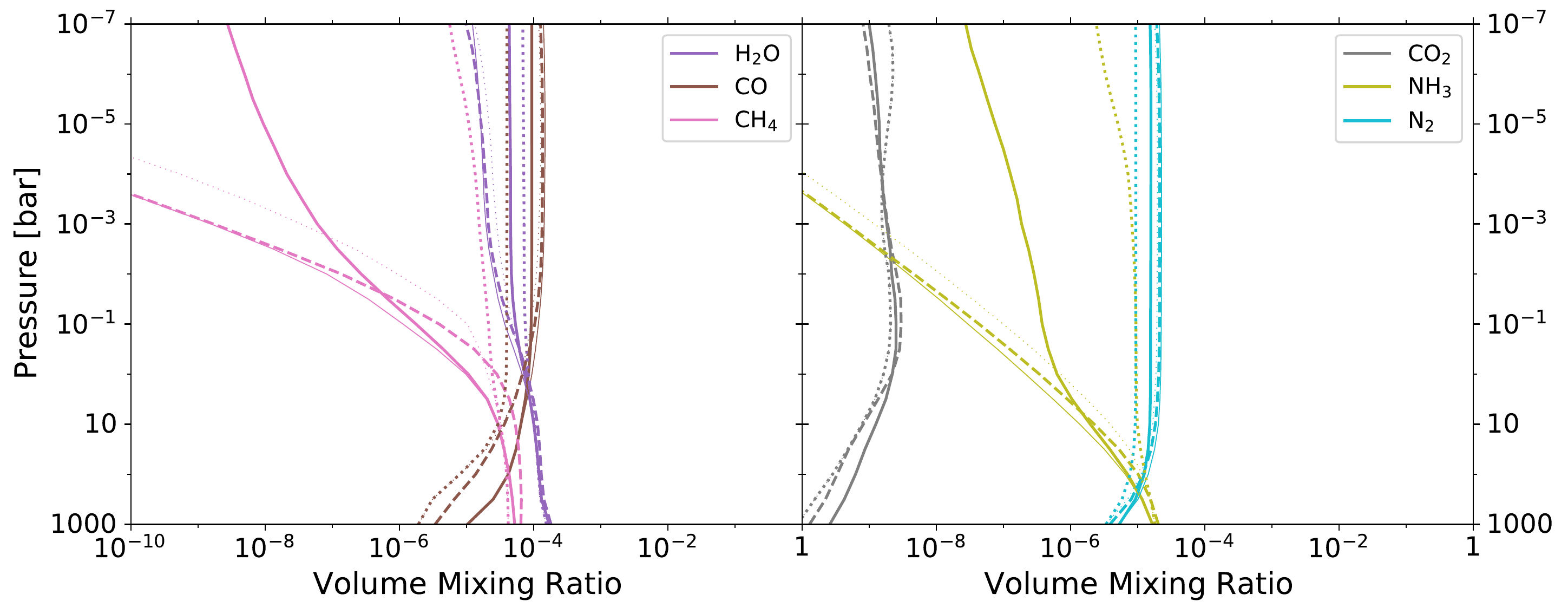}
   \caption{WASP-31b}
 \end{subfigure}
 \caption{Continued.}
\end{figure*}

\onecolumn
\section{Corner Plots} \label{append:corner}

\begin{figure*}[h]
\begin{minipage}{\hsize}
    \centering
    \includegraphics[width=\hsize]{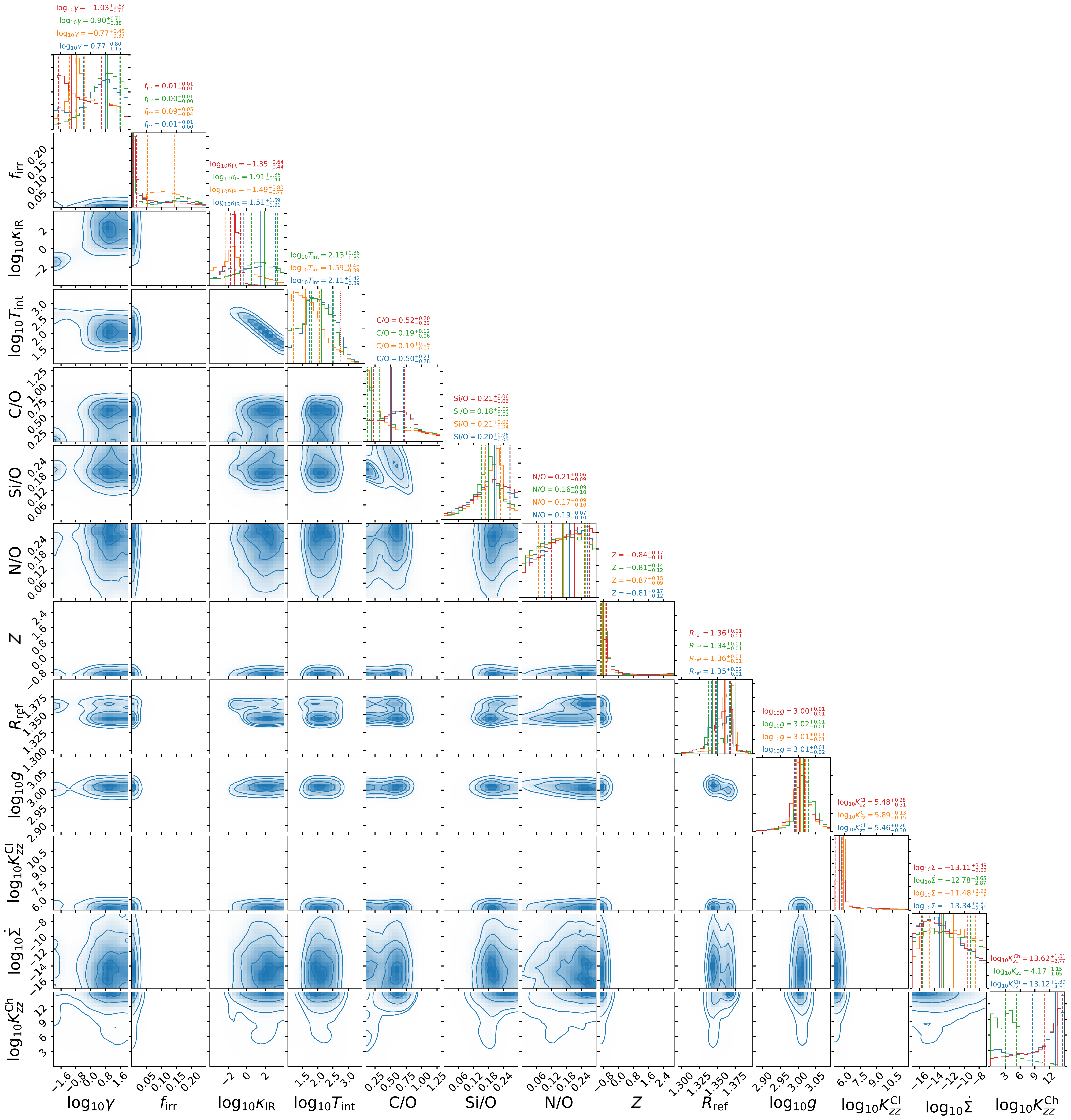}
    \subcaption{HD~209458b}
\end{minipage}

\caption{Corner plots for the standard disequilibrium retrieval (blue).
\edt{Also shown in the posterior probability distribution plots are the results of the equilibrium retrieval (orange) and disequilibrium retrievals with the same $K_{zz}$ (green) and with fixed $T_\mathrm{int}$ (red; only shown for planets with masses larger than half Jupiter mass).}
The vertical solid and dashed lines in the posterior %probability distribution 
plots indicate the median value and the 1$\sigma$ confidence interval for each parameter, respectively.
\hl{$K_{zz}$ retrieved from the same-$K_\mathrm{zz}$ retrieval is shown in the place of $K_{zz}^\mathrm{Ch}$, {and} the vertical red dotted line in the posterior plot of $\log_{10}{T_\mathrm{int}}$ {is} %indicates
the value assumed in the $T_\mathrm{int}$-fixed retrieval.
}
The unit for each parameter is the same as that used in Table~\ref{table:param}.}
\label{fig:corner}
\end{figure*}

\begin{figure*}
\ContinuedFloat
\begin{minipage}{\hsize}
    \centering
    \includegraphics[width=\hsize]{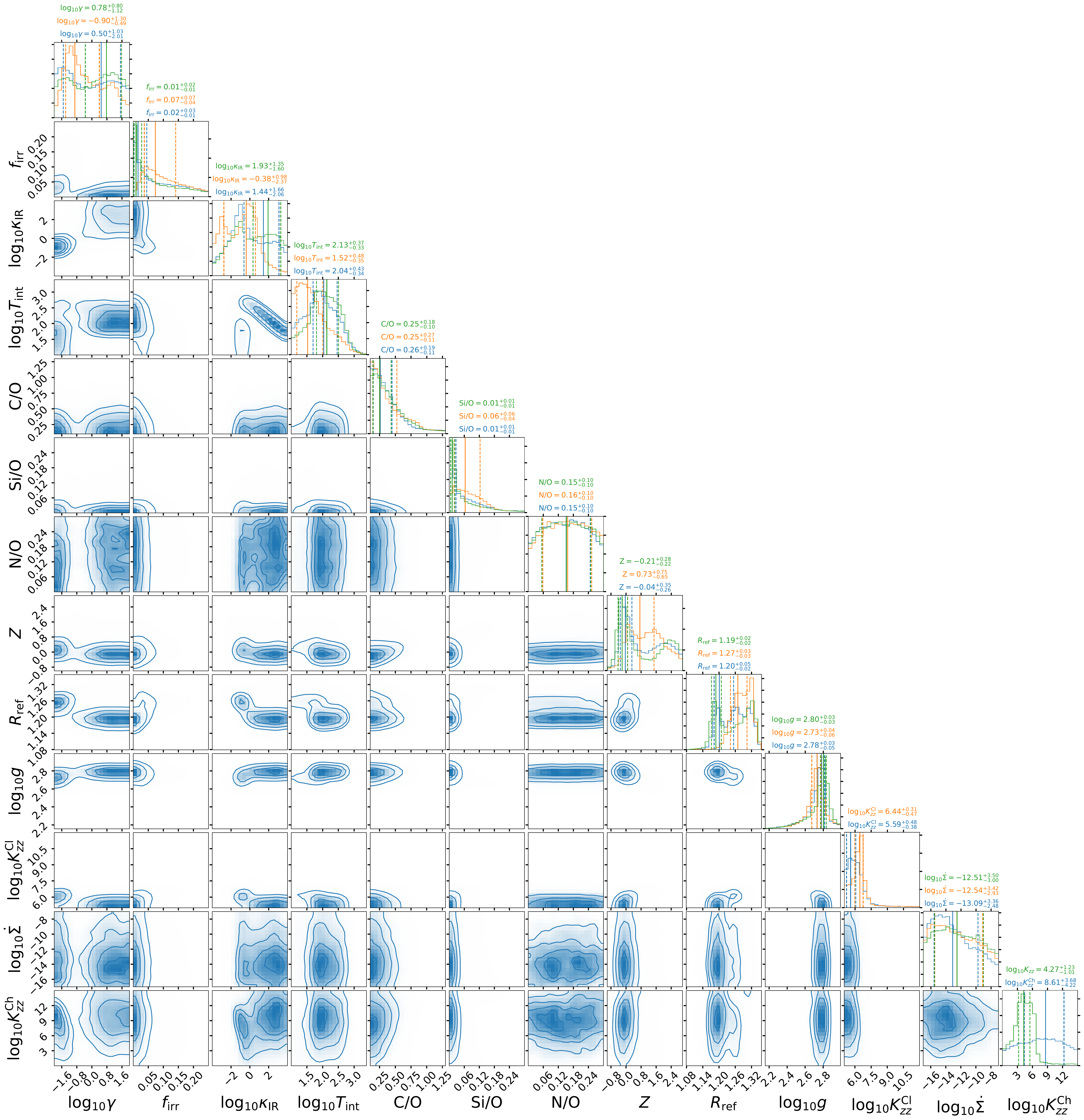}
    \subcaption{WASP-39b}
\end{minipage}
\caption{Continued}
\end{figure*}

\begin{figure*}
\ContinuedFloat
\begin{minipage}{\hsize}
    \centering
    \includegraphics[width=\hsize]{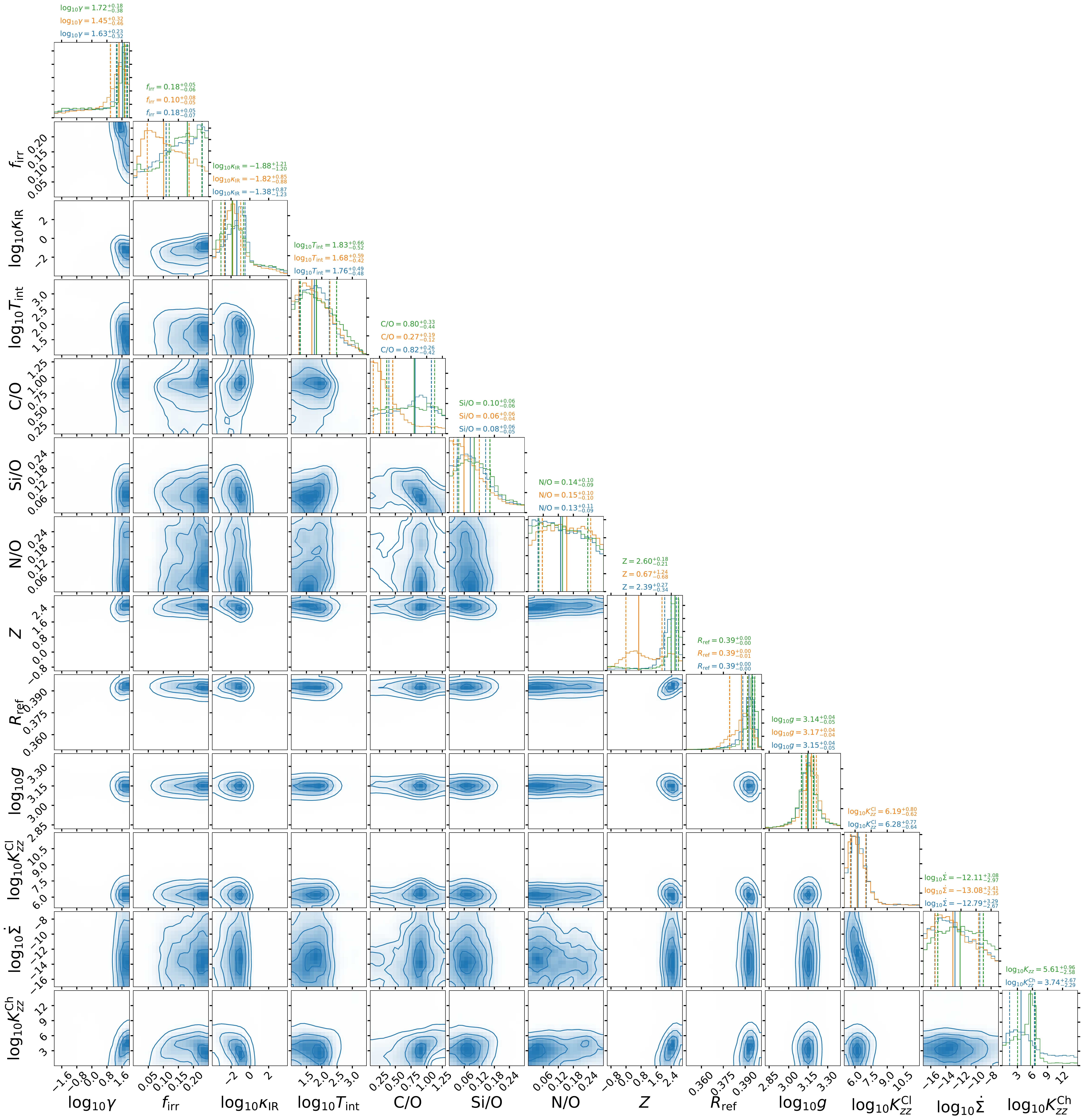}
    \subcaption{HAT-P-11b}
\end{minipage}
\caption{Continued}
\end{figure*}

\begin{figure*}
\ContinuedFloat
\begin{minipage}{\hsize}
    \centering
    \includegraphics[width=\hsize]{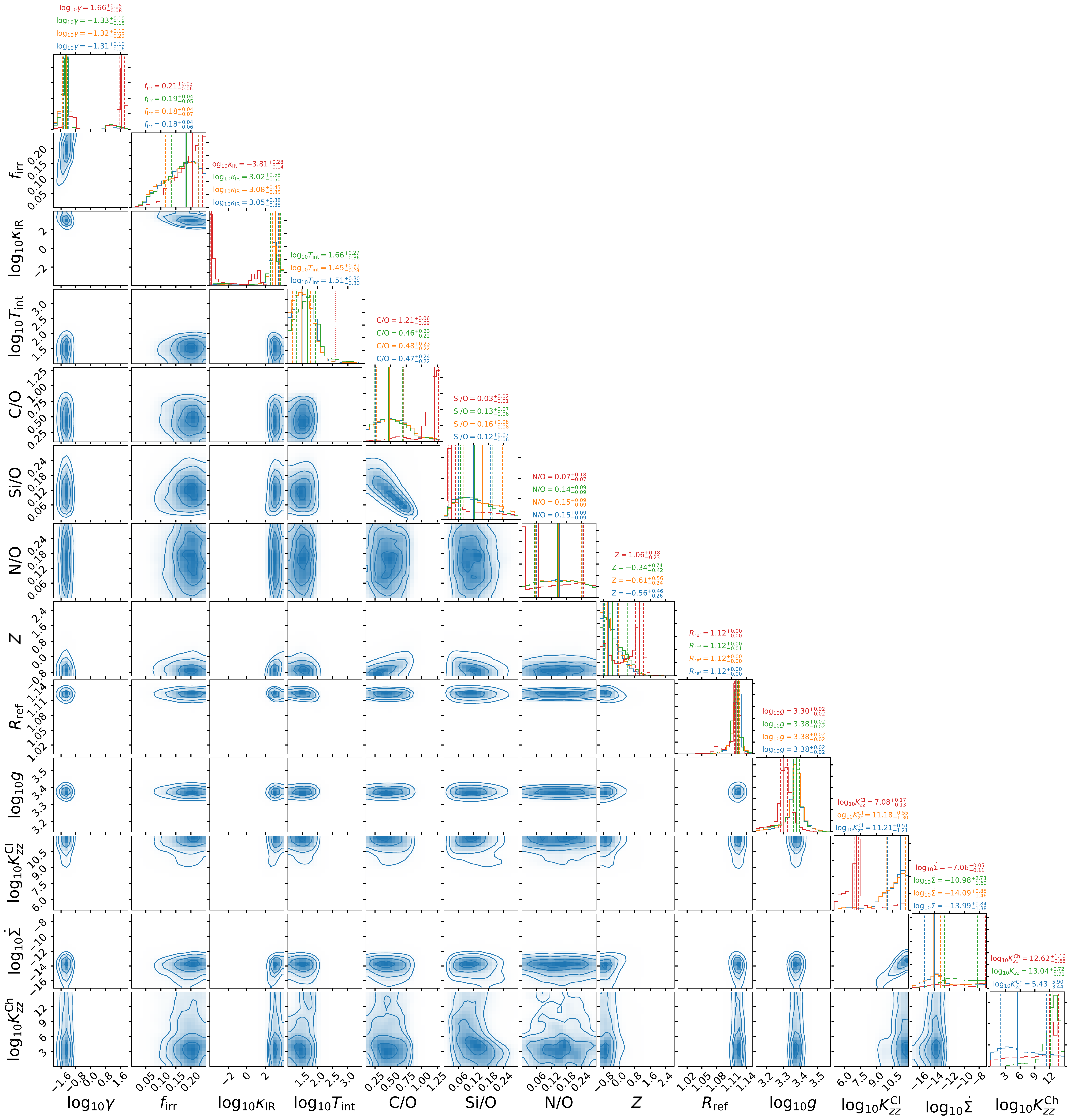}
    \subcaption{HD~189733b}
\end{minipage}
\caption{Continued}
\end{figure*}

\begin{figure*}
\ContinuedFloat
\begin{minipage}{\hsize}
    \centering
    \includegraphics[width=\hsize]{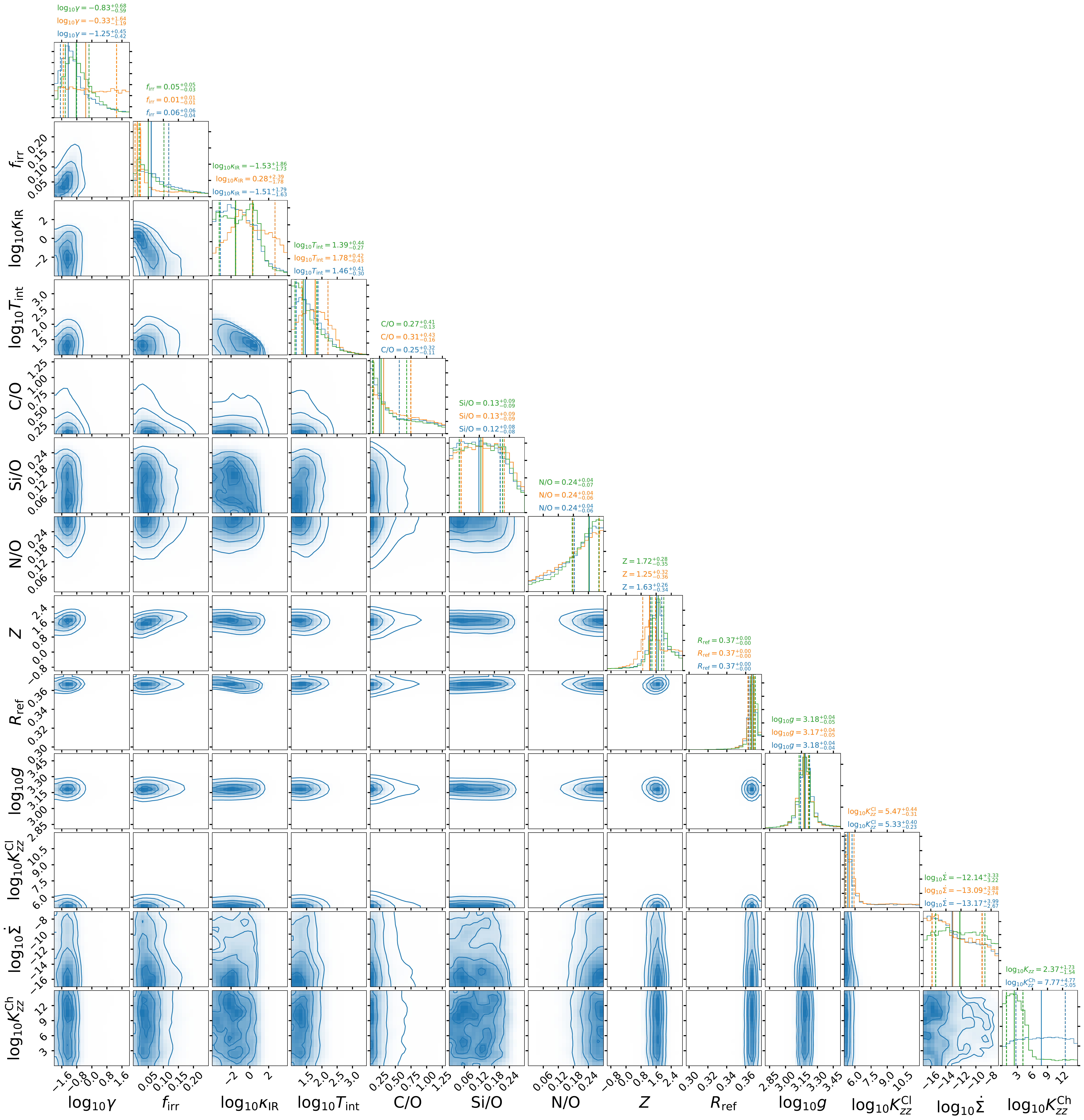}
    \subcaption{GJ~436b}
\end{minipage}
\caption{Continued}
\end{figure*}

\begin{figure*}
\ContinuedFloat
\begin{minipage}{\hsize}
    \centering
    \includegraphics[width=\hsize]{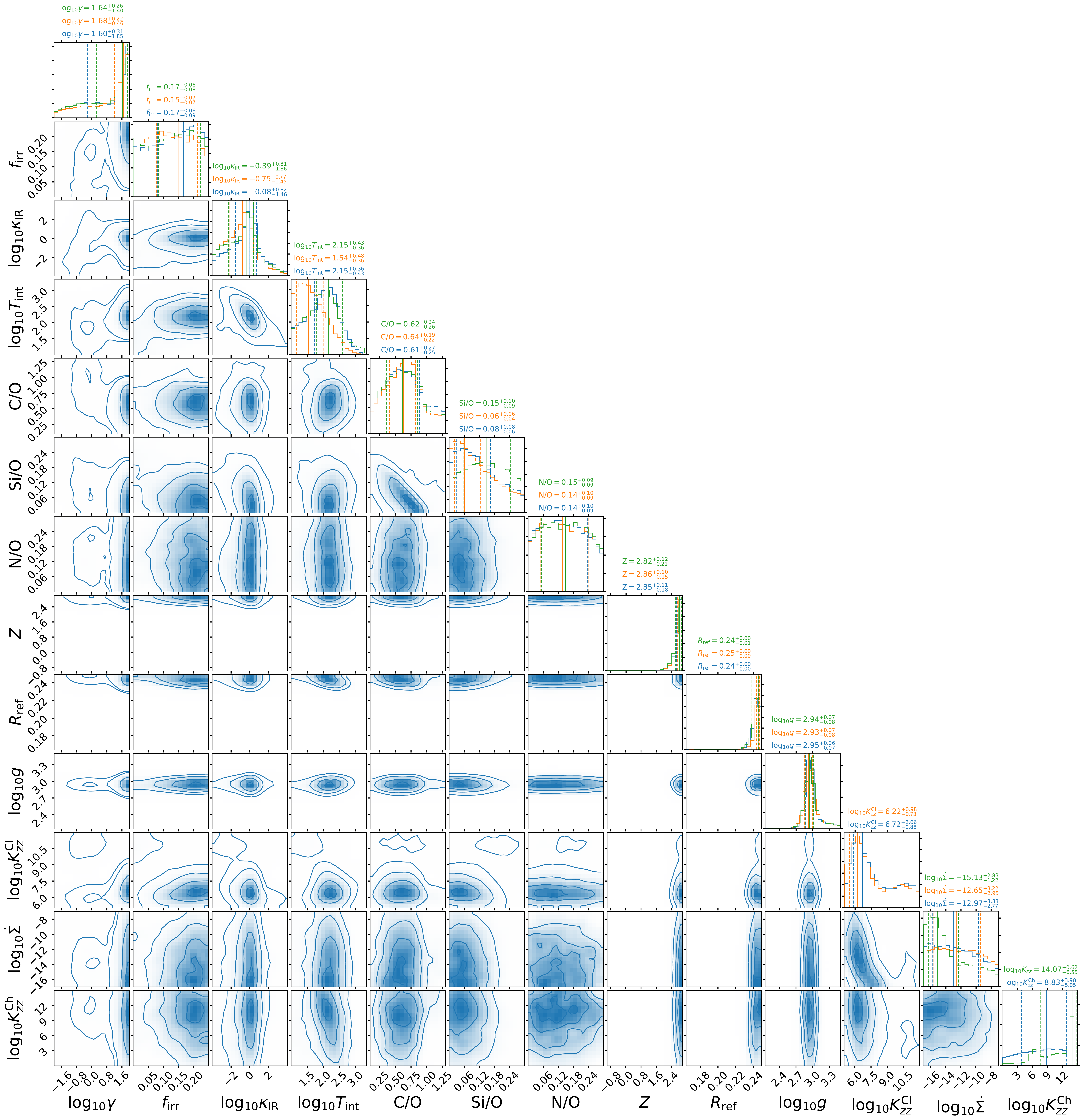}
    \subcaption{GJ~1214b}
\end{minipage}
\caption{Continued}
\end{figure*}

\begin{figure*}
\ContinuedFloat
\begin{minipage}{\hsize}
    \centering
    \includegraphics[width=\hsize]{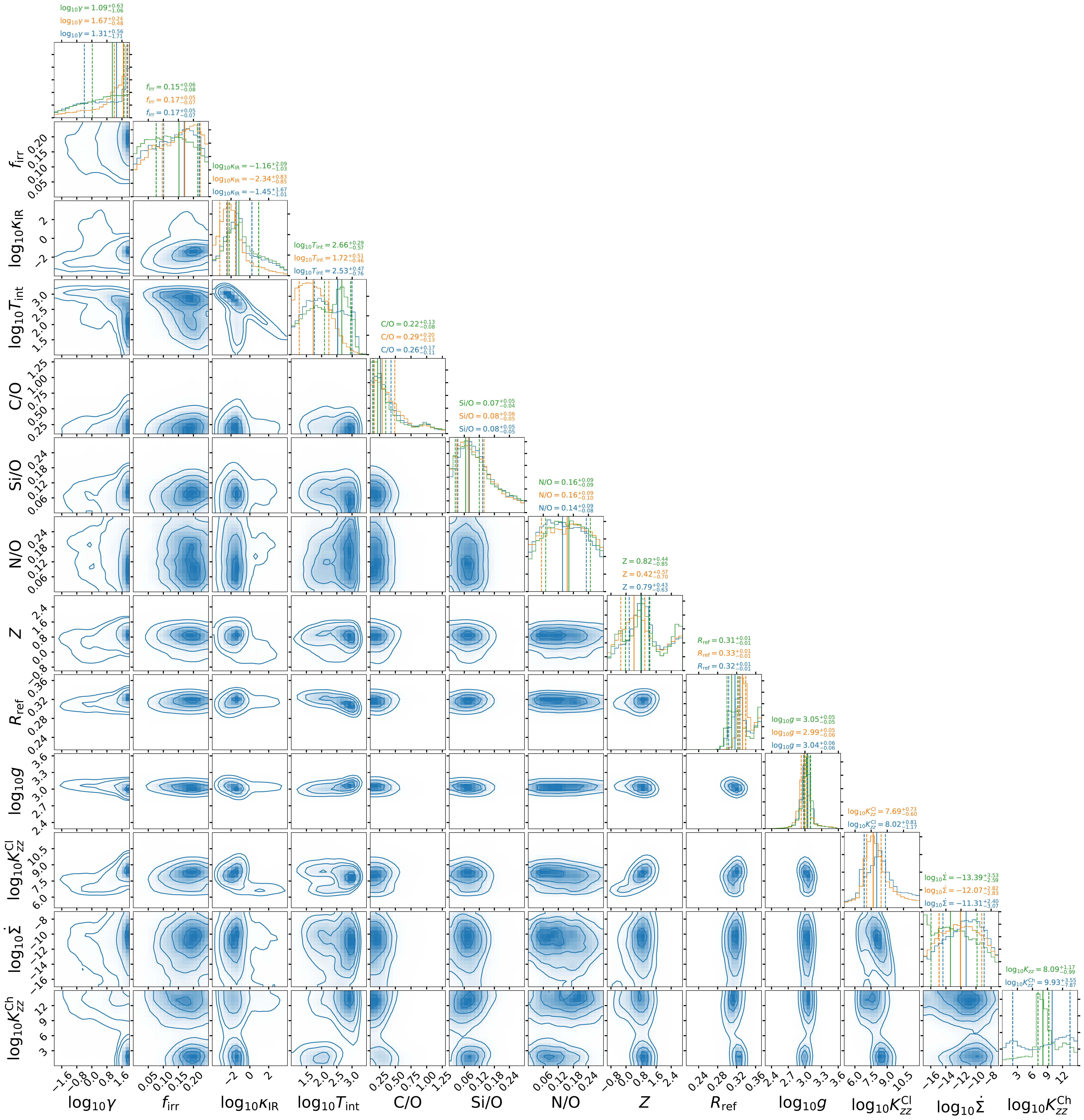}
    \subcaption{GJ~3470b}
\end{minipage}
\caption{Continued}
\end{figure*}

\begin{figure*}
\ContinuedFloat
\begin{minipage}{\hsize}
    \centering
    \includegraphics[width=\hsize]{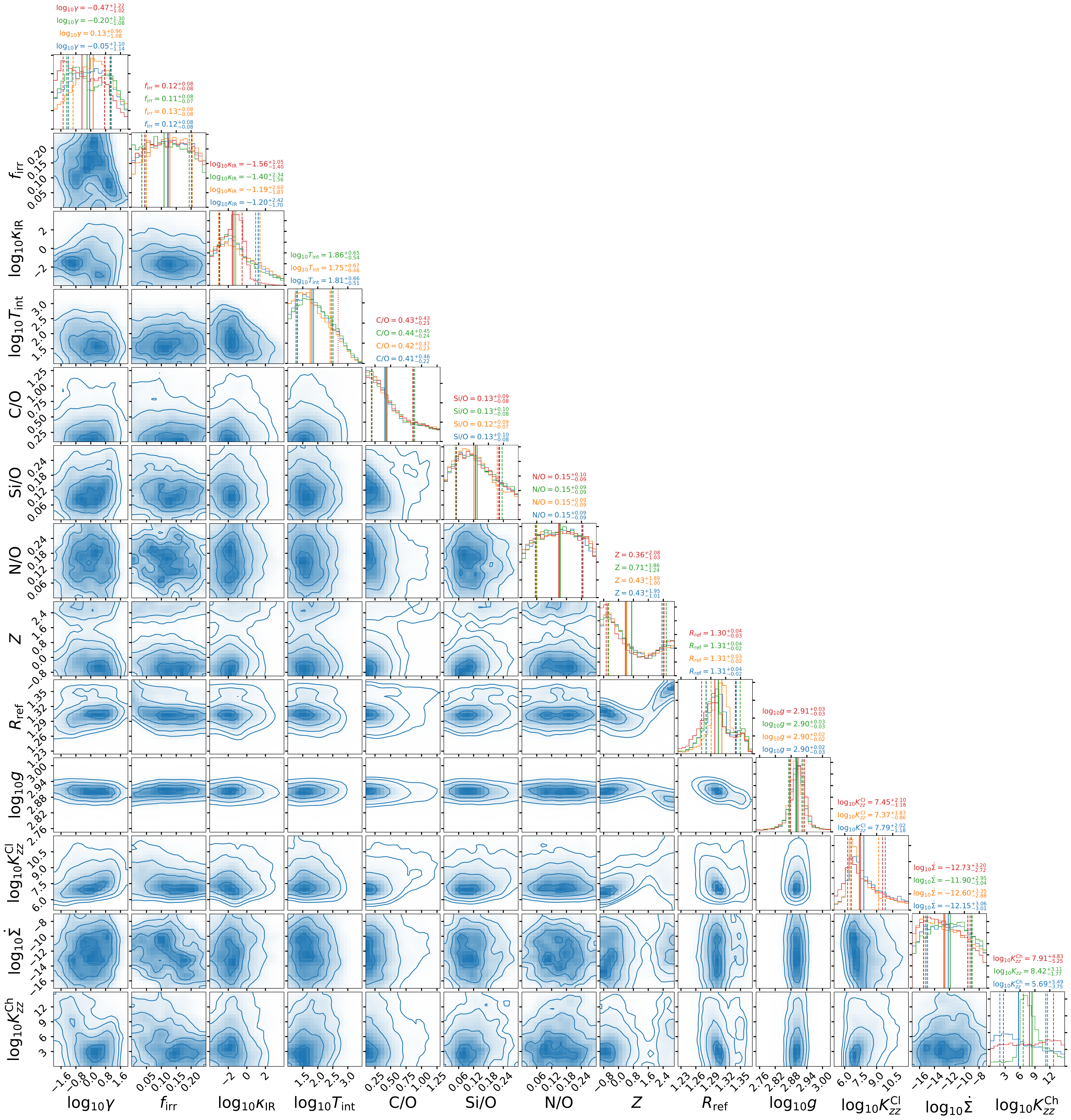}
    \subcaption{HAT-P-1b}
\end{minipage}
\caption{Continued}
\end{figure*}

\begin{figure*}
\ContinuedFloat
\begin{minipage}{\hsize}
    \centering
    \includegraphics[width=\hsize]{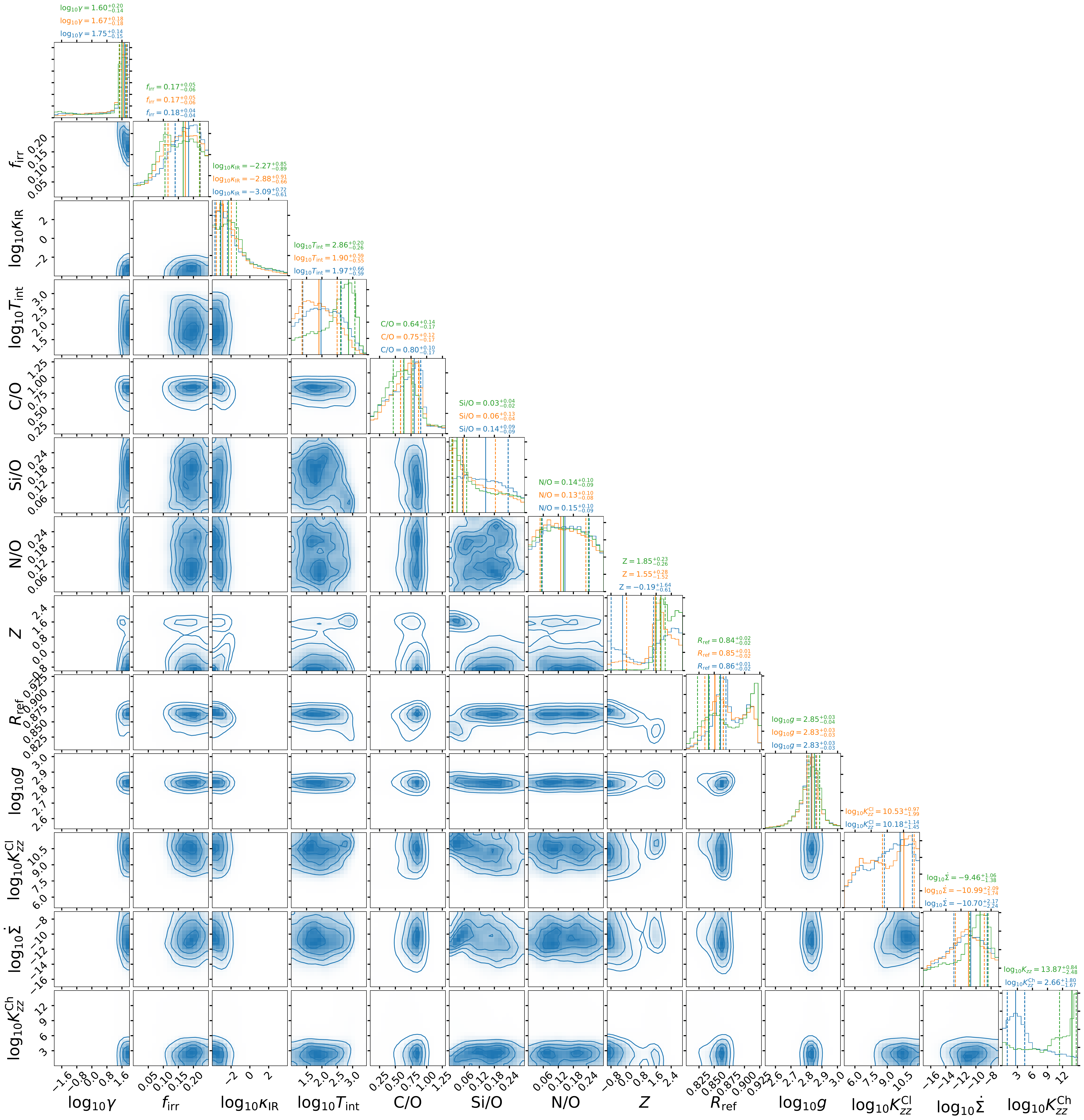}
    \subcaption{HAT-P-12b}
\end{minipage}
\caption{Continued}
\end{figure*}

\begin{figure*}
\ContinuedFloat
\begin{minipage}{\hsize}
    \centering
    \includegraphics[width=\hsize]{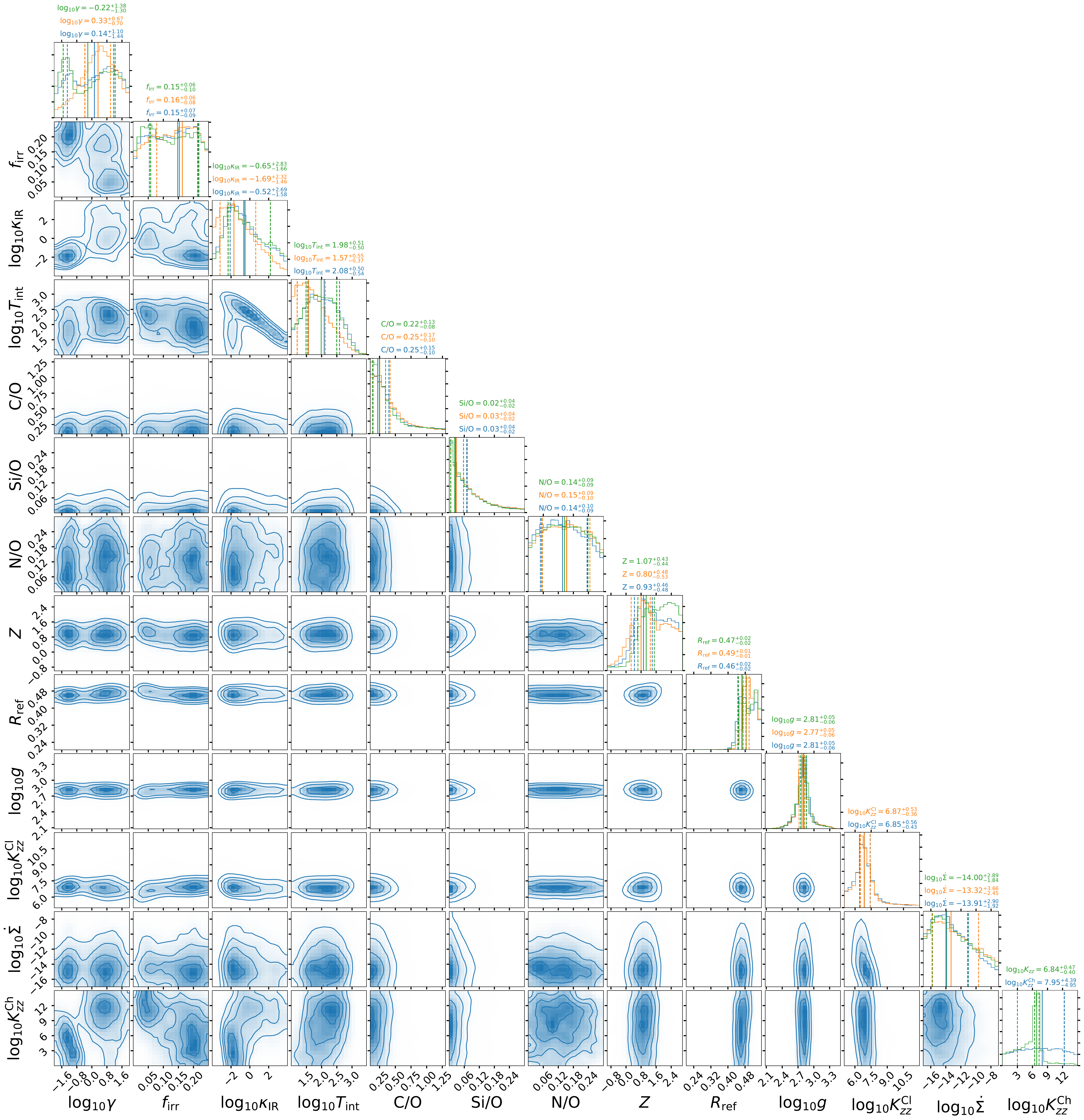}
    \subcaption{HAT-P-26b}
\end{minipage}
\caption{Continued}
\end{figure*}

\begin{figure*}
\ContinuedFloat
\begin{minipage}{\hsize}
    \centering
    \includegraphics[width=\hsize]{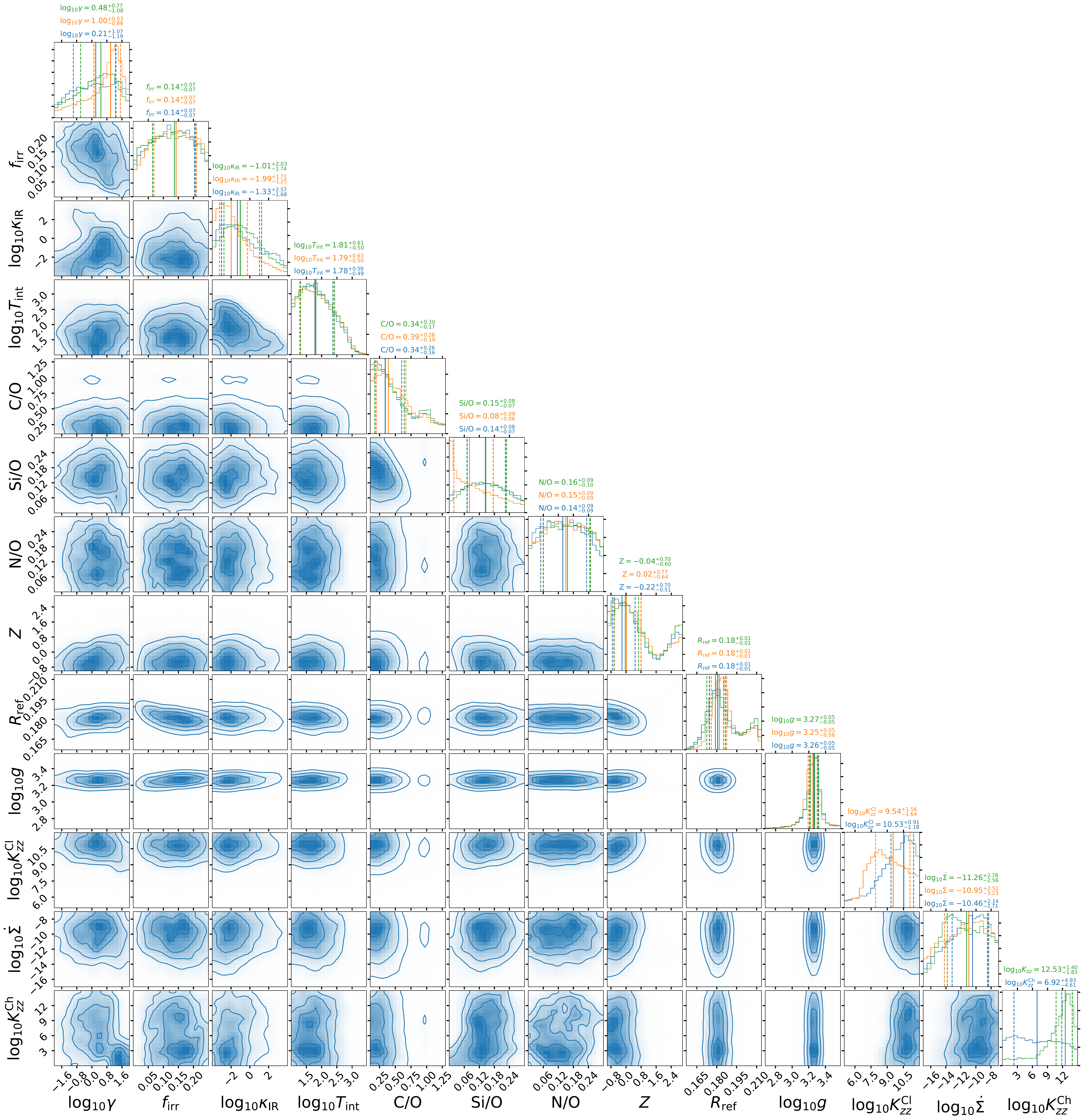}
    \subcaption{HD~97658b}
\end{minipage}
\caption{Continued}
\end{figure*}

\begin{figure*}
\ContinuedFloat
\begin{minipage}{\hsize}
    \centering
    \includegraphics[width=\hsize]{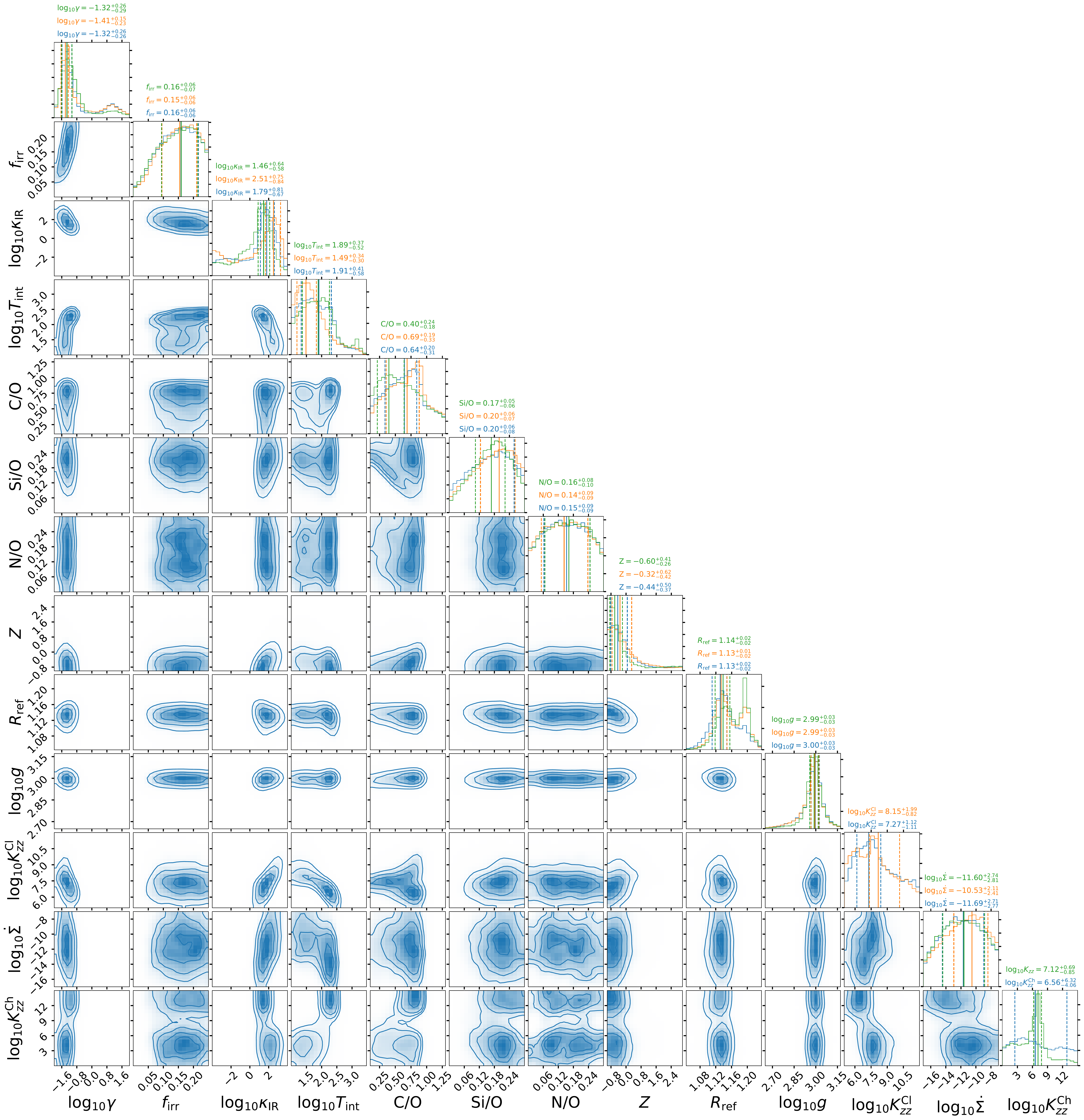}
    \subcaption{WASP-6b}
\end{minipage}
\caption{Continued}
\end{figure*}

\begin{figure*}
\ContinuedFloat
\begin{minipage}{\hsize}
    \centering
    \includegraphics[width=\hsize]{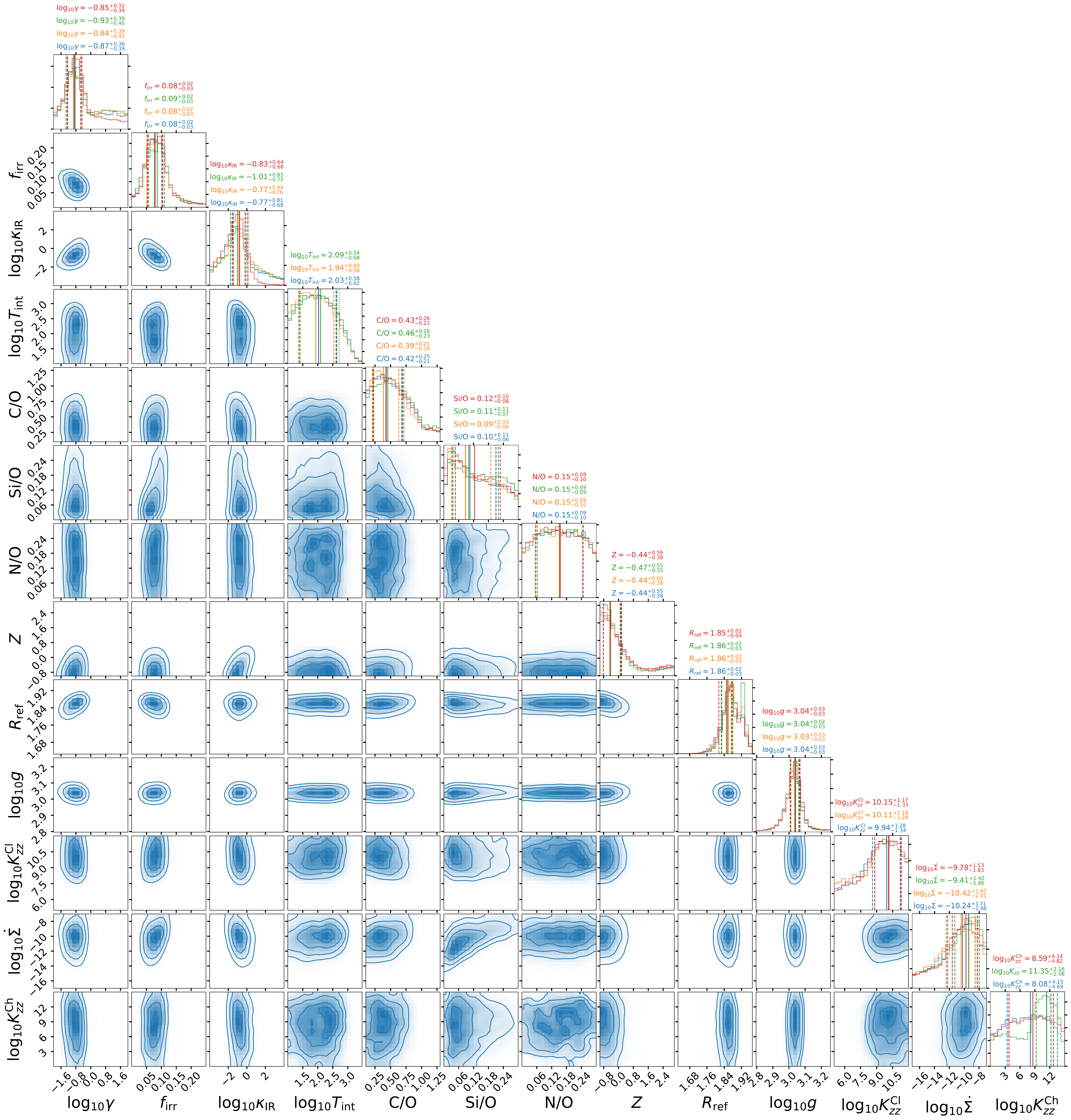}
    \subcaption{WASP-12b}
\end{minipage}
\caption{Continued}
\end{figure*}

\begin{figure*}
\ContinuedFloat
\begin{minipage}{\hsize}
    \centering
    \includegraphics[width=\hsize]{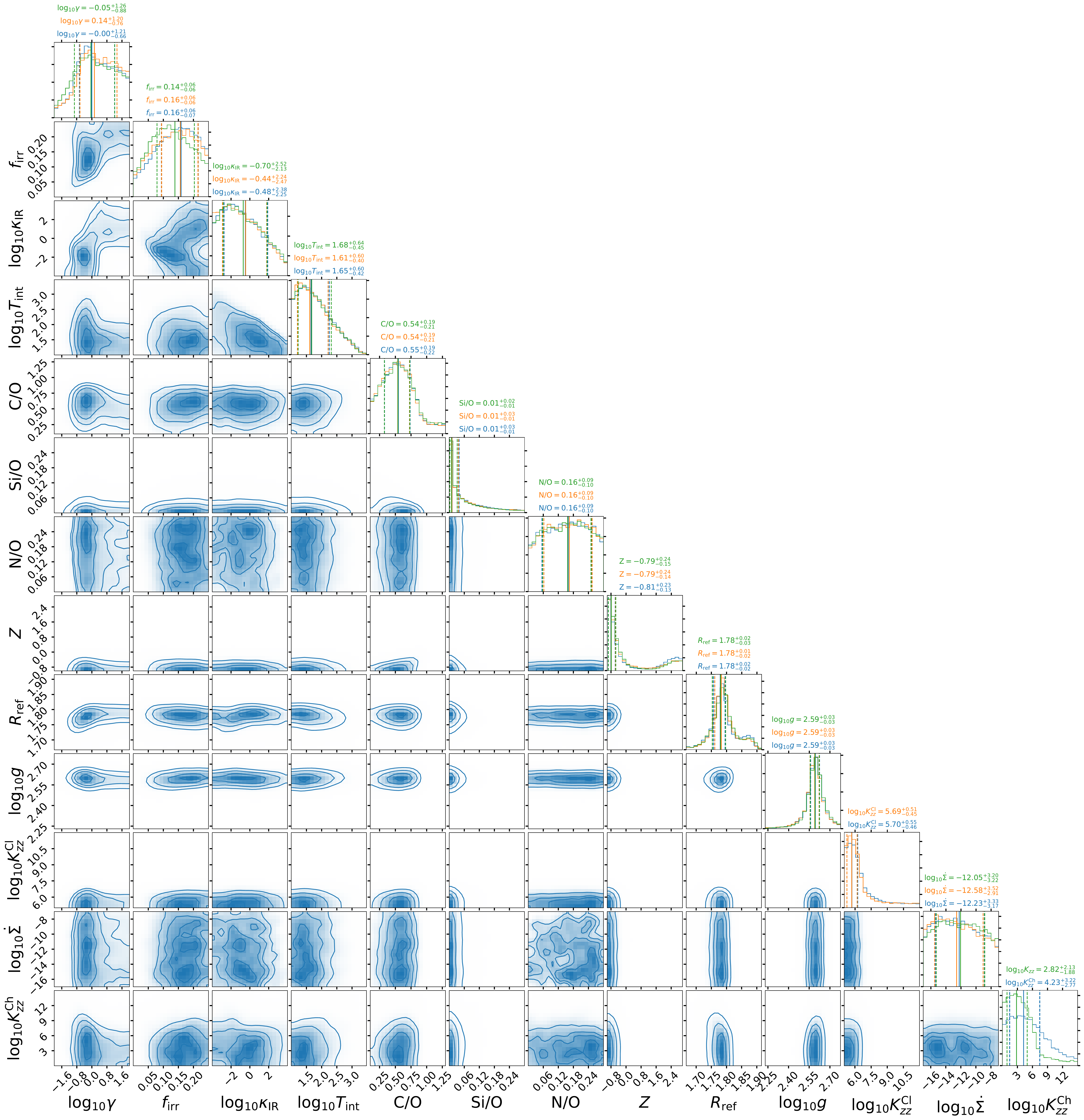}
    \subcaption{WASP-17b}
\end{minipage}
\caption{Continued}
\end{figure*}

\begin{figure*}
\ContinuedFloat
\begin{minipage}{\hsize}
    \centering
    \includegraphics[width=\hsize]{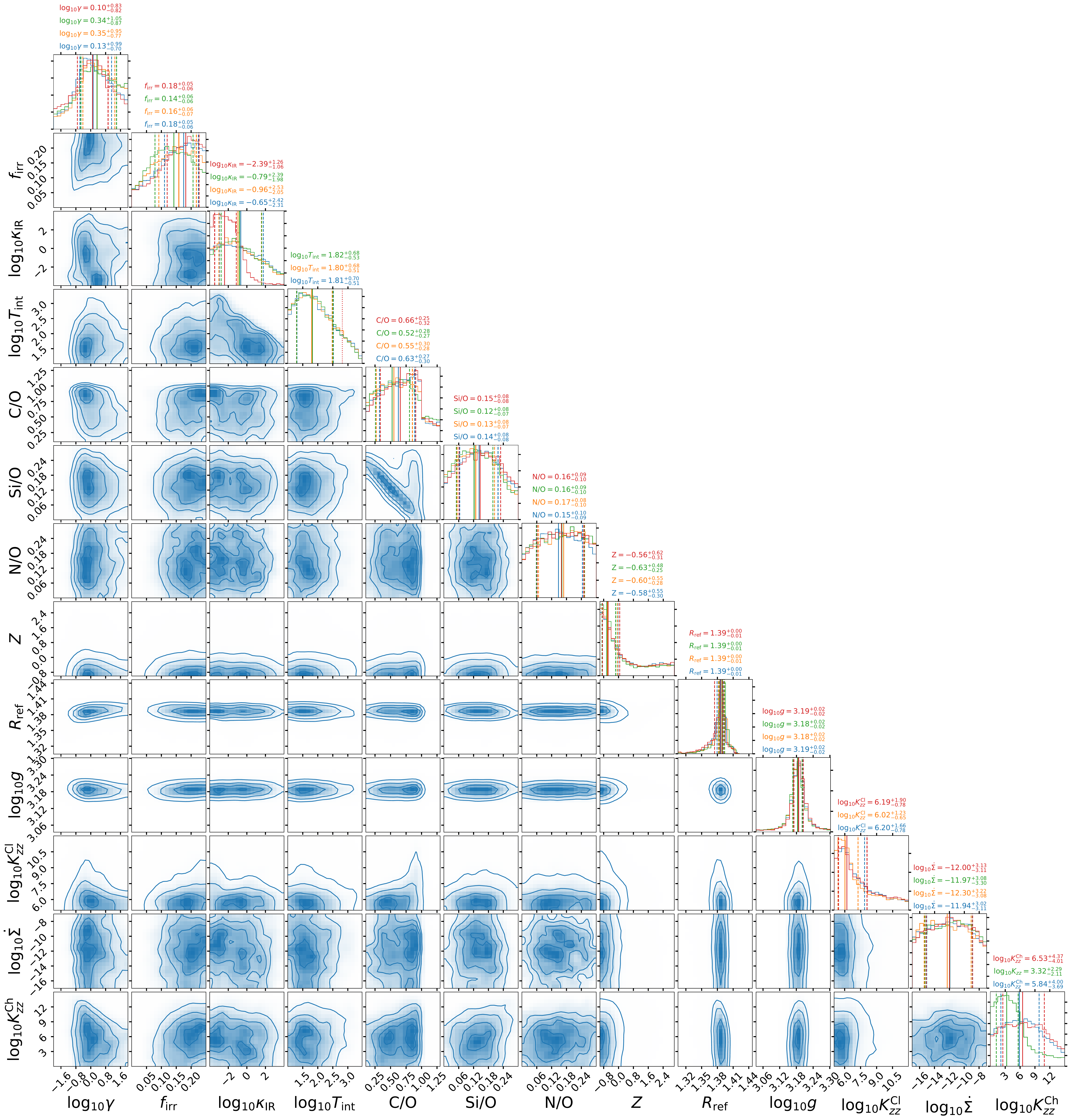}
    \subcaption{WASP-19b}
\end{minipage}
\caption{Continued}
\end{figure*}

\begin{figure*}
\ContinuedFloat
\begin{minipage}{\hsize}
    \centering
    \includegraphics[width=\hsize]{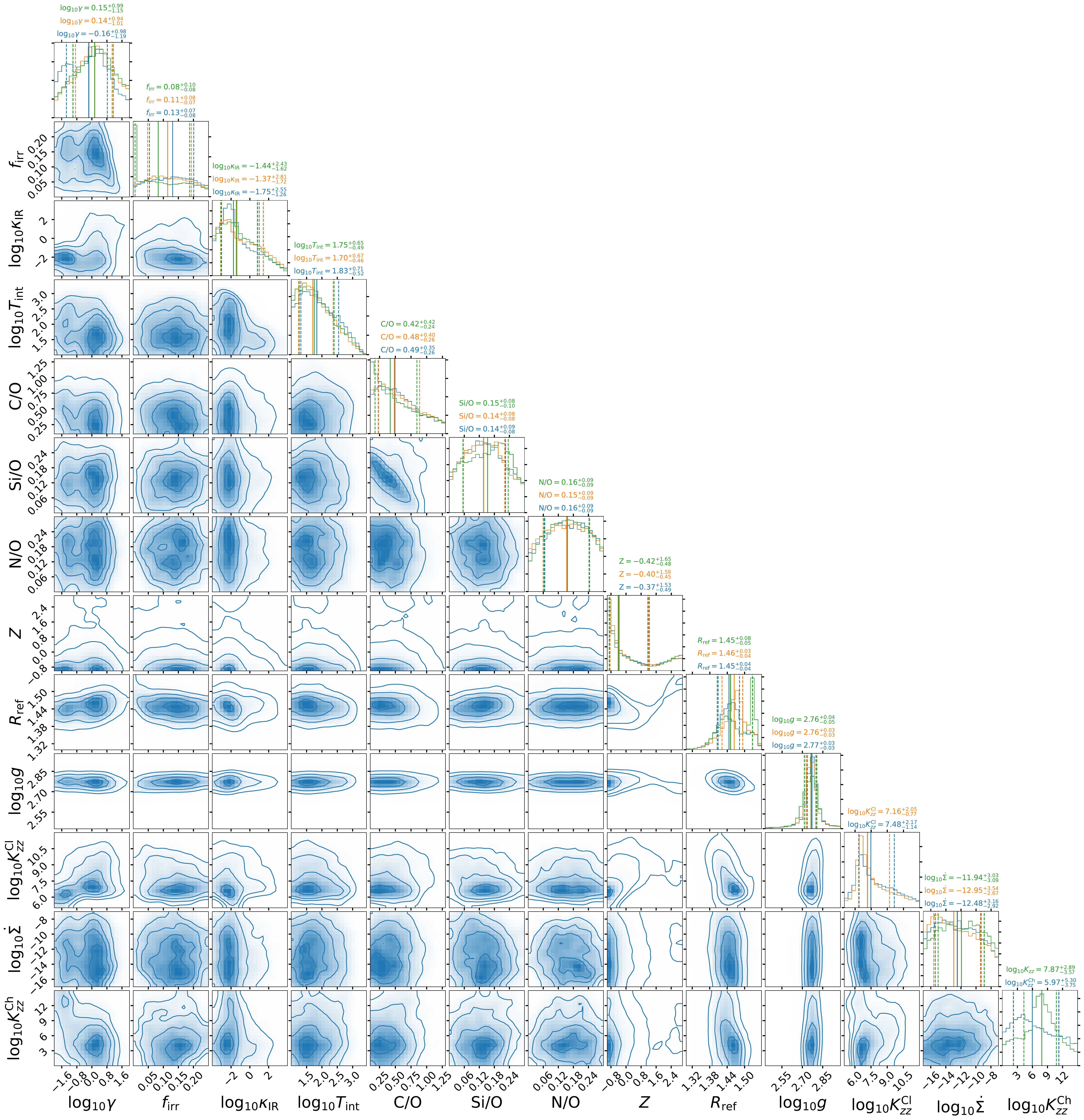}
    \subcaption{WASP-31b}
\end{minipage}
\caption{Continued}
\end{figure*}

\end{appendix}

\end{document}